[ english version / português abaixo ]

# **Antivax and off-label medication communities on brazilian Telegram:** between esotericism as a gateway and the monetization of false miraculous cures

*Ergon Cugler de Moraes Silva*

Brazilian Institute of Information in
Science and Technology (IBICT)
Brasília, Federal District, Brazil

contato@ergoncugler.com
www.ergoncugler.com

**Abstract**

Conspiracy theories, particularly those focused on anti-vaccine narratives and the promotion of off-label medications such as MMS and CDS, have proliferated on Telegram, including in Brazil, finding fertile ground among communities that share esoteric beliefs and distrust towards scientific institutions. In this context, this study seeks to answer **how Brazilian conspiracy theory communities on Telegram are characterized and articulated concerning anti-vaccine themes and off-label medications?** It is important to highlight that this study is part of a series of seven studies aimed at understanding and characterizing Brazilian conspiracy theory communities on Telegram. This series of seven studies is openly and originally available on the arXiv of Cornell University, applying a mirrored method across all studies, changing only the thematic object of analysis and providing replicable research, including proprietary and original codes developed, contributing to the culture of free and open-source software. Regarding the main findings of this study, it was observed: Themes such as the New World Order and Apocalypse and Survivalism act as significant gateways to anti-vaccine narratives, connecting them to theories of global control; Globalism and New World Order stand out as the main communities receiving invitations from anti-vaccine communities; Occultism and Esotericism emerge as the largest sources of invitations to off-label medication communities, creating a strong connection between esoteric beliefs and the promotion of non-scientific treatments; Anti-vaccine narratives experienced a 290% increase during the COVID-19 pandemic, evidencing a growing interconnectedness with other conspiracy theories; The overlap of themes between anti-vaccine and other conspiracy theories creates an interdependent disinformation network, where different narratives mutually reinforce each other.

## Key findings

- ➔ Themes such as the New World Order and Apocalypse and Survivalism are the main gateways to anti-vaccine narratives, demonstrating how these communities connect distrust in global institutions to theories of world control;

- ➔ Globalism, General Conspiracy, and the New World Order stand out as the main communities receiving invitations from anti-vaccine communities, expanding the disinformation network and reinforcing global conspiracy theories;

- ➔ Occultism and Esotericism communities emerge as the largest sources of invitations leading to off-label medication communities, reflecting a strong connection between esoteric beliefs and the promotion of non-scientific treatments;
- ➔ Occultism and Esotericism, along with Globalism and the New World Order, lead in the number of invitations received from off-label medication communities, creating a powerful intersection between disinformation and conspiracy theories;
- ➔ Anti-vaccine narratives experienced a 290% increase during the COVID-19 pandemic, demonstrating a growing interconnectedness with other conspiracy theories and the use of the health crisis to promote disinformation agendas;
- ➔ The overlap of themes between anti-vaccine and other conspiracy theories feeds an interdependent disinformation network, where different narratives mutually reinforce each other, creating a continuous cycle of false beliefs;
- ➔ The dishonest narrative linking vaccines to autism remains one of the most persistent, being frequently reintroduced in discussions within anti-vaccine and off-label medication communities, furthering distrust in conventional medicine;
- ➔ Anti-vaccine communities act as central hubs in the dissemination of disinformation, connecting multiple conspiracy narratives and amplifying their reach, making them powerful points of convergence within the disinformation network;
- ➔ The interconnectedness between anti-vaccine discussions and off-label medications reveals a cohesive and resilient ideological bubble, where the overlap of beliefs and ideological agendas facilitates the spread of disinformation and makes factual intervention challenging;
- ➔ The intersection between health-related disinformation and esoteric narratives creates a highly influential disinformation network, attracting new members by blending global conspiracy theories with alternative beliefs, increasing the reach of these communities. Additionally, it creates a network that monetizes the sale of unregulated drugs and chemical products containing chlorine dioxide, also known as “vaccine detox”, “MMS”, and “CDS”.

## 1. Introduction

After navigating through thousands of Brazilian conspiracy theory communities on Telegram, extracting tens of millions of pieces of content from these communities, created and/or shared by millions of users, this study aims to form part of a series of seven studies focused on the phenomenon of conspiracy theories on Telegram, using Brazil as a case study. Through the implemented identification approaches, it was possible to reach a total of 195 Brazilian conspiracy theory communities on Telegram, focused on anti-vaccine (antivax) and alternative treatments and medications (off-label), totaling 5,406,762 published content pieces from May 2017 (first publications) to August 2024 (the period of this study), with a combined total of 440,651 users across these communities. In this context, this study aims to understand and characterize the communities focused on anti-vaccine themes and alternative treatments and medications within this identified Brazilian conspiracy theory network on Telegram.

To achieve this, a mirrored method will be applied across all seven studies, changing only the thematic object of analysis and providing replicable research. Thus, we will address techniques for observing the connections, time series, content, and thematic overlaps of conspiracy theory communities. In addition to this study, the remaining six are openly and originally available on the arXiv of Cornell University. This series has been conducted with heightened attention to ensure data integrity and respect for user privacy, as provided by Brazilian legislation (Law No. 13,709/2018).

Therefore, the question arises: **How are Brazilian conspiracy theory communities on Telegram characterized and articulated concerning anti-vaccine (antivax) themes and alternative treatments and medications (off-label)?**

## 2. Materials and methods

The methodology of this study is organized into three subsections: **2.1. Data extraction**, which describes the process and tools used to collect information from Telegram communities; **2.2. Data processing**, which discusses the criteria and methods applied to classify and anonymize the collected data; and **2.3. Approaches to data analysis**, which details the techniques used to investigate the connections, temporal series, content, and thematic overlaps within conspiracy theory communities.

### 2.1. Data extraction

This project began in February 2023 with the publication of the first version of TelegramScrap (Silva, 2023), a proprietary, free, and open-source tool that utilizes Telegram's Application Programming Interface (API) by Telethon library and organizes data extraction cycles from groups and open channels on Telegram. Over the months, the database was expanded and refined using four approaches to identifying conspiracy theory communities:

**(i) Use of keywords:** at the project's outset, keywords were listed for direct identification in the search engine of Brazilian groups and channels on Telegram, such as "apocalypse", "survivalism", "climate change", "flat earth", "conspiracy theory", "globalism", "new world order", "occultism", "esotericism", "alternative cures", "qAnon" "reptilians", "revisionism", "aliens", among others. This initial approach provided some communities whose titles and/or descriptions of groups and channels explicitly contained terms related to conspiracy theories. However, over time, it was possible to identify many other communities that the listed keywords did not encompass, some of which deliberately used altered characters to make it difficult for those searching for them on the network.

**(ii) Telegram channel recommendation mechanism:** over time, it was identified that Telegram channels (except groups) have a recommendation tab called "similar channels", where Telegram itself suggests ten channels that have some similarity with the channel being observed. Through this recommendation mechanism, it was possible to find more Brazilian conspiracy theory communities, with these being recommended by the platform itself.

**(iii) Snowball approach for invitation identification:** after some initial communities were accumulated for extraction, a proprietary algorithm was developed to identify URLs containing "t.me/", the prefix for any invitation to Telegram groups and channels. Accumulating a frequency of hundreds of thousands of links that met this criterion, the unique addresses were listed, and their repetitions counted. In this way, it was possible to investigate new Brazilian groups and channels mentioned in the messages of those already investigated, expanding the network. This process was repeated periodically to identify new communities aligned with conspiracy theory themes on Telegram.

**(iv) Expansion to tweets published on X mentioning invitations:** to further diversify the sources of Brazilian conspiracy theory communities on Telegram, a proprietary search query was developed to identify conspiracy theory-themed keywords using tweets published on X (formerly Twitter) that, in addition to containing one of the keywords, also included "t.me/", the prefix for any invitation to Telegram groups and channels, "https://x.com/search?q=lang%3Apt%20%22t.me%2F%22%20SEARCH-TERM".

With the implementation of community identification approaches for conspiracy theories developed over months of investigation and method refinement, it was possible to build a project database encompassing a total of 855 Brazilian conspiracy theory communities on Telegram (including other themes not covered in this study). These communities have collectively published 27,227,525 pieces of content from May 2016 (the first publications) to August 2024 (the period of this study), with a combined total of 2,290,621 users across the Brazilian communities. It is important to note that this volume of users includes two elements: first, it is a variable figure, as users can join and leave communities daily, so this value represents what was recorded on the publication extraction date; second, it is possible that the same user is a member of more than one group and, therefore, is counted more than once. In this context, while the volume remains significant, it may be slightly lower when considering the deduplicated number of citizens within these Brazilian conspiracy theory communities.

### 2.2. Data processing

With all the Brazilian conspiracy theory groups and channels on Telegram extracted, a manual classification was conducted considering the title and description of the community. If there was an explicit mention in the title or description of the community related to a specific theme, it was classified into one of the following categories: (i) "Anti-Science"; (ii) "Anti-Woke and Gender"; (iii) "Antivax"; (iv) "Apocalypse and Survivalism"; (v) "Climate Changes"; (vi) "Flat Earth"; (vii) "Globalism"; (viii) "New World Order"; (ix) "Occultism and Esotericism"; (x) "Off Label and Quackery"; (xi) "QAnon"; (xii) "Reptilians and Creatures"; (xiii) "Revisionism and Hate Speech"; (xiv) "UFO and Universe". If there was no explicit mention related to the themes in the title or description of the community, it was classified as (xv) "General Conspiracy". In the following table, we can observe the metrics related to the classification of these conspiracy theory communities in Brazil.

**Table 01.** Conspiracy theory communities in Brazil (metrics up to August 2024)

| Categories | Groups | Users | Contents | Comments | Total |
|---|---|---|---|---|---|
| **Anti-Science** | 22 | 58,138 | 187,585 | 784,331 | 971,916 |
| **Anti-Woke and Gender** | 43 | 154,391 | 276,018 | 1,017,412 | 1,293,430 |
| **Antivax** | 111 | 239,309 | 1,778,587 | 1,965,381 | 3,743,968 |
| **Apocalypse and Survivalism** | 33 | 109,266 | 915,584 | 429,476 | 1,345,060 |
| **Climate Changes** | 14 | 20,114 | 269,203 | 46,819 | 316,022 |
| **Flat Earth** | 33 | 38,563 | 354,200 | 1,025,039 | 1,379,239 |
| **General Conspiracy** | 127 | 498,190 | 2,671,440 | 3,498,492 | 6,169,932 |
| **Globalism** | 41 | 326,596 | 768,176 | 537,087 | 1,305,263 |
| **NWO** | 148 | 329,304 | 2,411,003 | 1,077,683 | 3,488,686 |
| **Occultism and Esotericism** | 39 | 82,872 | 927,708 | 2,098,357 | 3,026,065 |
| **Off Label and Quackery** | 84 | 201,342 | 929,156 | 733,638 | 1,662,794 |
| **QAnon** | 28 | 62,346 | 531,678 | 219,742 | 751,420 |
| **Reptilians and Creatures** | 19 | 82,290 | 96,262 | 62,342 | 158,604 |
| **Revisionism and Hate Speech** | 66 | 34,380 | 204,453 | 142,266 | 346,719 |
| **UFO and Universe** | 47 | 58,912 | 862,358 | 406,049 | 1,268,407 |
| **Total** | **855** | **2,296,013** | **13,183,411** | **14,044,114** | **27,227,525** |

Source: Own elaboration (2024).

With this volume of extracted data, it was possible to segment and present in this study only the communities and content related to anti-vaccine themes and alternative treatments and medications (off-label). In parallel, other themes within Brazilian conspiracy theory communities have also been studied to characterize the extent and dynamics of the network, and these studies are openly and originally available on the arXiv of Cornell University.

Additionally, it should be noted that only open communities were extracted, meaning those that are not only publicly identifiable but also do not require any request to access the content, being available to any user with a Telegram account who needs to join the group or channel. Furthermore, in compliance with Brazilian legislation, particularly the General Data Protection Law (LGPD), or Law No. 13,709/2018 (Brazilian law from 2018), which deals with privacy control and the use/treatment of personal data, all extracted data were anonymized for the purposes of analysis and investigation. Therefore, not even the identification of the communities is possible through this study, thus extending the user's privacy by anonymizing their data beyond the community itself to which they submitted by being in a public and open group or channel on Telegram.

### 2.3. Approaches to data analysis

A total of 195 selected communities focused on anti-vaccine (ativax) themes and alternative treatments and medications (off-label), containing 5,406,762 publications and 440,651 combined users, will be analyzed. Four approaches will be used to investigate the conspiracy theory communities selected for the scope of this study. These metrics are detailed in the following table:

**Table 02.** Selected communities for analysis (metrics up to August 2024)

| Categories | Groups | Users | Contents | Comments | Total |
|---|---|---|---|---|---|
| **Antivacinas (*Antivax*)** | 111 | 239,309 | 1,778,587 | 1,965,381 | 3,743,968 |
| **Medicamentos *Off Label*** | 84 | 201,342 | 929,156 | 733,638 | 1,662,794 |
| **Total** | **195** | **440,651** | **2,707,743** | **2,699,019** | **5,406,762** |

Source: Own elaboration (2024).

**(i) Network:** by developing a proprietary algorithm to identify messages containing the term "t.me/" (inviting users to join other communities), we propose to present volumes and connections observed on how **(a)** communities within the anti-vaccine (antivax) themes and alternative treatments and medications (off-label) circulate invitations for their users to explore more groups and channels within the same theme, reinforcing shared belief systems; and how **(b)** these same communities circulate invitations for their users to explore groups and channels dealing with other conspiracy theories, distinct from their explicit purpose. This approach is valuable for observing whether these communities use themselves as a source of legitimation and reference and/or rely on other conspiracy theory themes, even opening doors for their users to explore other conspiracies. Furthermore, it is worth mentioning the study by Rocha *et al.* (2024), where a network identification approach was also applied in Telegram communities, but by observing similar content based on an ID generated for each unique message and its similar ones;

**(ii) Time series:** the "Pandas" library (McKinney, 2010) is used to organize the investigation data frames, observing **(a)** the volume of publications over the months; and **(b)** the volume of engagement over the months, considering metadata of views, reactions, and comments collected during extraction. In addition to volumetry, the "Plotly" library (Plotly Technologies Inc., 2015) enabled the graphical representation of this variation;

**(iii) Content analysis:** in addition to the general word frequency analysis, time series are applied to the variation of the most frequent words over the semesters—observing from May 2017 (initial publications) to August 2024 (when this study was conducted). With the "Pandas" (McKinney, 2010) and "WordCloud" (Mueller, 2020) libraries, the results are presented both volumetrically and graphically;

**(iv) Thematic agenda overlap:** following the approach proposed by Silva & Sátiro (2024) for identifying thematic agenda overlap in Telegram communities, we used the "BERTopic" model (Grootendorst, 2020). BERTopic is a topic modeling algorithm that facilitates the generation of thematic representations from large amounts of text. First, the algorithm extracts document embeddings using sentence transformer models, such as "all-MiniLM-L6-v2". These embeddings are then reduced in dimensionality using techniques like "UMAP", facilitating the clustering process. Clustering is performed using "HDBSCAN", a density-based technique that identifies clusters of different shapes and sizes, as well as outliers. Subsequently, the documents are tokenized and represented in a bag-of-words structure, which is normalized (L1) to account for size differences between clusters. The topic representation is refined using a modified version of "TF-IDF", called "Class-TF-IDF", which considers the importance of words within each cluster (Grootendorst, 2020). It is important to note that before applying BERTopic, we cleaned the dataset by removing Portuguese "stopwords" using "NLTK" (Loper & Bird, 2002). For topic modeling, we used the "loky" backend to optimize performance during data fitting and transformation.

In summary, the methodology applied ranged from data extraction using the own tool TelegramScrap (Silva, 2023) to the processing and analysis of the collected data, employing various approaches to identify and classify Brazilian conspiracy theory communities on Telegram. Each stage was carefully implemented to ensure data integrity and respect for user privacy, as mandated by Brazilian legislation. The results of this data will be presented below, aiming to reveal the dynamics and characteristics of the studied communities.

## 3. Results

The results are detailed below in the order outlined in the methodology, beginning with the characterization of the network and its sources of legitimation and reference, progressing to the time series, incorporating content analysis, and concluding with the identification of thematic agenda overlap among the conspiracy theory communities.

### 3.1. Network

The following figures provide a comprehensive analysis of the networks of communities that connect around conspiratorial narratives. These networks reveal the complexity of interactions and how narratives are strategically reinforced and amplified through cross-invitations between various thematic groups. The figures illustrate how certain communities function as central hubs, playing a crucial role in the introduction and dissemination of disinformation, connecting seemingly distinct topics that together form a cohesive ecosystem of conspiracy theories.

Specifically, we will analyze how themes such as Anti-vaccine, New World Order, Occultism and Esotericism, and Globalism not only centralize discussions but also act as springboards that guide members to a broader spectrum of disinformation, continually reinforcing beliefs and expanding the reach of these theories. Moreover, the flow of invitation links between these communities indicates a pattern of interdependence, where health-related disinformation is closely linked to narratives of global control, apocalyptic crises, and manipulation by hidden elites.

**Figure 01.** Internal network between anti-vaccine and off-label medication communities

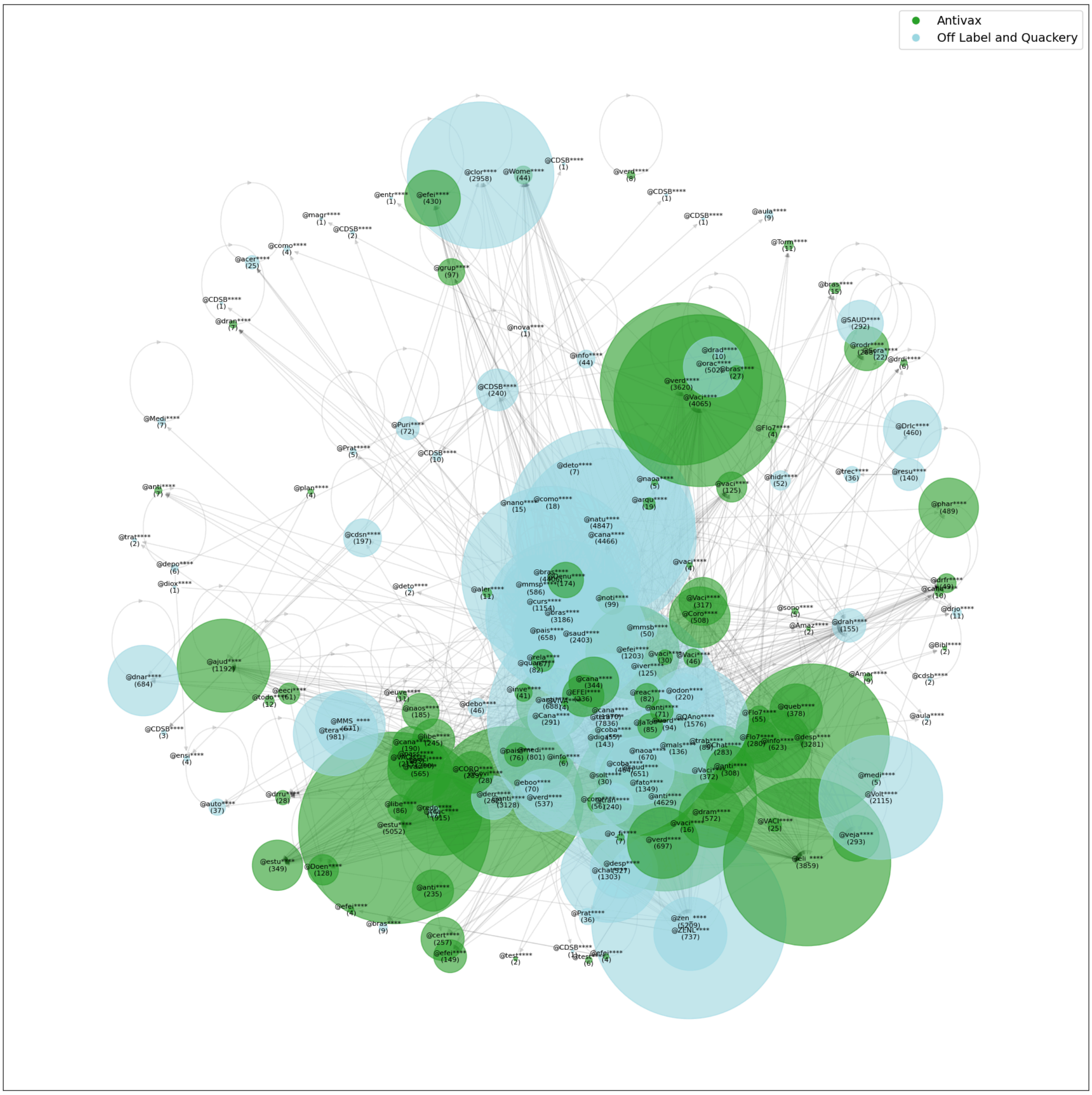

Source: Own elaboration (2024).

The figure presents a detailed analysis of the internal network connecting anti-vaccine communities with those promoting the use of off-label medications. The robustness of the connections within this network suggests a strong interdependence between these themes, where narratives about the dangers of vaccines are frequently reinforced by claims about the effectiveness of alternative and unconventional treatments. This interdependence creates an environment where followers are constantly exposed to information that validates their anti-vaccine beliefs while offering solutions in the form of off-label medications. The large nodes representing the main communities within this network indicate that certain communities act as opinion leaders or dissemination centers, where information and disinformation are widely shared and amplified. The density and cohesion of this internal network reflect the creation of a self-sustaining ecosystem, where health-related conspiratorial narratives feed off each other, creating a vicious cycle of disinformation. This dynamic suggests that once inside this network, it is difficult for followers to break away from these beliefs, as they are continuously exposed to narratives that reinforce their distorted worldview, making the community a closed space of radicalization and resistance to contrary or evidence-based information.

**Figure 02.** Network of communities that open doors to the theme (gateway)

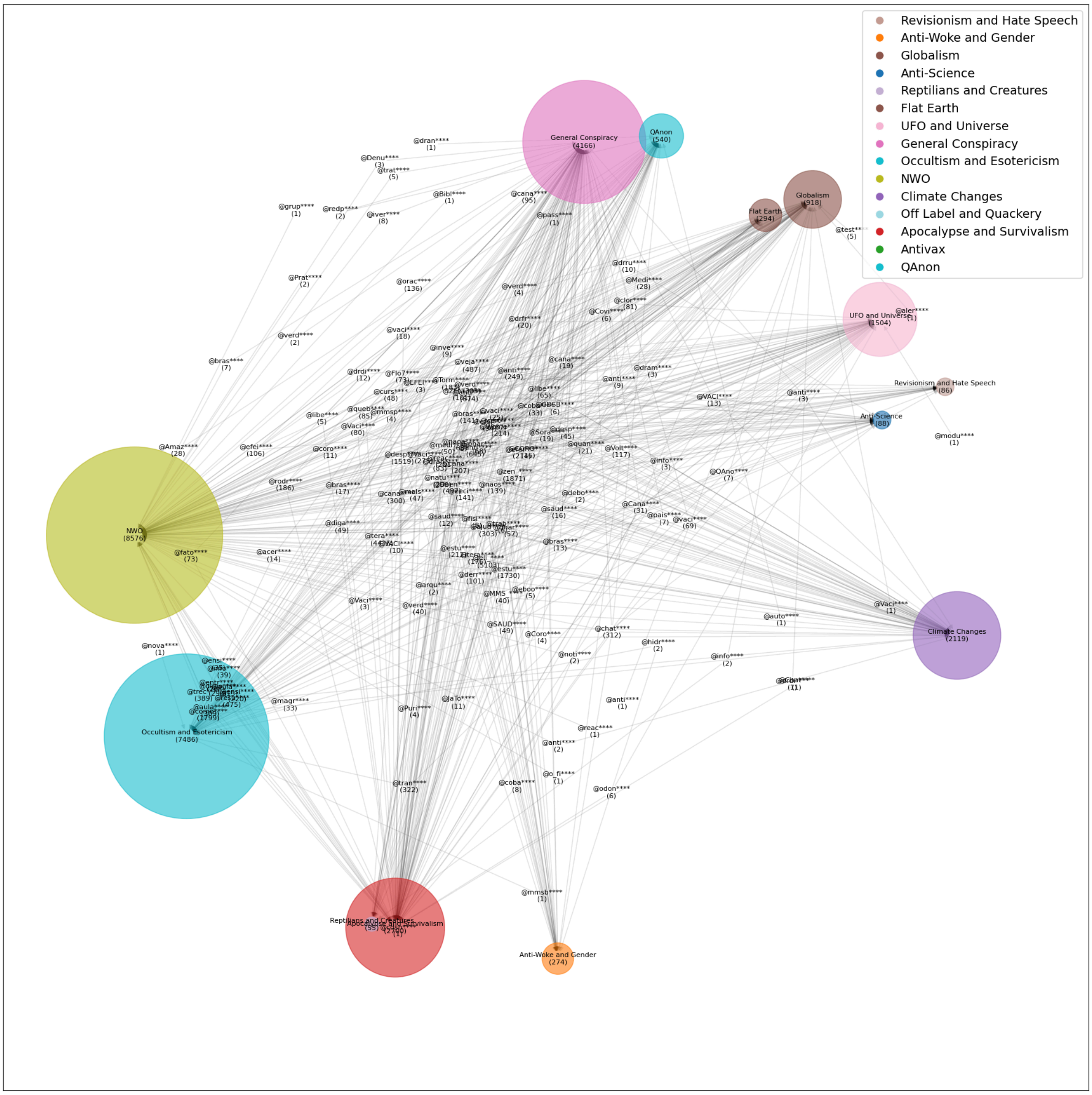

Source: Own elaboration (2024).

This figure depicts the complex web of communities that serve as entry points for individuals interested in conspiracy theories. By observing the network structure, it becomes apparent that larger nodes, such as New World Order, Occultism and Esotericism, and Climate Change, play a fundamental role in introducing new members to these themes. These communities function as hubs, centralizing and disseminating information that draws followers into a vast array of theories. The prominent position of these hubs suggests that they not only act as gateways but also as narrative aggregation centers, where various themes intertwine to create a cohesive and amplified worldview. This reflects the ability of these communities to offer a convergence point for individuals beginning to explore these theories, establishing a reference point that connects seemingly disparate themes but is presented within these communities as part of a larger picture of “hidden truths”. Additionally, the density of connections around these hubs indicates a highly interconnected network, where the exchange of narratives is intense and continuous, enhancing the reach and influence of these communities in introducing new adherents.

**Figure 03.** Network of communities whose theme opens doors (exit point)

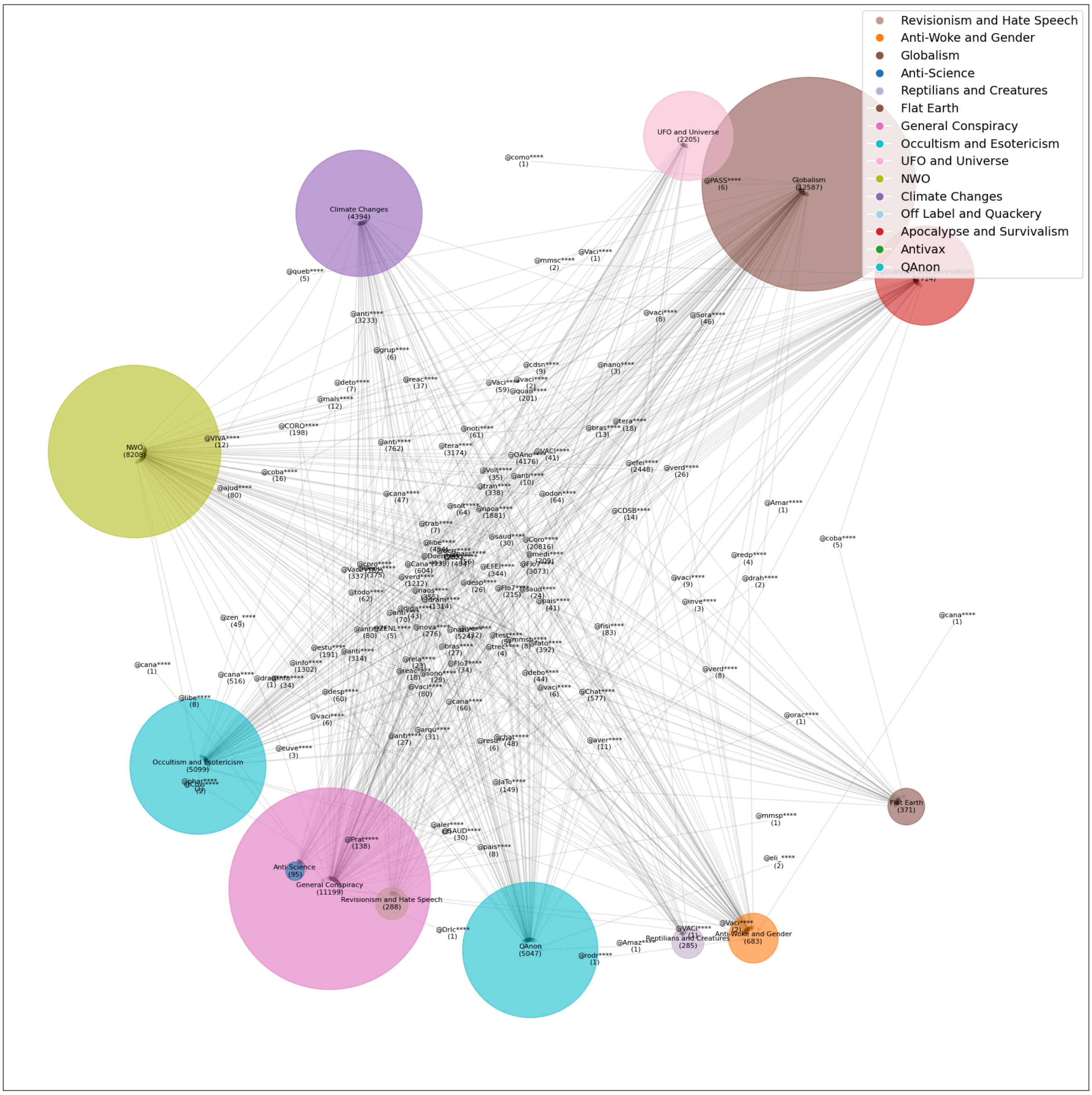

Source: Own elaboration (2024).

This figure highlights the communities that, based on their central themes, serve as springboards to other discussions within the conspiratorial universe. Communities like Globalism and Climate Change not only attract followers but also prepare them to transition to other related themes, such as the New World Order. The centrality of these communities in the network indicates that they play a crucial role in expanding conspiratorial narratives, acting as intermediaries that connect different areas. The network structure suggests that once inside these communities, individuals are exposed to a broader range of theories that expand their conspiratorial worldview. This "exit point" function not only implies a transition to other themes but also a deepening within the conspiratorial ecosystem, where the complexity and interconnection of narratives increase as individuals become more engaged.

**Figure 04.** Flow of invitation links between anti-vaccine communities

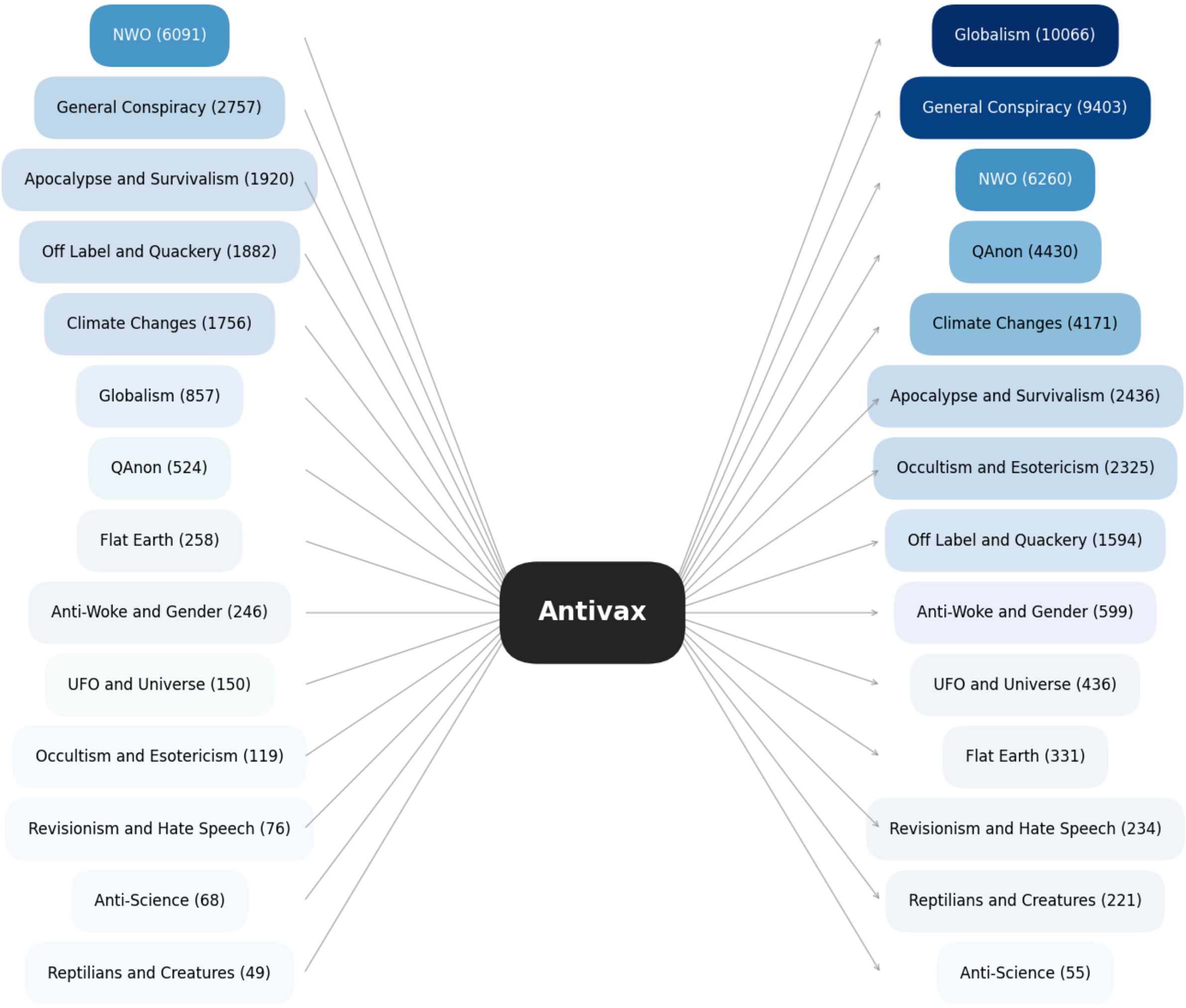

Source: Own elaboration (2024).

It is observed that the Anti-vaccine (antivax) community plays a central and highly interconnected role, serving as a node that distributes invitations to a wide variety of other thematic groups. On the left, we see how anti-vaccine discussions frequently connect with groups related to the New World Order (NWO), with 6,091 invitation links, followed by General Conspiracies with 2,757 links and Apocalypse and Survivalism with 1,920. These numbers reveal that the anti-vaccine narrative is often intertwined with major conspiracy theories, suggesting that those who join anti-vaccine communities are quickly exposed to a larger network of disinformation and fear. On the right, the groups that most receive invitations from anti-vaccine communities include Globalism, with 10,066 links, General Conspiracies with 9,403 links, and NWO with 6,260 links. This suggests a strong tendency for member crossover between these themes, showing that the ideas propagated in anti-vaccine communities are not limited to health but expand to theories of global domination, manipulation, and apocalyptic crises produced by a supposed elite. This pattern highlights how anti-vaccine theories function as a gateway to a vast disinformation network, where each narrative reinforces the others, creating a continuous cycle of radicalization.

**Figure 05.** Flow of invitation links between off-label medication communities

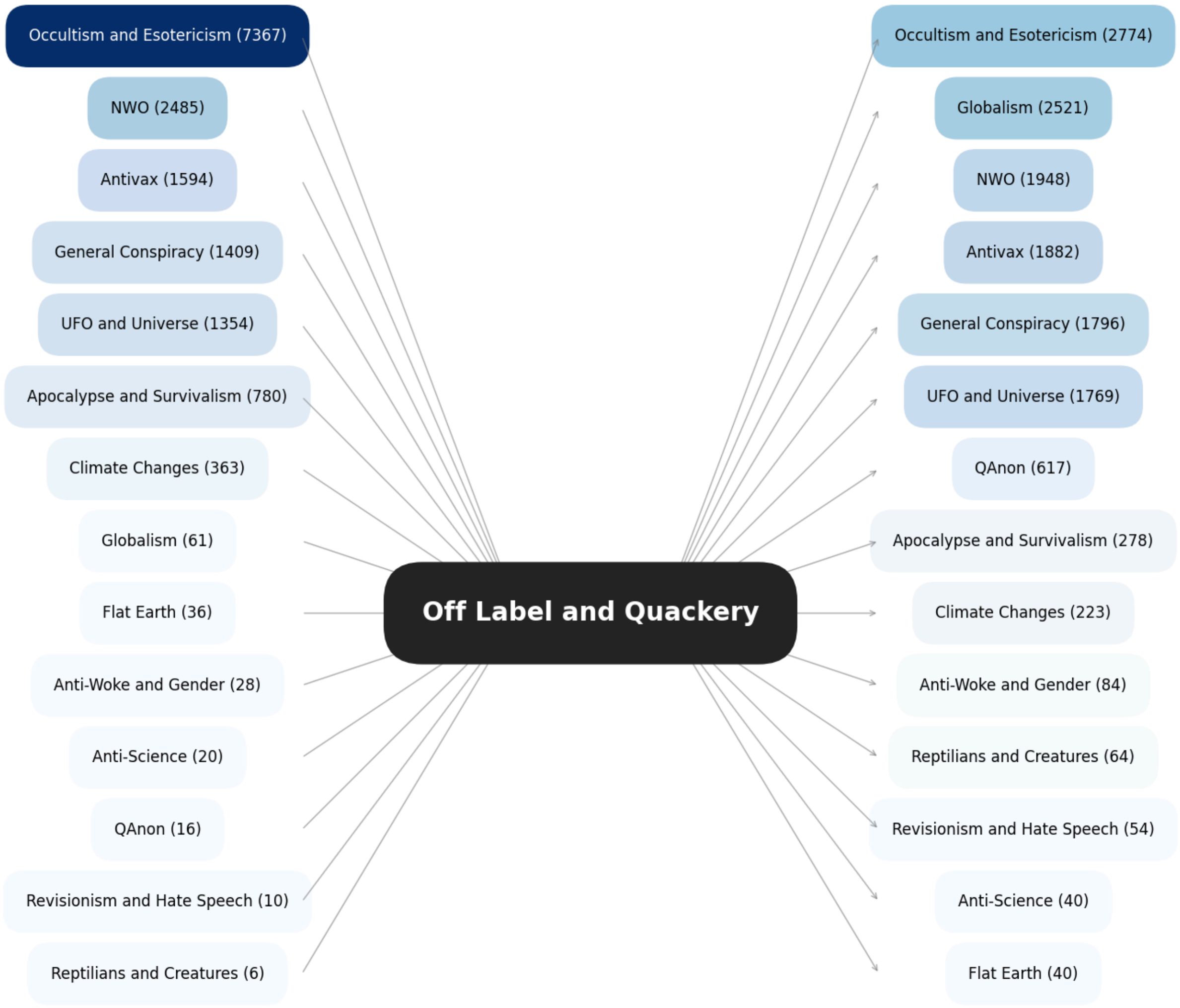

Source: Own elaboration (2024).

Similar to the previous figure, off-label medications are a central theme that connects several other conspiratorial narratives. On the left, Occultism and Esotericism emerge as the most strongly connected community with 7,367 links, suggesting an association between the promotion of alternative treatments and esoteric beliefs. This may indicate a tendency among members of these communities to seek “alternative cures” that are often based on non-scientific or esoteric practices, which can be harmful to society's health. Other significant connections include NWO with 2,485 links and Anti-vaccine with 1,594, showing that those drawn to off-label medication discussions are also likely to be involved in anti-vaccine debates and global conspiracy theories. On the right, the communities that most receive invitations from groups discussing off-label medications include Occultism and Esotericism with 2,774 links, Globalism with 2,521 links, and NWO with 1,948 links. This flow of invitations reinforces the interconnection between health-related disinformation and a range of other conspiracy theories, especially those related to a hidden world order or global manipulation. The link pattern suggests that the promotion of off-label medications is part of a broader disinformation strategy, where different narratives mutually reinforce each other, increasing the credibility of these beliefs among members of these communities.

### 3.2. Time series

The following figures present a comprehensive analysis of the evolution of discussions related to anti-vaccine themes and off-label medications over recent years, with a particular focus on the period marked by the COVID-19 pandemic. Through various graphical representations—including line graphs, absolute area charts, and relative area charts—it is possible to observe how these topics gained prominence in public discourse, reflecting social tensions, disinformation, and the influence of global events on collective perceptions of health and science. These visualizations highlight not only the substantial increase in the volume of debates and shares about these themes during the most critical moments of the pandemic but also illustrate how these narratives remained persistent and adaptable even after the peak of the health crisis.

**Figure 06.** Line graph over the period

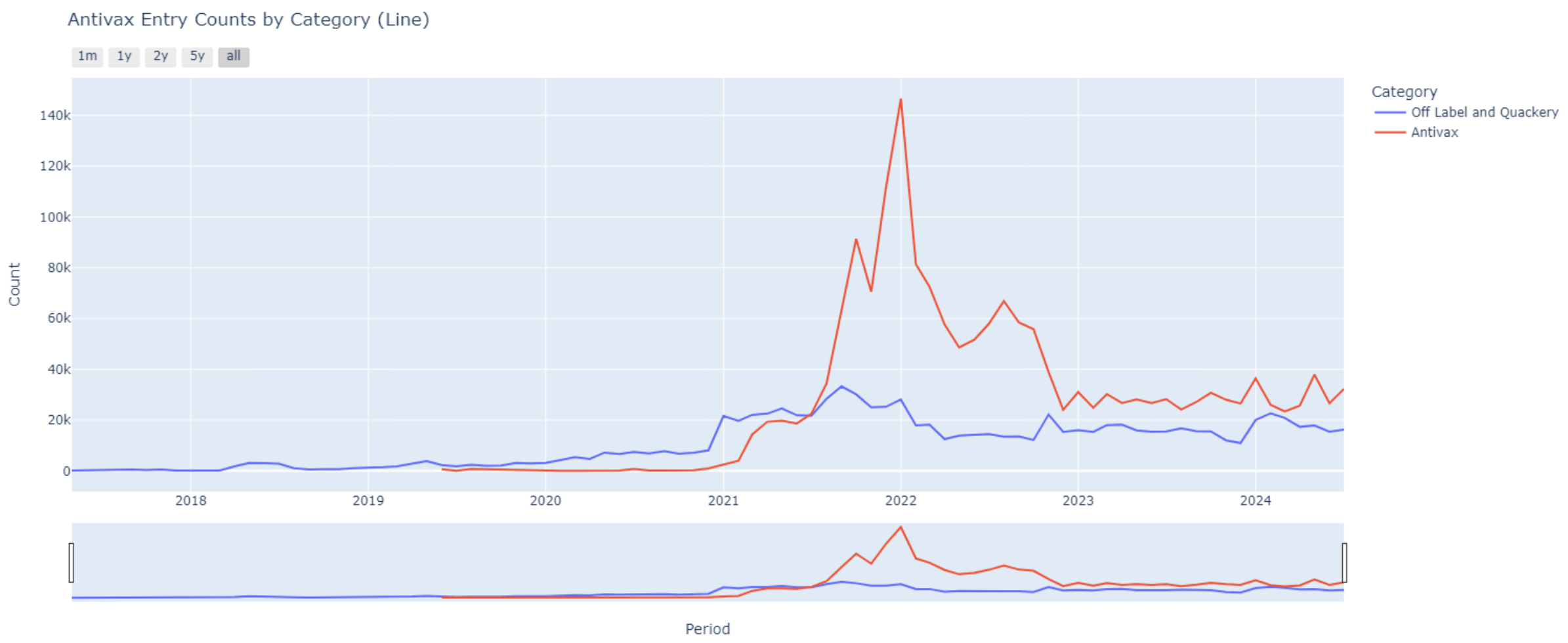

Source: Own elaboration (2024).

[ english version / português abaixo ]

The line graph over the period reveals the fluctuations in discussions about anti-vaccine and off-label medications over the years, with particular emphasis on the profound impact that the COVID-19 pandemic had on the spread of these narratives. Before 2020, both themes maintained a relatively stable flow, with few significant variations. However, starting in early 2020, with the global advance of the coronavirus and subsequent public health measures such as mass vaccination, there was a sharp increase in mentions of the anti-vaccine theme, culminating in a notable peak in 2021 and 2022. This increase can be linked to the proliferation of disinformation surrounding COVID-19 vaccines. Conspiracy theories related to the efficacy and safety of vaccines gained traction during this period, fueled by pre-existing anti-vaccine movements outside social networks. Concurrently, the use of off-label medications such as hydroxychloroquine and ivermectin also grew, reflecting the desperation of a portion of the population and the irresponsible promotion of unproven treatments by public figures. After the peak, the graph indicates a trend of stabilization, but still at elevated levels compared to the pre-pandemic period, suggesting that these narratives have become ingrained in certain segments of society, maintaining a persistent discourse against vaccination and pro-off-label.

**Figure 07.** Absolute area chart over the period

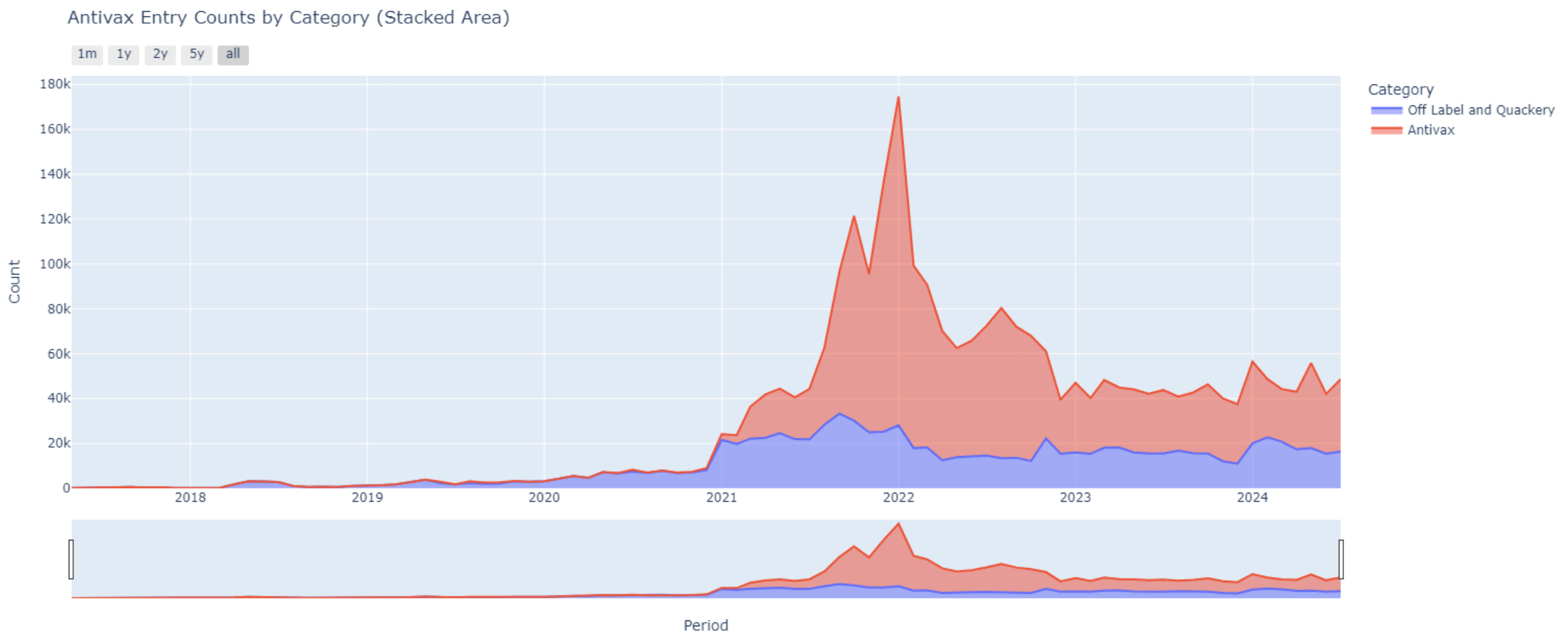

Source: Own elaboration (2024).

The absolute area chart over the period expands the previous analysis, clearly visualizing the total volume of discussions over time, divided between anti-vaccine and off-label medications. The filled area of each category illustrates not only the drastic increase in mentions during the COVID-19 pandemic but also the lasting impact of these narratives in the public sphere. From 2020 onwards, the area representing anti-vaccine themes increases significantly, highlighting how vaccine resistance became one of the main concerns during the health crisis. The chart also shows a parallel, though smaller-scale, growth in discussions about off-label medications, which became popular as people sought alternatives to conventional treatments recommended by health authorities. The peak in 2022 suggests a period of intense controversy and polarization, when these narratives reached their zenith in popularity before beginning to decline. However, the considerable residual area after the peak indicates that while the intensity of discussions has decreased, these themes continue to occupy a significant space in public discourse, demonstrating the staying power of these beliefs and their ability to resist factual correction.

**Figure 08.** Relative area chart over the period

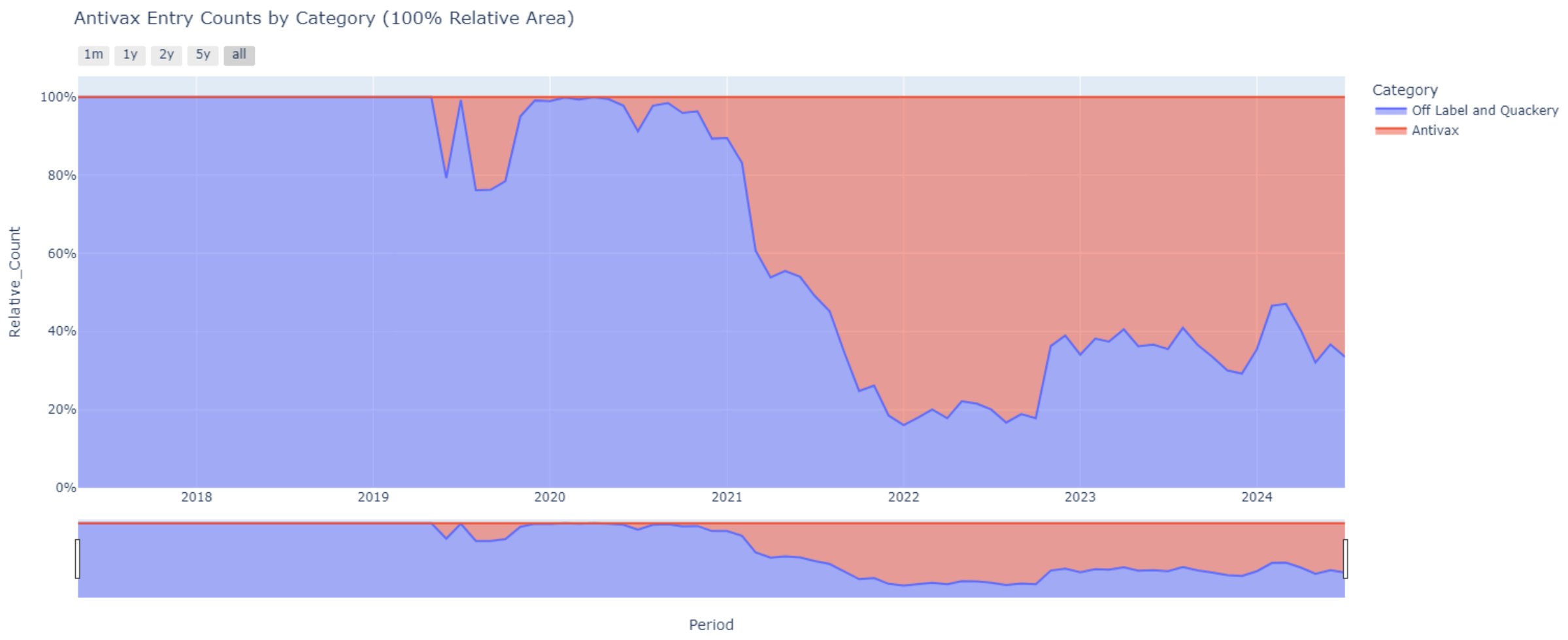

Source: Own elaboration (2024).

The relative area chart over the period provides a perspective on the relative dominance of each theme over time, showing how the anti-vaccine narrative gained prominence during the COVID-19 pandemic, significantly surpassing discussions about off-label medications. Before 2020, both themes occupied relatively balanced proportions, with a slight predominance of discussions about alternative treatments. However, with the advent of the pandemic and the subsequent global rush for vaccination, mentions of anti-vaccine themes quickly surpassed those of off-label medications, reaching an almost absolute dominance in the period from 2021 to 2022. This dominance reflects the centrality of the vaccine debate in public controversies and the mobilization of anti-vaccine movements, which managed to channel fears and distrust regarding vaccines to a broader audience. This chart vividly illustrates the shifts in discursive priorities over time, showing how global crises can reshape the landscape of conspiratorial narratives and disinformation.

## 3.3. Content analysis

The following word clouds provide a visual analysis of the main emerging narratives in online communities discussing anti-vaccine and off-label medications over the years. These representations reveal how certain keywords have consolidated and evolved within these discourses, reflecting the intersection of health, politics, and individual beliefs. The analysis of these words not only exposes the central themes that dominated conversations during the COVID-19 pandemic, but also illustrates how these narratives have rooted themselves and adapted to the constantly changing social and political context. The persistence of terms related to “truth”, “freedom”, and “vaccine” suggests a continuous battle for narrative control, where discussions about public health intertwine with issues of identity, power, and resistance to institutional measures.

**Figure 09.** Consolidated word cloud for anti-vaccine and off-label medications

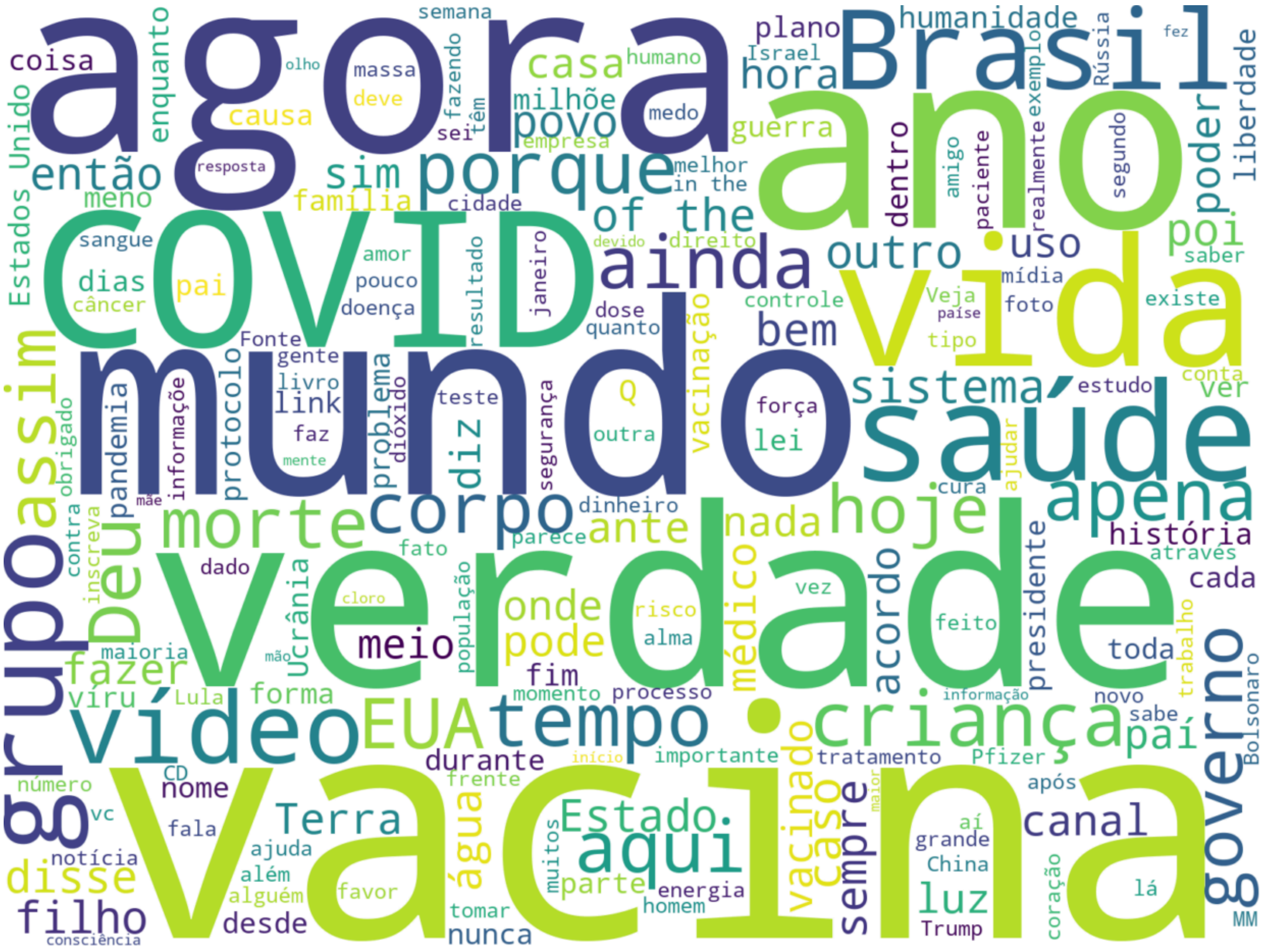

Source: Own elaboration (2024).

The consolidated word cloud reveals the predominance and interrelation of narratives around anti-vaccine and off-label medications. Terms such as “now”, “truth”, “vaccine”, “world”, “life”, and “covid” stand out, reflecting how these words serve as pillars of a narrative that mixes urgency, distrust, and challenges to scientific authority. Observing the publications for more context, the word “now” suggests a sense of immediacy and time pressure, often used in emotionally charged posts or calls to action, while “truth” reflects the search for or affirmation of an alternative reality, typical in conspiratorial discussions. The recurrence of “vaccine” and “covid” highlights the profound impact of the pandemic on the polarization of the health debate, where “life” and “death” appear as symbolic poles. The mention of “system”, “government”, and “Brazil” indicates the insertion of these discussions into a broader political context, where public health measures are often seen as part of a larger agenda, fueling distrust in institutions. This word cloud suggests that the anti-vaccine narrative is not isolated but is interconnected with a broader discourse on power, arguments about individual and/or collective freedoms, and national identity, where public health becomes an ideological battleground.

[ english version / português abaixo ]

**Chart 01.** Temporal word cloud series for anti-vaccine narratives

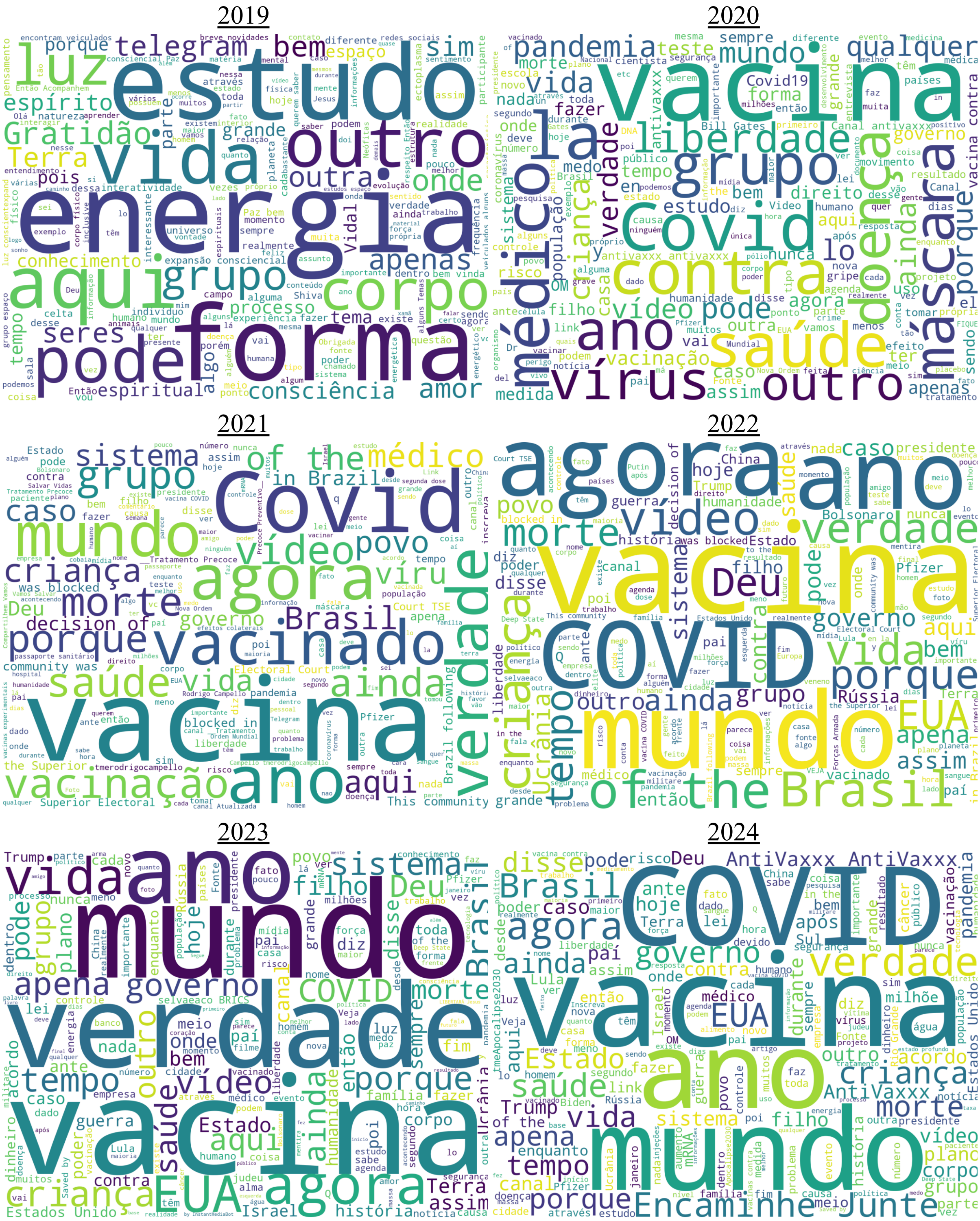

Source: Own elaboration (2024).

The temporal word cloud series for anti-vaccine narratives reveals the evolution of discussions over the years. In 2019, words like “energy” and “form” suggest a focus on more esoteric or alternative themes, which then dramatically expand to “vaccine”, “covid”, and “freedom” in 2020, with the advent of the pandemic. The word “control”, in 2020, reflects the widespread fear that vaccines were being used as a form of manipulation or domination, a theory that gained traction with the introduction of proposals such as quarantines and lockdowns. In 2021, “now” and “truth” become central, indicating an intensification in the narrative struggle, where anti-vaccination is presented not just as a health choice but as a fight for truth against a system perceived as oppressively promoting collective health through vaccination. In 2022 and 2023, words like “death”, “world”, and “Brazil” indicate a nationalization of the debate, where resistance to vaccines is linked to political and cultural identities. The consistency of “vaccine” over the years, especially in 2024, shows that even after the pandemic's peak, the theme remains a central concern, revealing the persistence of disinformation and the difficulty of eradicating these deeply rooted beliefs.

**Chart 02.** Temporal word cloud series for off-label medications

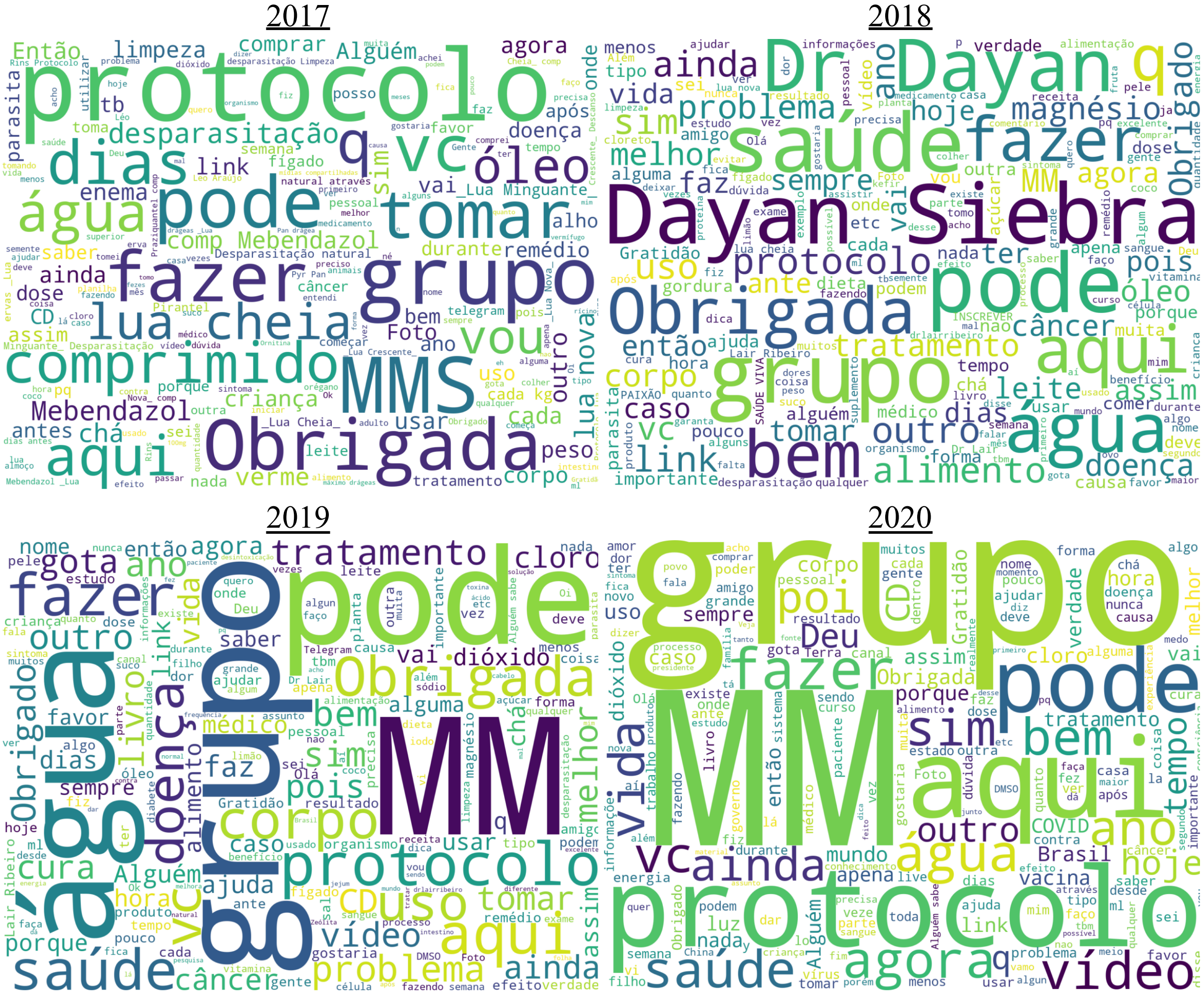

[ english version / português abaixo ]

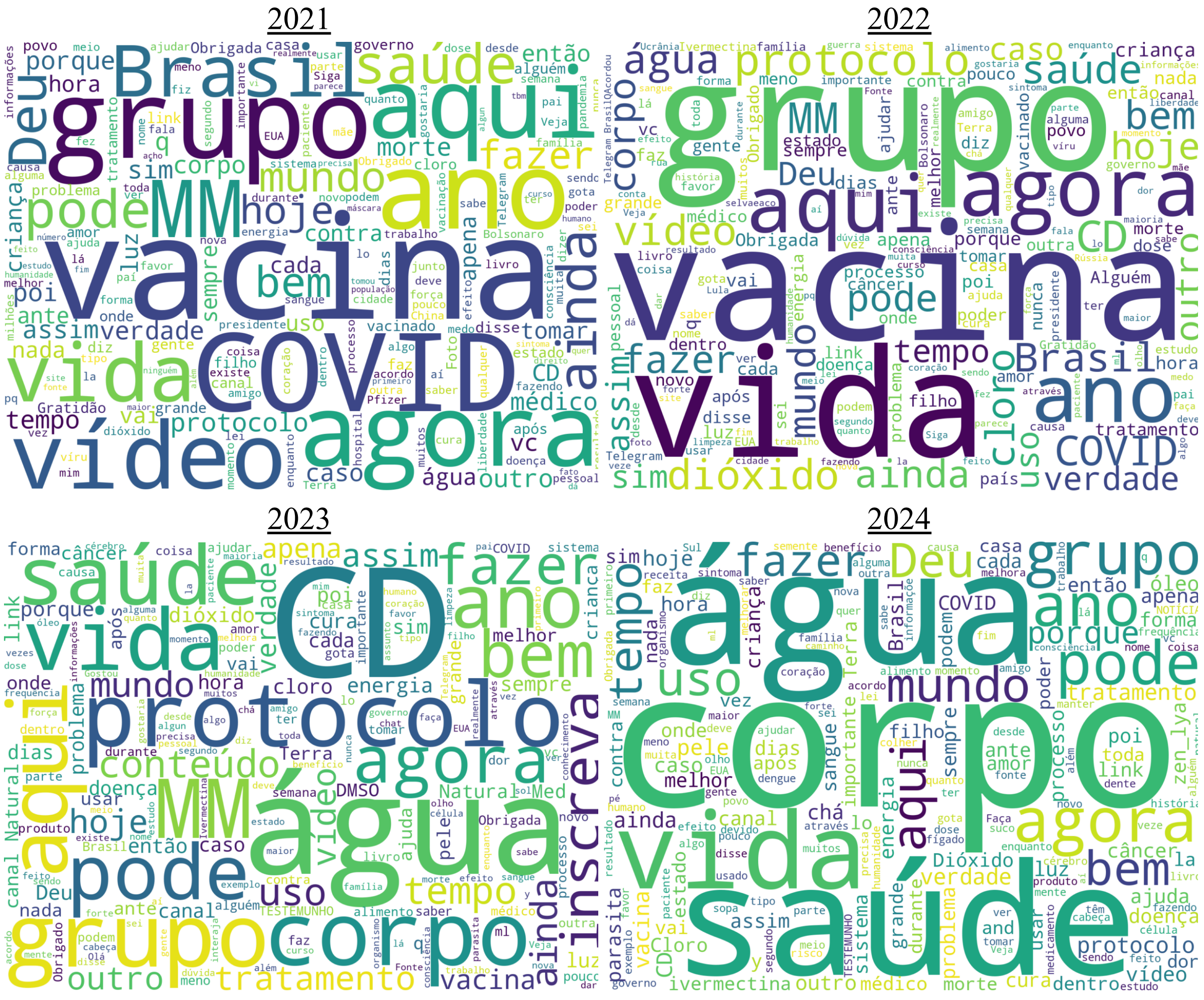

Source: Own elaboration (2024).

The temporal word cloud series for off-label medications highlights the evolution of discussions around alternative treatments, particularly those related to chlorine dioxide (CDS) and MMS (Miracle Mineral Solution), which were widely promoted as “vaccine detox” during the COVID-19 pandemic. It is important to note that many products are marketed in communities as supposed miracle cures. In addition to profiting from the sale of potentially harmful bottles to people's health, there is also the monetization of infoproducts, such as e-books and online courses on practices for producing “vaccine detox”, for example.

In the early years, such as 2017 and 2018, words like “Dayan Siebra” and “MMS” already indicate the influence of public figures and the popularization of these products, which gained traction among followers seeking alternatives to conventional treatments. From 2019 onwards, terms like “group” and “protocol” suggest the formation of communities dedicated to sharing information about these treatments, establishing support networks that promote the use of CDS as an allegedly effective solution against various diseases. During the COVID-19 pandemic in 2020 and 2021, mentions of “vaccine” and “covid” increase significantly, integrating these substances into the anti-vaccine debate, where chlorine dioxide was often promoted as an alternative to “detoxify” the body from the supposed harmful effects of vaccines. In 2022 and 2023, “MMS”, “group”, and “protocol” remain central, reflecting the persistence of these networks and the continued focus on treatments that promise to “purify” the body. The consistency of these terms in 2024 suggests that even with the pandemic's decline, the mistaken belief in the efficacy of MMS and CDS remains alive in various communities, showing the resilience of these disinformation narratives and the difficulty of combating them within these belief bubbles.

### 3.4. Thematic agenda overlap

The figures below not only show the centrality of health discussions within these communities, but also how these topics interconnect with a wide range of other conspiracy theories and cultural disputes. The visual analysis of the figures allows us to identify how each theme acts as a convergence point for different disinformation narratives, amplifying the reach of these communities and reinforcing their core beliefs. By integrating varied topics such as geopolitics, esotericism, and culture wars, these communities are able to create a cohesive discourse that attracts and retains their members, making intervention and factual correction more challenging.

**Figure 10.** Vaccination and Pandemic themes

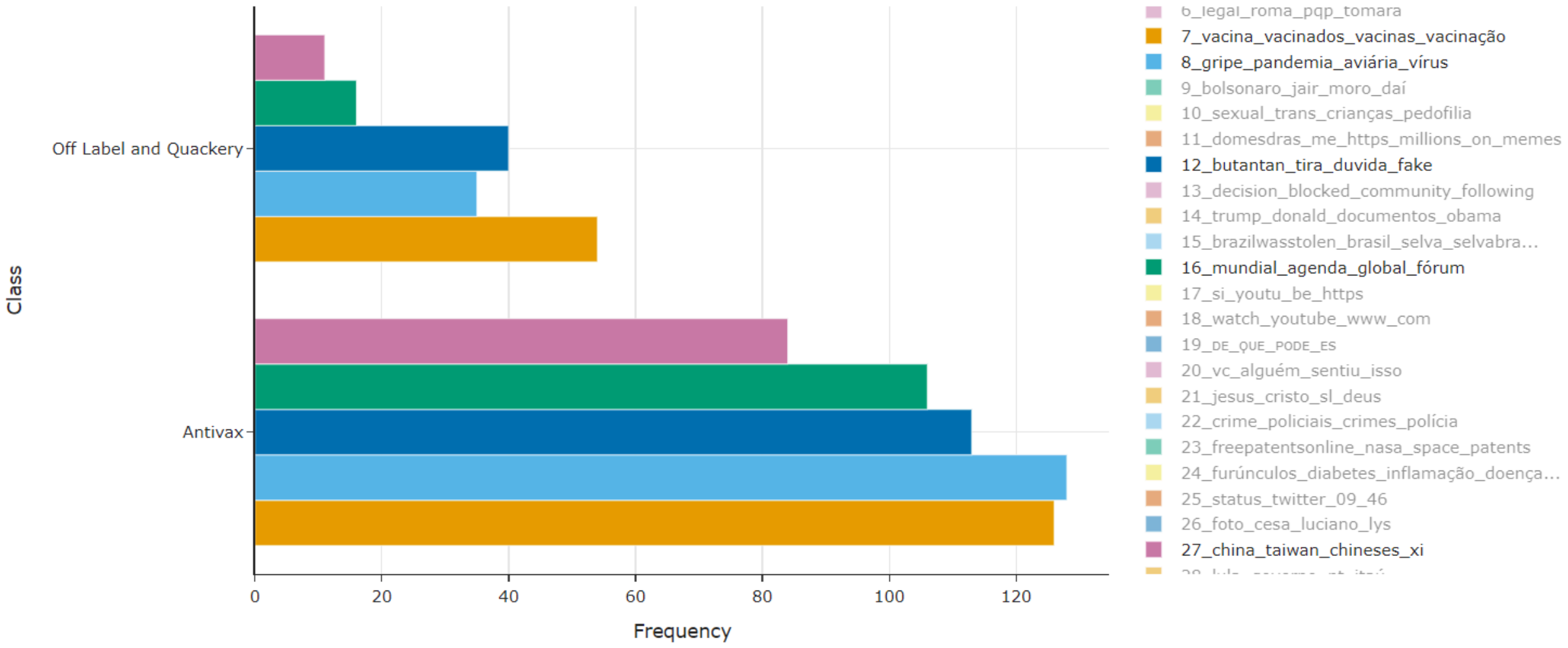

Source: Own elaboration (2024).

This figure illustrates how themes related to vaccination and the pandemic are intertwined with other narratives, broadening the scope of the anti-vaccine discourse. Notably, topics such as “Butantan”, “fake” and “vaccine”, “vaccinated”, “vaccines”, “vaccination” stand out, suggesting that these communities not only question the safety and efficacy of vaccines but also promote widespread distrust in public health institutions like the Butantan Institute. The reference to the narrative “world”, “agenda”, “global”, “forum” indicates an overlap with global conspiracy theories, where vaccination is seen as part of a hidden agenda for population control. This integration of external narratives reflects the strategy of these communities to link their core beliefs to widely discussed topics, using the pandemic as a catalyst to reinforce and disseminate disinformation theories.

**Figure 11.** Geopolitical disputes themes

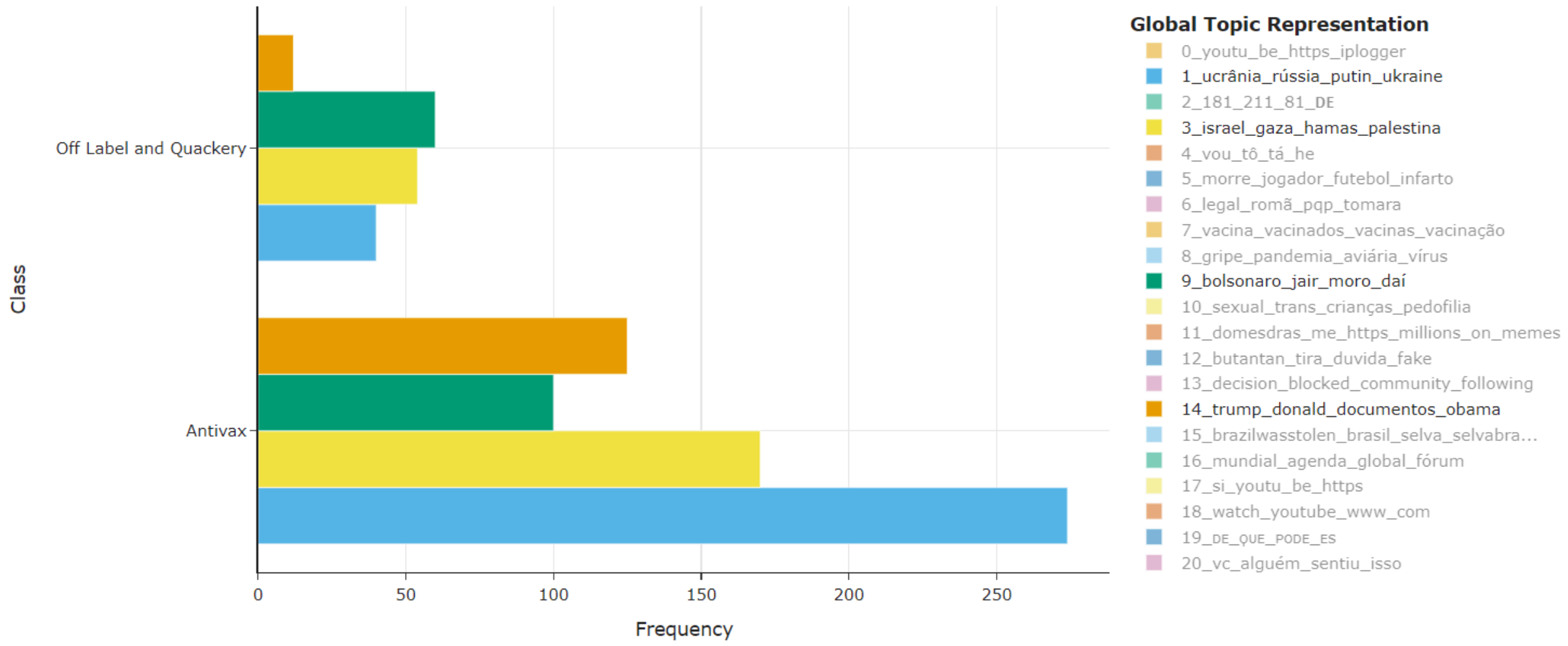

Source: Own elaboration (2024).

This figure highlights how anti-vaccine and off-label medication discussions are often associated with geopolitical narratives, reflecting the politicization of the discourse. The prominence of the topic “Ukraine”, “Russia”, “Putin”, “Ukraine” suggests that these communities use geopolitical conflicts as elements and evidence to validate their conspiracy theories by connecting themes. The inclusion of “Bolsonaro”, “Jair”, and “Moro” shows the internal politicization of these discussions, where local politicians are positioned as central figures influencing or being influenced by this supposed global conspiracy. These overlaps indicate that anti-vaccine communities are not limited to health issues but incorporate political and international disputes to strengthen their narratives and expand their reach, using global events as tools to reinforce distrust in institutions and health policies.

**Figure 12.** Off-label medications themes

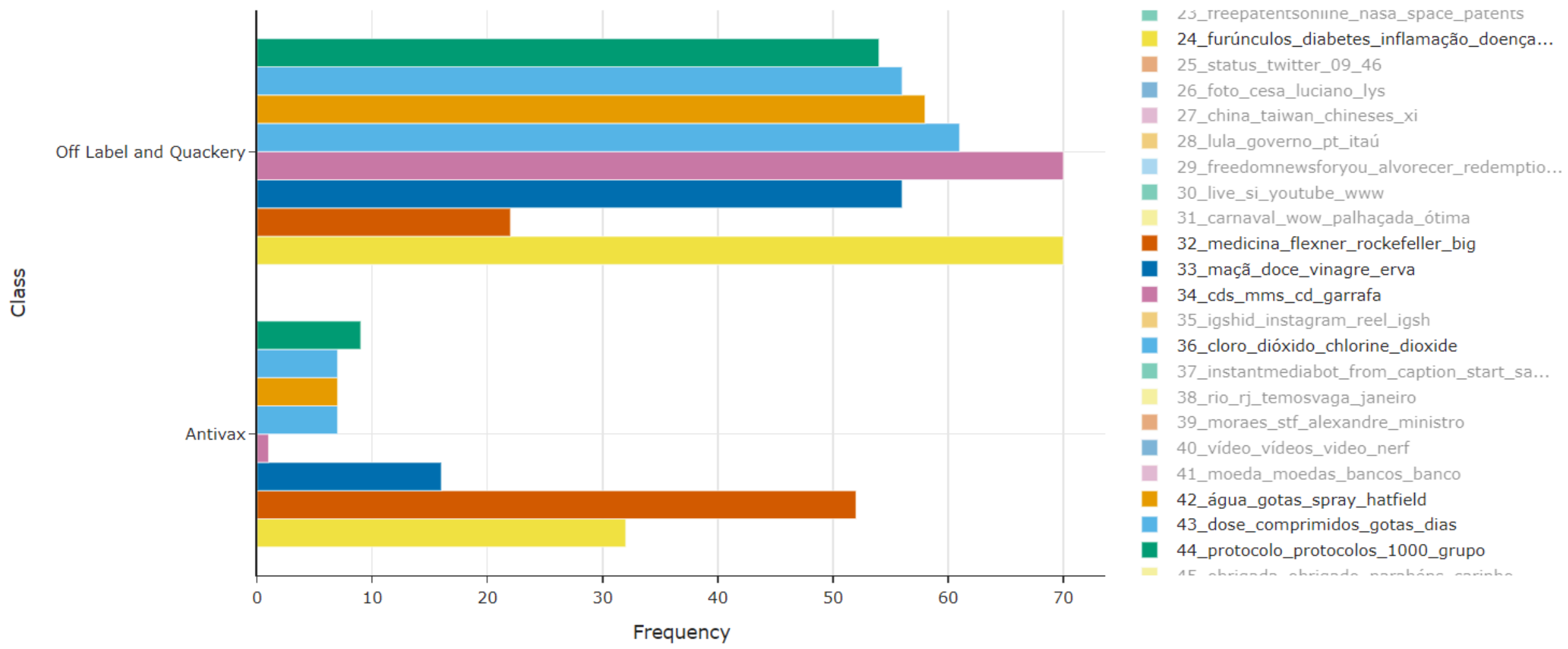

Source: Own elaboration (2024).

In this figure, off-label medication topics show how the promotion of alternative treatments, such as MMS and CDS, is connected to other narratives that question conventional medicine. The topic “medicine”, “Flexner”, “Rockefeller”, “Big” reflects a critique of the foundations, often portrayed in these communities as tools of control by elites. “Chloro”, “dioxide”, “chlorine”, “dioxide” is directly linked to the promotion of CDS as an alternative solution to vaccines, reinforcing distrust in traditional science. These narrative overlaps show how anti-vaccine and off-label medication communities construct a worldview where conventional treatments are seen as dangerous or manipulative, and “natural” alternatives are promoted as the true solutions, even without scientific basis.

**Figure 13.** Culture wars and intersection with vaccination themes

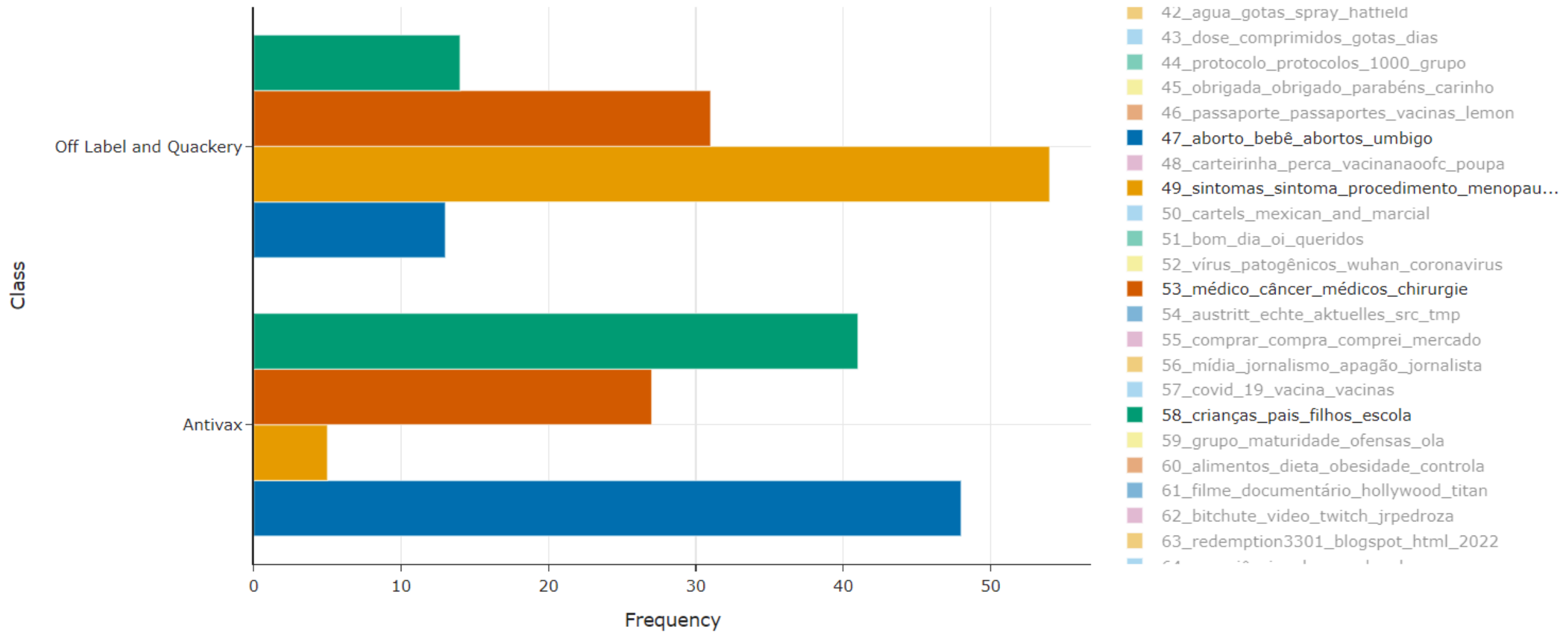

Source: Own elaboration (2024).

This figure highlights how culture war themes are intertwined with anti-vaccine and off-label medication narratives. The topics “abortion”, “baby”, “abortions”, “navel” and “children”, “parents”, “sons”, “school” indicate how these communities often integrate culture war narratives, such as the debate on reproductive rights and child education, into the anti-vaccine discourse. In some cases, they go as far as claiming that aborted fetal cells are used to make vaccines, or that vaccines could cause spontaneous abortion in women. This not only broadens the scope of disinformation but also reinforces the idea that vaccines are part of a broader agenda aimed at corrupting or controlling traditional and family values. The intersection of these narratives allows resistance to vaccines to be seen as part of a broader defense of moral and cultural values, creating common ground among different groups that oppose perceived social and political changes. Thus, health discussions become inseparable from cultural debates, solidifying the commitment of community members to the anti-vaccine cause while also engaging in a broader crusade.

**Figure 14.** Other general conspiracy theories themes

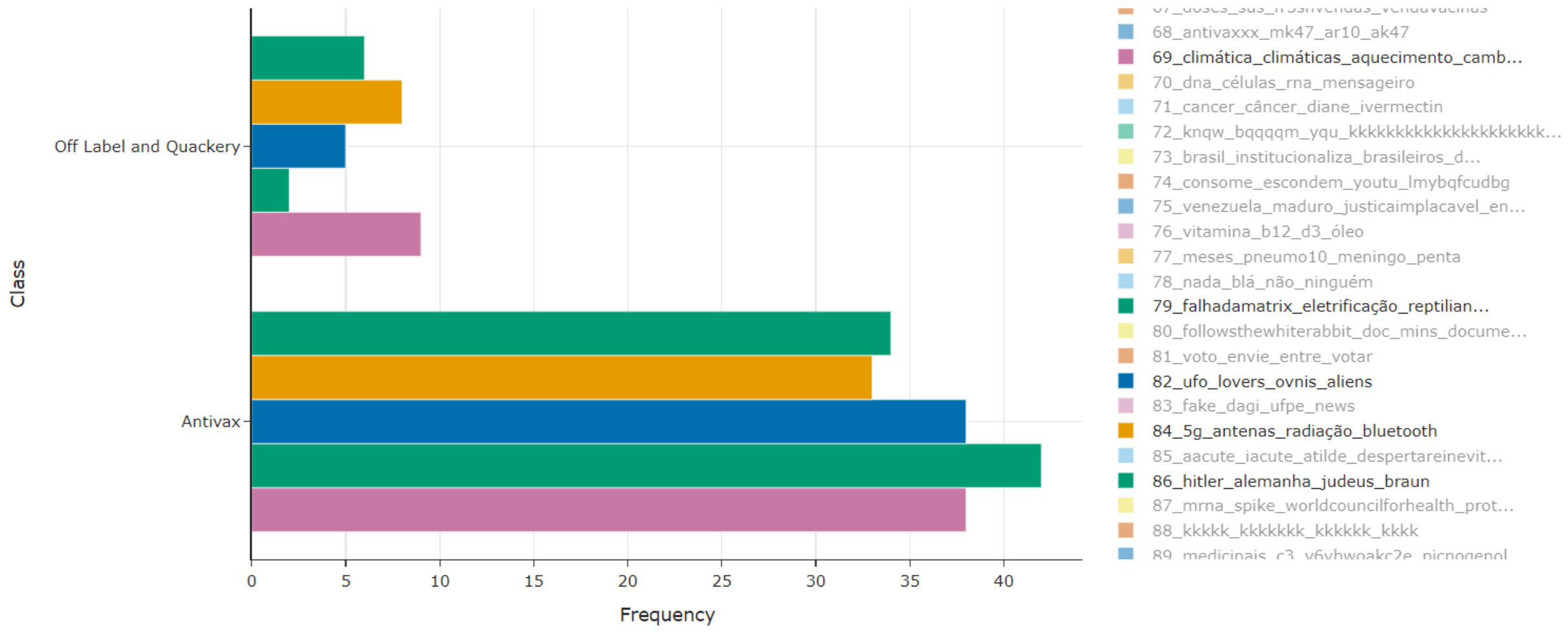

Source: Own elaboration (2024).

The figure on general conspiracy theories reveals how anti-vaccine and off-label medication communities intertwine with a wide range of other theories. The prominence of the topic “falhamatrix”, “electrification”, “reptilian” indicates the inclusion of more extreme theories, such as the belief in reptilians or the existence of a matrix, within the anti-vaccine discourse. This demonstrates how these communities expand their core narratives by integrating theories that, while seemingly disconnected, reinforce a worldview where vaccines are part of a broader plan for global domination. By associating vaccines with ideas of control by hidden elites, these communities can attract individuals already predisposed to believe in other forms of conspiracy, creating an interconnected network of disinformation that sustains and reinforces itself.

**Figure 15.** Themes of a false link between vaccines and autism

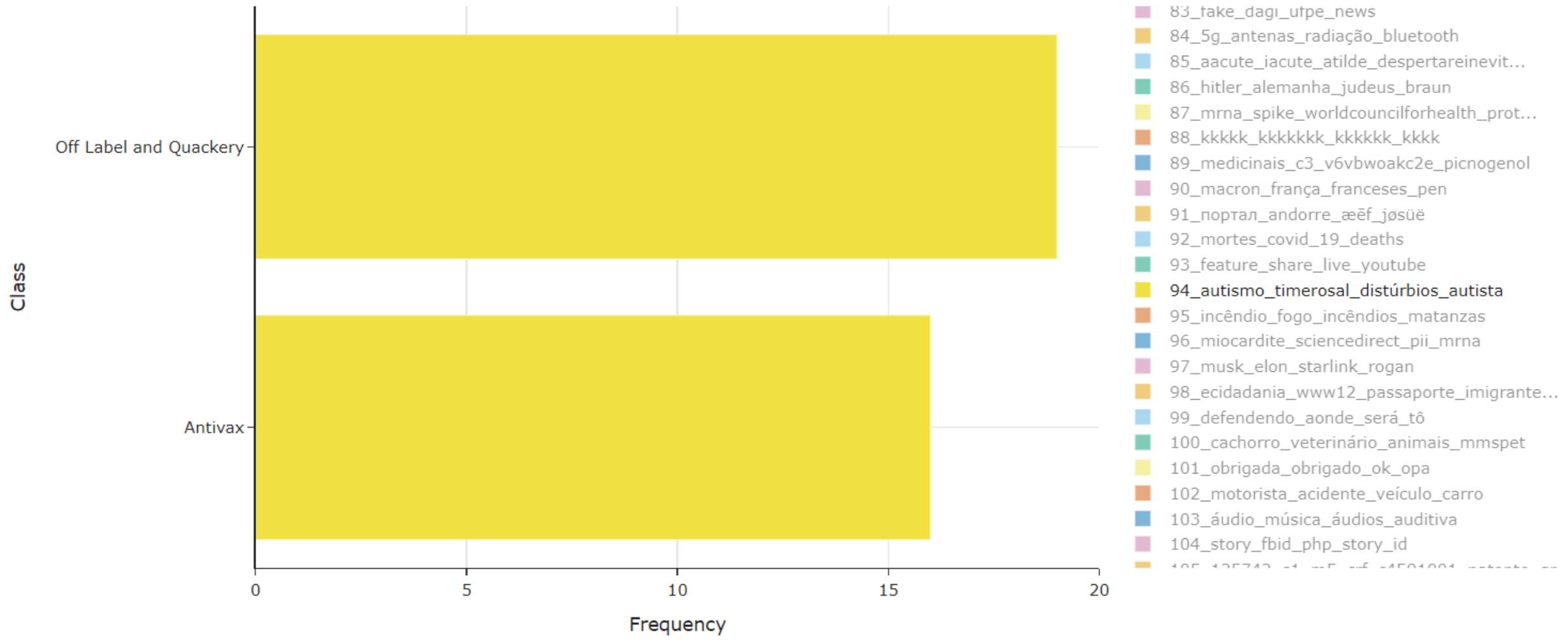

Source: Own elaboration (2024).

The figure addressing the themes of a false link between vaccines and autism highlights how this specific narrative continues to be a rhetorical device within anti-vaccine and off-label medication communities. The topic “autism”, “thimerosal”, “disorders”, “autistic” appears with significant frequency in both anti-vaccine and alternative treatment discussions. This persistence reflects the resilience of this conspiracy theory, which has been widely debunked by the scientific community, but continues to be promoted by these communities as a hidden truth. The idea that vaccines cause autism serves as a convergence point for different disinformation narratives, creating a common foundation that connects health themes to broader issues of distrust in conventional medicine and authorities. This thematic overlap shows how anti-vaccine and off-label medication communities use old and debunked theories to sustain their core beliefs while also recruiting new members to their ranks, amplifying the cycle of disinformation.

## 4. Reflections and future works

To answer the research question, **“How are Brazilian conspiracy theory communities characterized and articulated regarding anti-vaccine (antivax) themes and alternative treatments and medications (off-label) on Telegram?”**, this study adopted techniques mirrored across a series of seven publications that seek to characterize and describe the phenomenon of conspiracy theories on Telegram, using Brazil as a case study. After months of investigation, a total of 195 Brazilian conspiracy theory communities on Telegram were extracted, focusing on anti-vaccine (antivax) themes and alternative treatments and medications (off-label). These communities generated 5,406,762 pieces of content published between May 2017 (first publications) and August 2024 (when this study was conducted), with a total of 440,651 users within the communities.

Four main approaches were adopted: **(i)** Network, which involved the creation of an algorithm to map connections between communities through invitations circulated among groups and channels; **(ii)** Time series, which used libraries like “Pandas” (McKinney, 2010) and “Plotly” (Plotly Technologies Inc., 2015) to analyze the evolution of publications and engagements over time; **(iii)** Content analysis, where textual analysis techniques were applied to identify patterns and word frequencies in the communities over the semesters; and **(iv)** Thematic agenda overlap, which utilized the BERTopic model (Grootendorst, 2020) to group and interpret large volumes of text, generating coherent topics from the analyzed publications. The main reflections are detailed below, followed by suggestions for future works.

### 4.1. Main reflections

**Themes such as New World Order and Apocalypse and Survivalism are the main gateways for anti-vaccine narratives**: The main communities that serve as gateways for anti-vaccine narratives include New World Order and Apocalypse and Survivalism. These communities directed a significant number of links to anti-vaccine communities, with NWO generating 6,091 links and Apocalypse and Survivalism 1,920 links. This indicates that these conspiracy themes fuel vaccine distrust by linking them to ideas of global control;

[ english version / português abaixo ]

**Globalism, General Conspiracy, and New World Order are the primary communities that receive invitations from anti-vaccine communities**: Among the communities that receive the most invitations from anti-vaccine communities are Globalism, with 10,066 links, General Conspiracy with 9,403 links, and New World Order with 6,260 links. These numbers show that once inside anti-vaccine communities, members are quickly exposed to a broader network of disinformation, where theories related to globalism and generalized conspiracies are heavily promoted, strengthening the anti-vaccine discourse with other conspiracy theories;

**Occultism and Esotericism are the largest sources of invitations for off-label medication communities**: In the context of off-label medication communities, Occultism and Esotericism stand out as the main source of invitations, generating 7,367 links for these communities. This suggests a strong interconnection between esoteric beliefs and the promotion of off-label medications, where alternative and non-scientific practices find fertile ground for propagation by manipulating users' esoteric beliefs;

**Occultism and Esotericism, Globalism, and New World Order receive the most invitations from off-label medication communities**: The main communities receiving invitations from off-label medication communities include Occultism and Esotericism with 2,774 links, Globalism with 2,521 links, and New World Order with 1,948 links. These data indicate that the promotion of off-label medications is closely tied to esoteric narratives and global conspiracy theories, creating a cohesive and interconnected disinformation network;

**Anti-vaccine narratives experienced a 290% increase during the pandemic, demonstrating growing interconnectivity with other conspiracy theories**: During the peak of the COVID-19 pandemic, discussions about anti-vaccine narratives saw a 290% increase compared to the pre-pandemic period. This expansion was not limited to health topics but also reinforced interconnectivity with other theories, such as Globalism and New World Order, demonstrating how the health crisis was used to promote a broader disinformation agenda;

**The strong overlap of agendas between anti-vaccine and other conspiracy theories feeds an interdependent disinformation network**: Anti-vaccine and off-label medication communities do not operate in isolation; there is significant overlap with other conspiracy theories, such as Globalism and NWO. For example, Globalism was responsible for 10,066 invitation links to anti-vaccine groups, indicating an interdependent network where different narratives reinforce each other, creating a continuous cycle of disinformation;

**The false and dishonest narrative linking vaccines to autism remains one of the most persistent and frequently reintroduced discussions**: Even after repeated scientific refutations, the false narrative linking vaccines to autism continues to be one of the most discussed topics in anti-vaccine and off-label medication communities, where supposed autism cures are presented. The analysis revealed that this narrative was consistently reintroduced into discussions to reinforce general distrust in conventional medicine, highlighting its persistence in the analyzed communities;

**Anti-vaccine communities act as central hubs in the dissemination of disinformation, connecting and amplifying multiple narratives**: The study identified that some anti-vaccine communities function as central hubs within the disinformation network, with a disproportionate ability to influence discourse. These communities connect and amplify multiple conspiracy narratives, functioning as convergence points where different theories meet and reinforce each other;

**The strong interconnectivity between anti-vaccine and off-label medication discussions reveals a cohesive and resistant ideological bubble**: The interconnectivity between anti-vaccine and off-label medication discussions, such as MMS and CDS, is notable. The same communities that promote disinformation about vaccines are often at the forefront of promoting dangerous alternative treatments. This creates an overlap of beliefs within a cohesive ideological bubble, facilitating the spread of disinformation;

**The intersection between health disinformation and esoteric narratives creates a highly influential and attractive network for new members and monetizes the sale of irregular drugs and chemicals with chlorine dioxide**: The study reveals that the combination of health disinformation, such as anti-vaccine and off-label narratives, with esoteric themes like Occultism and Esotericism, not only strengthens cohesion within these communities but also makes them more attractive to new members. By mixing global conspiracy theories with alternative beliefs, these communities create a disinformation network that is both complex and appealing, increasing its reach and the difficulty of dismantling these narratives.

### 4.2. Future works

Based on the main findings of this study, several directions can be suggested for future research. First, the exploration of conspiracy bridges between global themes and anti-vaccine narratives deserves attention. Knowing that themes such as New World Order and Apocalypse and Survivalism are significant gateways for anti-vaccine narratives, future investigations could examine how these theories evolve and intertwine with health narratives in different cultural contexts. Additionally, it would be interesting to explore how emerging conspiracy theories, such as “Great Reset” or “5G” as waves of alleged global mind control, could play similar roles in perpetuating disinformation about vaccines.

Another relevant point is the mapping of network structures and the dissemination of disinformation. Given that communities like Globalism, General Conspiracy, and New World Order receive a large number of invitations from anti-vaccine communities, future studies could focus on more detailed mapping of the internal dynamics of these communities. Understanding how these structures facilitate the dissemination of disinformation and identifying individuals or groups that act as “super spreaders” of these narratives could provide valuable insights for disrupting these networks.

The intersection between esotericism and off-label medications is also a promising area for future research. Considering that Occultism and Esotericism are the main sources of invitations for off-label medication communities, investigations could focus on the psychology behind the attraction to esoteric and alternative practices. Studies examining the effectiveness of factual correction campaigns in these specific contexts would be essential for developing more efficient strategies to combat disinformation among populations that adhere to non-scientific practices.

Future studies could also explore the resistance and persistence of false narratives over time. The false narrative linking vaccines to autism, which continues to be persistently reintroduced into discussions, indicates the need for strategies to dismantle these deeply entrenched beliefs. Research could focus on how these narratives, even after being debunked, manage to resurface and gain traction again, providing support for more effective interventions on digital platforms.

Additionally, the analysis of interconnectivity and the creation of cohesive ideological bubbles are critical aspects for understanding how anti-vaccine and off-label medication communities interact. Future studies could investigate how these bubbles form and persist, as well as explore how different communities collaborate or compete for influence within the larger disinformation network. Understanding the barriers that hinder the entry of scientific information into these bubbles would be fundamental for developing more effective interventions in combating harmful disinformation.

Studying the feedback loops between conspiracy narratives is also essential. The overlap of agendas between anti-vaccine and other conspiracy theories suggests that there is a disinformation feedback loop. Future research could focus on how these narratives reinforce each other and how these cycles can be interrupted, offering strategies to slow the spread of these erroneous beliefs.

Finally, it is important to explore new methods for identifying and neutralizing disinformation hubs. Future studies could focus on developing techniques to detect and limit the influence of these hubs before they gain traction. This could include creating algorithms or tools that allow real-time monitoring and intervention, especially during global crises, where disinformation can have severe consequences.

## 6. Author biography

**Ergon Cugler de Moraes Silva** has a Master's degree in Public Administration and Government (FGV), Postgraduate MBA in Data Science & Analytics (USP) and Bachelor's degree in Public Policy Management (USP). He is associated with the Bureaucracy Studies Center (NEB FGV), collaborates with the Interdisciplinary Observatory of Public Policies (OIPP USP), with the Study Group on Technology and Innovations in Public Management (GETIP USP) with the Monitor of Political Debate in the Digital Environment (Monitor USP) and with the Working Group on Strategy, Data and Sovereignty of the Study and Research Group on International Security of the Institute of International Relations of the University of Brasília (GEPSI UnB). He is also a researcher at the Brazilian Institute of Information in Science and Technology (IBICT), where he works for the Federal Government on strategies against disinformation. Brasília, Federal District, Brazil. Web site: https://ergoncugler.com/.

[ versão português / english above ]

# **Comunidades de antivacinas (*antivax*) e medicamentos *off label* no Telegram brasileiro:** entre o esoterismo como porta de entrada e a monetização de falsas curas milagrosas

*Ergon Cugler de Moraes Silva*

Instituto Brasileiro de Informação
em Ciência e Tecnologia (IBICT)
Brasília, Distrito Federal, Brasil

contato@ergoncugler.com
www.ergoncugler.com

## Resumo

As teorias da conspiração, especialmente aquelas focadas em narrativas antivacinas (*antivax*) e na promoção de medicamentos *off label*, como MMS e CDS, têm se proliferado no Telegram, inclusive no Brasil, encontrando terreno fértil entre comunidades que compartilham crenças esotéricas e desconfiança em relação às instituições científicas. Nesse cenário, esse estudo busca responder **como são caracterizadas e articuladas as comunidades de teorias da conspiração brasileiras sobre temáticas de antivacinas (*antivax*) e medicamentos *off label* no Telegram?** Vale ressaltar que este estudo faz parte de uma série de um total de sete estudos que possuem como objetivo principal compreender e caracterizar as comunidades brasileiras de teorias da conspiração no Telegram. Esta série de sete estudos está disponibilizada abertamente e originalmente no arXiv da Cornell University, aplicando um método espelhado nos sete estudos, mudando apenas o objeto temático de análise e provendo uma replicabilidade de investigação, inclusive com códigos próprios e autorais elaborados, somando-se à cultura de software livre e de código aberto. No que diz respeito aos principais achados deste estudo, observa-se: Temáticas como Nova Ordem Mundial e Apocalipse e Sobrevivência atuam como portas de entrada significativas para narrativas antivacinas, conectando-as a teorias de controle global; Globalismo e Nova Ordem Mundial destacam-se como as principais comunidades que recebem convites das comunidades antivacinas; Ocultismo e Esoterismo emergem como as maiores fontes de convites para comunidades de medicamentos *off label*, criando uma forte conexão entre crenças esotéricas e a promoção de tratamentos não científicos; Narrativas antivacinas experimentaram um aumento de 290% durante a Pandemia da COVID-19, evidenciando uma interconectividade crescente com outras teorias conspiratórias; A sobreposição de pautas entre antivacinas e outras teorias de conspiração cria uma rede interdependente de desinformação, onde diferentes narrativas se reforçam.

### Principais descobertas

➔ Temáticas como Nova Ordem Mundial e Apocalipse e Sobrevivência são as principais portas de entrada para narrativas antivacinas, demonstrando como essas comunidades conectam a desconfiança em instituições globais a teorias de controle mundial;

➔ Globalismo, Conspirações Gerais e Nova Ordem Mundial se destacam como as principais comunidades que recebem convites que partem das comunidades antivacinas, ampliando a rede de desinformação e reforçando teorias conspiratórias globais;

- ➔ As comunidades de Ocultismo e Esoterismo emergem como as maiores fontes de convites que levam para as comunidades de medicamentos *off label*, refletindo uma forte conexão entre crenças esotéricas e a promoção de tratamentos não científicos;
- ➔ Ocultismo e Esoterismo, junto com Globalismo e Nova Ordem Mundial, lideram em número de convites recebidos a partir das comunidades de medicamentos *off label*, criando uma interseção poderosa entre desinformação e teorias de conspiração;
- ➔ As narrativas antivacinas experimentaram um aumento de 290% durante a Pandemia da COVID-19, demonstrando uma interconectividade crescente com outras teorias conspiratórias e o uso da crise sanitária para promover agendas de desinformação;
- ➔ A sobreposição de pautas entre antivacinas e outras teorias de conspiração alimenta uma rede interdependente de desinformação, onde diferentes narrativas se reforçam mutuamente, criando um ciclo contínuo de crenças equivocadas;
- ➔ A narrativa desonesta que associa vacinas ao autismo continua a ser uma das mais persistentes, sendo frequentemente reintroduzida nas discussões das comunidades antivacinas e de medicamentos *off label*, reforçando a desconfiança na medicina convencional;
- ➔ Comunidades antivacinas atuam como *hubs* centrais na disseminação de desinformação, conectando múltiplas narrativas conspiratórias e amplificando seu alcance, o que as torna pontos de convergência poderosos dentro da rede de desinformação;
- ➔ A interconectividade entre discussões antivacinas e de medicamentos *off label* revela uma bolha ideológica coesa e resistente, onde a sobreposição de crenças e agendas ideológicas facilita a disseminação de desinformação e torna difícil a intervenção factual;
- ➔ A interseção entre desinformação sobre saúde e narrativas esotéricas cria uma rede de desinformação altamente influente, que atrai novos membros ao misturar teorias de conspiração globais com crenças alternativas, aumentando o alcance dessas comunidades. Além disso, cria uma rede que monetiza com a venda de fármacos irregulares e produtos químicos com o dióxido de cloro, conhecido também como “detox vacinal”, “mms” e “cds”.

## 1. Introdução

Após percorrer milhares de comunidades brasileiras de teorias da conspiração no Telegram, extrair dezenas de milhões de conteúdos dessas comunidades, elaborados e/ou compartilhados por milhões de usuários que as compõem, este estudo tem o objetivo de compor uma série de um total de sete estudos que tratam sobre o fenômeno das teorias da conspiração no Telegram, adotando o Brasil como estudo de caso. Com as abordagens de identificação implementadas, foi possível alcançar um total de 195 comunidades de teorias da conspiração brasileiras no Telegram sobre temáticas de antivacinas (*antivax*) e tratamentos e medicamentos alternativos (*off label*), estas somando 5.406.762 de conteúdos publicados entre maio de 2017 (primeiras publicações) até agosto de 2024 (realização deste estudo), com 440.651 usuários somados dentre as comunidades. Dessa forma, este estudo tem como objetivo compreender e caracterizar as comunidades sobre temáticas de antivacinas (*antivax*) e tratamentos e medicamentos alternativos (*off label*) presentes nessa rede brasileira de teorias da conspiração identificada no Telegram.

Para tal, será aplicado um método espelhado em todos os sete estudos, mudando apenas o objeto temático de análise e provendo uma replicabilidade de investigação. Assim, abordaremos técnicas para observar as conexões, séries temporais, conteúdos e sobreposições temáticas das comunidades de teorias da conspiração. Além desse estudo, é possível encontrar os seis demais disponibilizados abertamente e originalmente no arXiv da Cornell University. Essa série contou com a atenção redobrada para garantir a integridade dos dados e o respeito à privacidade dos usuários, conforme a legislação brasileira prevê (Lei nº 13.709/2018).

Portanto questiona-se: **como são caracterizadas e articuladas as comunidades de teorias da conspiração brasileiras sobre temáticas de antivacinas (*antivax*) e tratamentos e medicamentos alternativos (*off label*) no Telegram?**

## 2. Materiais e métodos

A metodologia deste estudo está organizada em três subseções, sendo: **2.1. Extração de dados**, que descreve o processo e as ferramentas utilizadas para coletar as informações das comunidades no Telegram; **2.2. Tratamento de dados**, onde são abordados os critérios e métodos aplicados para classificar e anonimizar os dados coletados; e **2.3. Abordagens para análise de dados**, que detalha as técnicas utilizadas para investigar as conexões, séries temporais, conteúdos e sobreposições temáticas das comunidades de teorias da conspiração.

### 2.1. Extração de dados

Este projeto teve início em fevereiro de 2023, com a publicação da primeira versão do TelegramScrap (Silva, 2023), uma ferramenta própria e autoral, de software livre e de código aberto, que faz uso da Application Programming Interface (API) da plataforma Telegram por meio da biblioteca Telethon e organiza ciclos de extração de dados de grupos e canais abertos no Telegram. Ao longo dos meses, a base de dados pôde ser ampliada e qualificada fazendo uso de quatro abordagens de identificação de comunidades de teorias da conspiração:

**(i) Uso de palavras chave:** no início do projeto, foram elencadas palavras-chave para identificação diretamente no buscador de grupos e canais brasileiros no Telegram, tais como “apocalipse”, “sobrevivencialismo”, “mudanças climáticas”, “terra plana”, “teoria da conspiração”, “globalismo”, “nova ordem mundial”, “ocultismo”, “esoterismo”, “curas alternativas”, “qAnon”, “reptilianos”, “revisionismo”, “alienígenas”, dentre outras. Essa primeira abordagem forneceu algumas comunidades cujos títulos e/ou descrições dos grupos e canais contassem com os termos explícitos relacionados a teorias da conspiração. Contudo, com o tempo foi possível identificar outras diversas comunidades cujas palavras-chave elencadas não davam conta de abarcar, algumas inclusive propositalmente com caracteres trocados para dificultar a busca de quem a quisesse encontrar na rede;

**(ii) Mecanismo de recomendação de canais do Telegram:** com o tempo, foi identificado que canais do Telegram (exceto grupos) contam com uma aba de recomendação chamada de “canais similares”, onde o próprio Telegram sugere dez canais que tenham alguma similaridade com o canal que se está observando. A partir desse mecanismo de recomendação do próprio Telegram, foi possível encontrar mais comunidades de teorias da conspiração brasileiras, com estas sendo recomendadas pela própria plataforma;

**(iii) Abordagem de bola de neve para identificação de convites:** após algumas comunidades iniciais serem acumuladas para a extração, foi elaborado um algoritmo próprio autoral de identificação de urls que contivessem "t.me/", sendo o prefixo de qualquer convite para grupos e canais do Telegram. Acumulando uma frequência de centenas de milhares de links que atendessem a esse critério, foram elencados os endereços únicos e contabilizadas as suas repetições. Dessa forma, foi possível fazer uma investigação de novos grupos e canais brasileiros mencionados nas próprias mensagens dos já investigados, ampliando a rede. Esse processo foi sendo repetido periodicamente, buscando identificar novas comunidades que tivessem identificação com as temáticas de teorias da conspiração no Telegram;

**(iv) Ampliação para tweets publicados no X que mencionassem convites:** com o objetivo de diversificar ainda mais a fonte de comunidades de teorias da conspiração brasileiras no Telegram, foi elaborada uma query de busca própria que pudesse identificar as palavras-chave de temáticas de teorias da conspiração, porém usando como fonte tweets publicados no X (antigo Twitter) e que, além de conter alguma das palavras-chave, contivesse também o "t.me/", sendo o prefixo de qualquer convite para grupos e canais do Telegram, "https://x.com/search?q=lang%3Apt%20%22t.me%2F%22%20TERMO-DE-BUSCA".

Com as abordagens de identificação de comunidades de teorias da conspiração implementadas ao longo de meses de investigação e aprimoramento de método, foi possível construir uma base de dados do projeto com um total de 855 comunidades de teorias da conspiração brasileiras no Telegram (considerando as demais temáticas também não incluídas nesse estudo), estas somando 27.227.525 de conteúdos publicados entre maio de 2016 (primeiras publicações) até agosto de 2024 (realização deste estudo), com 2.290.621 usuários somados dentre as comunidades brasileiras. Há de se considerar que este volume de usuários conta com dois elementos, o primeiro é que trata-se de uma variável, pois usuários podem entrar e sair diariamente, portanto este valor representa o registrado na data de extração de publicações da comunidade; além disso, é possível que um mesmo usuário esteja em mais de um grupo e, portanto, é contabilizado mais de uma vez. Nesse sentido, o volume ainda sinaliza ser expressivo, mas pode ser levemente menor quando considerado o volume de cidadãos deduplicados dentro dessas comunidades brasileiras de teorias da conspiração.

### 2.2. Tratamento de dados

Com todos os grupos e canais brasileiros de teorias da conspiração no Telegram extraídos, foi realizada uma classificação manual considerando o título e a descrição da comunidade. Caso houvesse menção explícita no título ou na descrição da comunidade a alguma temática, esta foi classificada entre: (i) "Anticiência"; (ii) "Anti-Woke e Gênero"; (iii) "Antivax"; (iv) "Apocalipse e Sobrevivencialismo"; (v) "Mudanças Climáticas"; (vi) Terra Plana; (vii) "Globalismo"; (viii) "Nova Ordem Mundial"; (ix) "Ocultismo e Esoterismo"; (x) "Off Label e Charlatanismo"; (xi) "QAnon"; (xii) "Reptilianos e Criaturas"; (xiii) "Revisionismo e Discurso de Ódio"; (xiv) "OVNI e Universo". Caso não houvesse nenhuma menção explícita relacionada às temáticas no título ou na descrição da comunidade, esta foi classificada como (xv) "Conspiração Geral". Na Tabela a seguir, podemos observar as métricas relacionadas à classificação dessas comunidades de teorias da conspiração no Brasil.

**Tabela 01.** Comunidades de teorias da conspiração no Brasil (métricas até agosto de 2024)

| Categorias | Grupos | Usuários | Publicações | Comentários | Total |
|---|---|---|---|---|---|
| **Anticiência** | 22 | 58.138 | 187.585 | 784.331 | 971.916 |
| **Anti-*Woke* e Gênero** | 43 | 154.391 | 276.018 | 1.017.412 | 1.293.430 |
| **Antivacinas (*Antivax*)** | 111 | 239.309 | 1.778.587 | 1.965.381 | 3.743.968 |
| **Apocalipse e Sobrevivência** | 33 | 109.266 | 915.584 | 429.476 | 1.345.060 |
| **Mudanças Climáticas** | 14 | 20.114 | 269.203 | 46.819 | 316.022 |
| **Terraplanismo** | 33 | 38.563 | 354.200 | 1.025.039 | 1.379.239 |
| **Conspirações Gerais** | 127 | 498.190 | 2.671.440 | 3.498.492 | 6.169.932 |
| **Globalismo** | 41 | 326.596 | 768.176 | 537.087 | 1.305.263 |
| **Nova Ordem Mundial (NOM)** | 148 | 329.304 | 2.411.003 | 1.077.683 | 3.488.686 |
| **Ocultismo e Esoterismo** | 39 | 82.872 | 927.708 | 2.098.357 | 3.026.065 |
| **Medicamentos *off label*** | 84 | 201.342 | 929.156 | 733.638 | 1.662.794 |
| **QAnon** | 28 | 62.346 | 531.678 | 219.742 | 751.420 |
| **Reptilianos e Criaturas** | 19 | 82.290 | 96.262 | 62.342 | 158.604 |
| **Revisionismo e Ódio** | 66 | 34.380 | 204.453 | 142.266 | 346.719 |
| **OVNI e Universo** | 47 | 58.912 | 862.358 | 406.049 | 1.268.407 |
| **Total** | **855** | **2.296.013** | **13.183.411** | **14.044.114** | **27.227.525** |

Fonte: Elaboração própria (2024).

Com esse volume de dados extraídos, foi possível segmentar para apresentar neste estudo apenas comunidades e conteúdos referentes às temáticas de antivacinas (*antivax*) e tratamentos e medicamentos alternativos (*off label*). Em paralelo, as demais temáticas de comunidades brasileiras de teorias da conspiração também contaram com estudos elaborados para a caracterização da extensão e da dinâmica da rede, estes sendo disponibilizados abertamente e originalmente no arXiv da Cornell University.

Além disso, cabe citar que apenas foram extraídas comunidades abertas, isto é, não apenas identificáveis publicamente, mas também sem necessidade de solicitação para acessar ao conteúdo, estando aberto para todo e qualquer usuário com alguma conta do Telegram sem que este necessite ingressar no grupo ou canal. Além disso, em respeito à legislação brasileira e especialmente da Lei Geral de Proteção de Dados Pessoais (LGPD), ou Lei nº 13.709/2018, que trata do controle da privacidade e do uso/tratamento de dados pessoais, todos os dados extraídos foram anonimizados para a realização de análises e investigações. Dessa forma, nem mesmo a identificação das comunidades é possível por meio deste estudo, estendendo aqui a privacidade do usuário ao anonimizar os seus dados para além da própria comunidade à qual ele se submeteu ao estar em um grupo ou canal público e aberto no Telegram.

### 2.3. Abordagens para análise de dados

Totalizando 195 comunidades selecionadas nas temáticas de antivacinas (*antivax*) e tratamentos e medicamentos alternativos (*off label*), contendo 5.406.762 publicações e 440.651 usuários somados, quatro abordagens serão utilizadas para investigar as comunidades de teorias da conspiração selecionadas para o escopo do estudo. Tais métricas são detalhadas na Tabela a seguir:

**Tabela 02.** Comunidades selecionadas para análise (métricas até agosto de 2024)

| Categorias | Grupos | Usuários | Publicações | Comentários | Total |
|---|---|---|---|---|---|
| **Antivacinas (*Antivax*)** | 111 | 239.309 | 1.778.587 | 1.965.381 | 3.743.968 |
| **Medicamentos *Off Label*** | 84 | 201.342 | 929.156 | 733.638 | 1.662.794 |
| **Total** | **195** | **440.651** | **2.707.743** | **2.699.019** | **5.406.762** |

Fonte: Elaboração própria (2024).

**(i) Rede:** com a elaboração de um algoritmo próprio para a identificação de mensagens que contenham o termo de “t.me/” (de convite para entrarem em outras comunidades), propomos apresentar volumes e conexões observadas sobre como **(a)** as comunidades da temática de antivacinas (*antivax*) e tratamentos e medicamentos alternativos (*off label*) circulam convites para que os seus usuários conheçam mais grupos e canais da mesma temática, reforçando os sistemas de crença que comungam; e como **(b)** essas mesmas comunidades circulam convites para que os seus usuários conheçam grupos e canais que tratem de outras teorias da conspiração, distintas de seu propósito explícito. Esta abordagem é interessante para observar se essas comunidades utilizam a si próprias como fonte de legitimação e referência e/ou se embasam-se em demais temáticas de teorias da conspiração, inclusive abrindo portas para que seus usuários conheçam outras conspirações. Além disso, cabe citar o estudo de Rocha *et al.* (2024) em que uma abordagem de identificação de rede também foi aplicada em comunidades do Telegram, porém observando conteúdos similares a partir de um ID gerado para cada mensagem única e suas similares;

**(ii) Séries temporais:** utiliza-se da biblioteca “Pandas” (McKinney, 2010) para organizar os data frames de investigação, observando **(a)** o volume de publicações ao longo dos meses; e **(b)** o volume de engajamento ao longo dos meses, considerando metadados de visualizações, reações e comentários coletados na extração; Além da volumetria, a biblioteca “Plotly” (Plotly Technologies Inc., 2015) viabilizou a representação gráfica dessa variação;

**(iii) Análise de conteúdo:** além da análise geral de palavras com identificação das frequências, são aplicadas séries temporais na variação das palavras mais frequentes ao longo dos semestres — observando entre maio de 2017 (primeiras publicações) até agosto de 2024 (realização deste estudo). E com as bibliotecas “Pandas” (McKinney, 2010) e “WordCloud” (Mueller, 2020), os resultados são apresentados tanto volumetricamente quanto graficamente;

**(iv) Sobreposição de agenda temática:** seguindo a abordagem proposta por Silva & Sátiro (2024) para identificação de sobreposição de agenda temática em comunidades do Telegram, utilizamos o modelo “BERTopic” (Grootendorst, 2020). O BERTopic é um algoritmo de modelagem de tópicos que facilita a geração de representações temáticas a partir de grandes quantidades de textos. Primeiramente, o algoritmo extrai embeddings dos documentos usando modelos transformadores de sentenças, como o “all-MiniLM-L6-v2”. Em seguida, essas embeddings têm sua dimensionalidade reduzida por técnicas como “UMAP”, facilitando o processo de agrupamento. A clusterização é realizada usando “HDBSCAN”, uma técnica baseada em densidade que identifica clusters de diferentes formas e tamanhos, além de detectar outliers. Posteriormente, os documentos são tokenizados e representados em uma estrutura de bag-of-words, que é normalizada (L1) para considerar as diferenças de tamanho entre os clusters. A representação dos tópicos é refinada usando uma versão modificada do “TF-IDF”, chamada “Class-TF-IDF”, que considera a importância das palavras dentro de cada cluster (Grootendorst, 2020). Cabe destacar que, antes de aplicar o BERTopic, realizamos a limpeza da base removendo “stopwords” em português, por meio da biblioteca “NLTK” (Loper & Bird, 2002). Para a modelagem de tópicos, utilizamos o backend “loky” para otimizar o desempenho durante o ajuste e a transformação dos dados.

Em síntese, a metodologia aplicada compreendeu desde a extração de dados com a ferramenta própria autoral TelegramScrap (Silva, 2023), até o tratamento e a análise de dados coletados, utilizando diversas abordagens para identificar e classificar comunidades de teorias da conspiração brasileiras no Telegram. Cada uma das etapas foi cuidadosamente implementada para garantir a integridade dos dados e o respeito à privacidade dos usuários, conforme a legislação brasileira prevê. A seguir, serão apresentados os resultados desses dados, com o intuito de revelar as dinâmicas e as características das comunidades estudadas.

## 3. Resultados

A seguir, os resultados são detalhados na ordem prevista na metodologia, iniciando com a caracterização da rede e suas fontes de legitimação e referência, avançando para as séries temporais, incorporando a análise de conteúdo e concluindo com a identificação de sobreposição de agenda temática dentre as comunidades de teorias da conspiração.

### 3.1. Rede

As figuras a seguir oferecem uma análise abrangente das redes de comunidades que se conectam em torno de narrativas conspiratórias. Essas redes revelam a complexidade das interações e como as narrativas são estrategicamente reforçadas e ampliadas por meio de convites cruzados entre diversos grupos temáticos. As figuras ilustram como certas comunidades funcionam como *hubs* centrais, desempenhando um papel crucial na introdução e disseminação de desinformação, conectando tópicos aparentemente distintos, mas que juntos compõem um ecossistema coeso de teorias conspiratórias.

Especificamente, analisaremos como temas como Antivacinas, Nova Ordem Mundial, Ocultismo e Esoterismo, e Globalismo não apenas centralizam discussões, mas também atuam como trampolins que guiam os membros para um espectro mais amplo de desinformação, reforçando continuamente crenças e ampliando o alcance dessas teorias. Além disso, os fluxos de links de convites entre essas comunidades indicam um padrão de interdependência, onde a desinformação sobre saúde é intimamente ligada a narrativas de controle global, crises apocalípticas e manipulação por elites ocultas.

**Figura 01.** Rede interna entre antivacinas e medicamentos *off label*

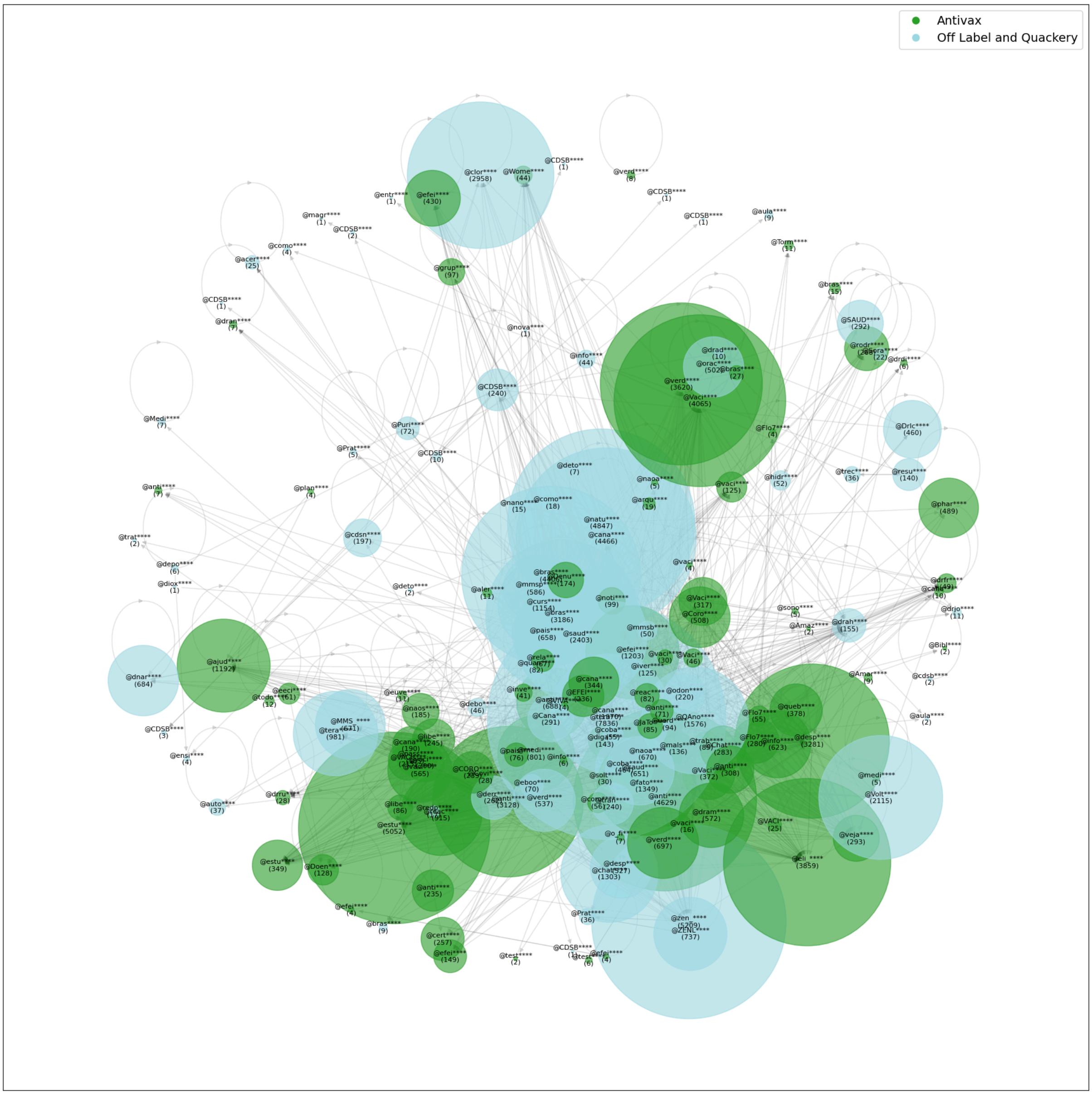

Fonte: Elaboração própria (2024).

A figura apresenta uma análise detalhada da rede interna que conecta comunidades antivacinas e aquelas que promovem o uso de medicamentos *off label*. A robustez das conexões dentro dessa rede sugere uma forte interdependência entre essas temáticas, onde narrativas sobre os perigos das vacinas são frequentemente reforçadas por alegações sobre a eficácia de tratamentos alternativos e não convencionais. Essa interdependência cria um ambiente no qual os seguidores são constantemente expostos a informações que validam suas crenças antivacinas, ao mesmo tempo em que oferecem soluções em forma de medicamentos *off label*. Os grandes nós representando as principais comunidades dentro dessa rede indicam que há certas comunidades que atuam como líderes de opinião ou centros de disseminação, onde informações e desinformações são amplamente compartilhadas e amplificadas. A densidade e a coesão dessa rede interna refletem a criação de um ecossistema próprio, onde as narrativas conspiratórias relacionadas à saúde se retroalimentam, criando um ciclo vicioso de desinformação. Essa dinâmica sugere que uma vez dentro dessa rede, é difícil para os seguidores se desvencilharem dessas crenças, pois estão continuamente expostos a narrativas que reforçam sua visão de mundo distorcida, tornando a comunidade um espaço fechado de radicalização e resistência a informações contrárias ou baseadas em evidências científicas.

**Figura 02.** Rede de comunidades que abrem portas para a temática (porta de entrada)

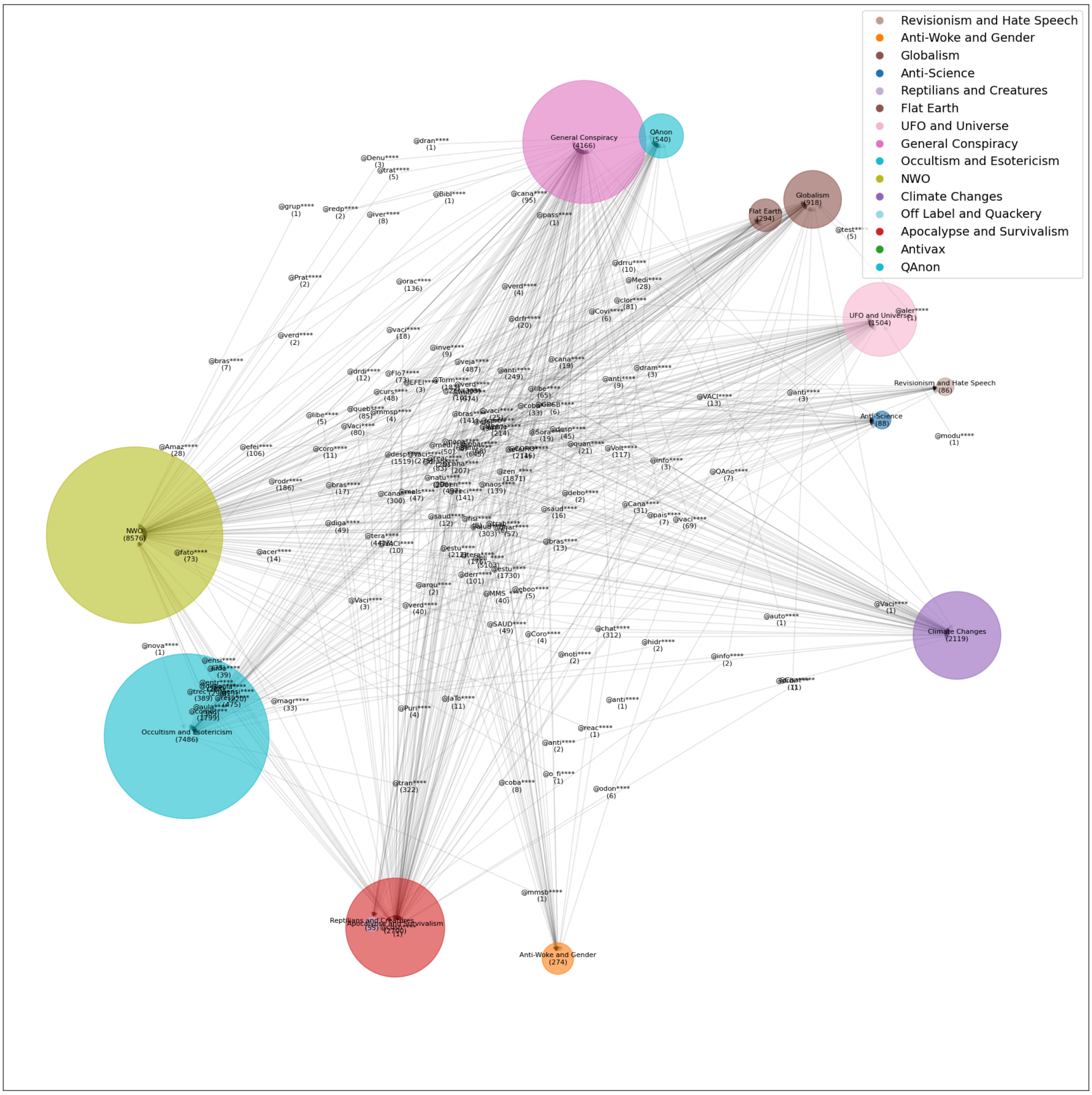

Fonte: Elaboração própria (2024).

Esta figura retrata a complexa teia de comunidades que servem como pontos de entrada para indivíduos interessados em teorias da conspiração. Ao observar a estrutura da rede, percebe-se que os nós maiores, como Nova Ordem Mundial, Ocultismo e Esoterismo e Mudanças Climáticas, exercem um papel fundamental na introdução de novos membros a esses temas. Essas comunidades funcionam como *hubs*, centralizando e disseminando informações que atraem seguidores para uma vasta gama de teorias. A posição de destaque desses *hubs* sugere que eles não apenas atuam como portas de entrada, mas também como centros de aglutinação de narrativas, onde diversas temáticas se entrelaçam e criam uma visão de mundo coesa e amplificada. Isso reflete a capacidade dessas comunidades de oferecer um ponto de convergência para indivíduos que estão começando a explorar essas teorias, estabelecendo um ponto de referência que conecta temas aparentemente díspares, mas que, dentro dessas comunidades, são apresentados como parte de um quadro maior de "verdades ocultas". Além disso, a densidade das conexões ao redor desses *hubs* evidencia uma rede altamente interligada, onde a troca de narrativas é intensa e contínua, potencializando o alcance e a influência dessas comunidades na introdução de novos adeptos.

**Figura 03.** Rede de comunidades cuja temática abre portas (porta de saída)

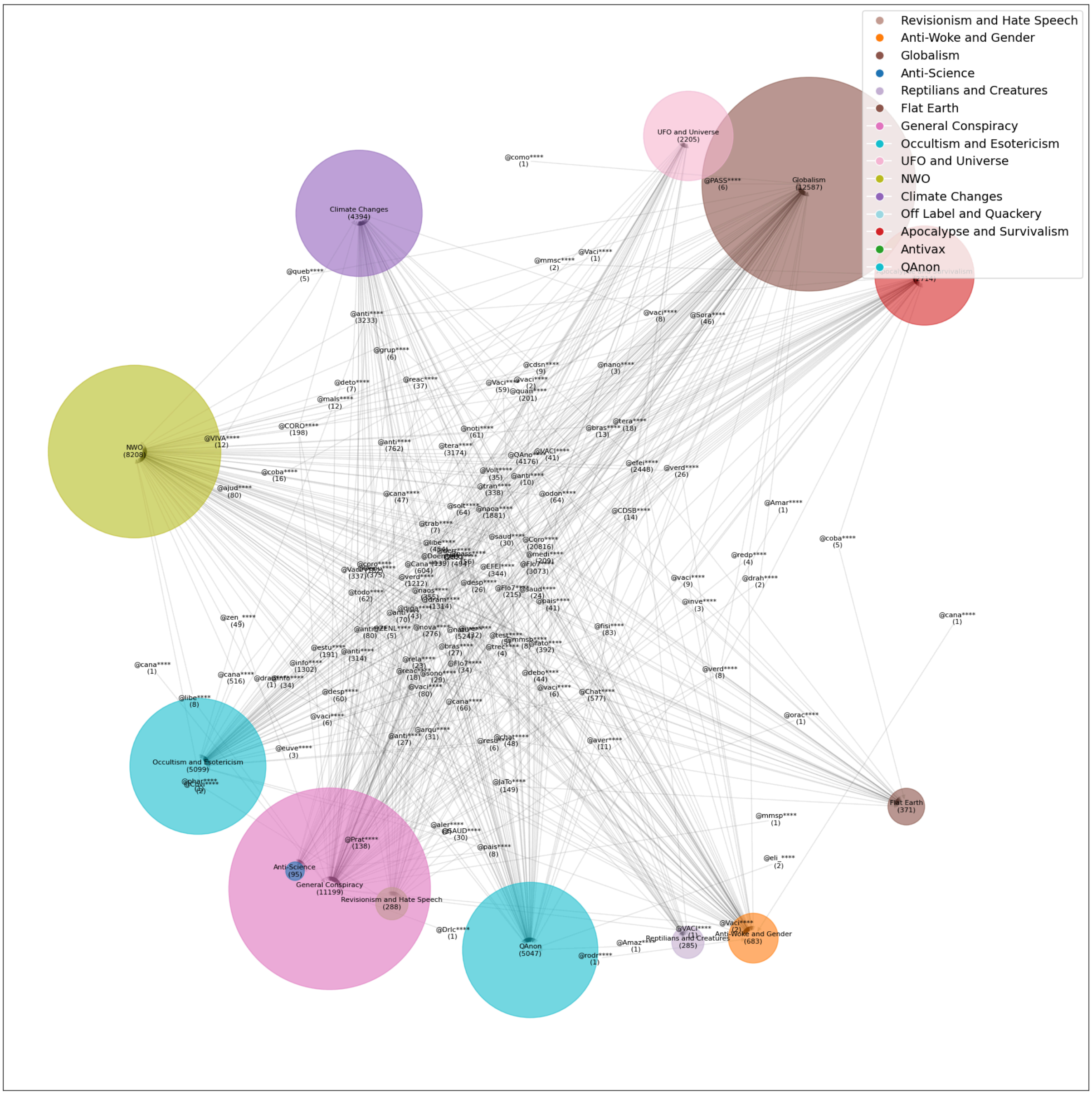

Fonte: Elaboração própria (2024).

Esta figura destaca as comunidades que, a partir de suas temáticas centrais, servem como trampolins para outras discussões dentro do universo conspiratório. Comunidades como Globalismo e Mudanças Climáticas não apenas atraem seguidores, mas também os preparam para transitar para outros temas correlatos, como Nova Ordem Mundial. A centralidade dessas comunidades na rede indica que elas desempenham um papel crucial na expansão das narrativas conspiratórias, agindo como intermediárias que conectam diferentes áreas de interesse. A estrutura da rede sugere que, uma vez dentro dessas comunidades, os indivíduos são expostos a um leque maior de teorias, que ampliam sua visão de mundo conspiratória.

Essa função de "porta de saída" não implica apenas em uma transição para outras temáticas, mas em um aprofundamento dentro do ecossistema conspiratório, onde a complexidade e a interconexão das narrativas aumentam à medida que os indivíduos se engajam ainda mais.

**Figura 04.** Fluxo de links de convites entre comunidades de antivacinas

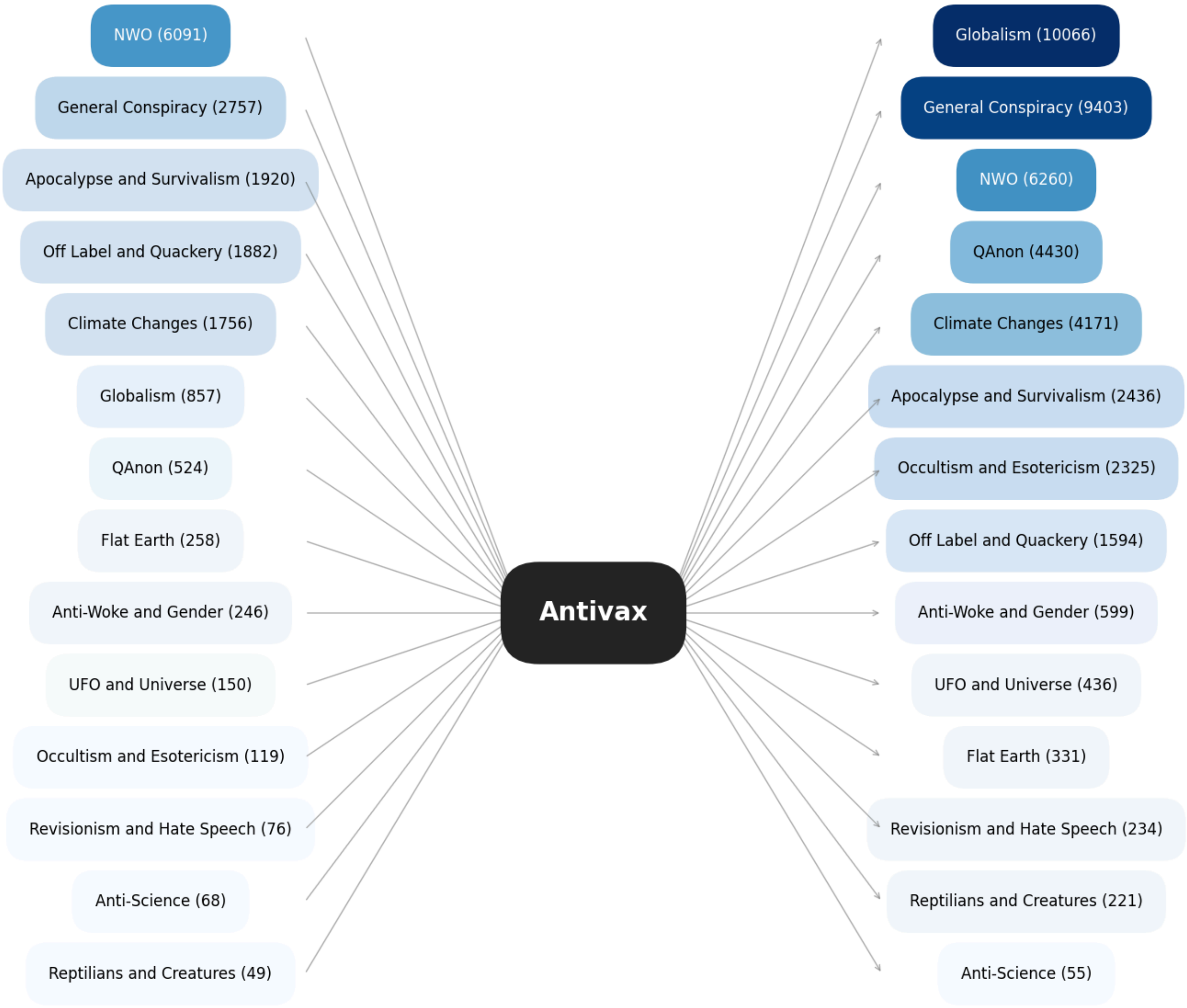

Fonte: Elaboração própria (2024).

Observa-se que a comunidade Antivacinas (*antivax*) desempenha um papel central e altamente interconectado, servindo como um nó que distribui convites para uma ampla variedade de outros grupos temáticos. À esquerda, vemos como as discussões antivacinas se conectam frequentemente com grupos relacionados à Nova Ordem Mundial (NOM), com 6.091 links de convites, seguidos por Conspirações Gerais com 2.757 links e Apocalipse e Sobrevivência com 1.920. Esses números revelam que a narrativa antivacinas é frequentemente entrelaçada com grandes teorias conspiratórias, sugerindo que aqueles que ingressam em comunidades antivacinas estão rapidamente expostos a uma rede maior de desinformação e medo. À direita, os grupos que mais recebem convites de comunidades antivacinas incluem Globalismo, com 10.066 links, Conspirações Gerais com 9.403 links, e NOM com 6.260 links. Isso sugere uma forte tendência de cruzamento de membros entre essas temáticas, mostrando que as ideias propagadas nas comunidades antivacinas não se limitam apenas à saúde, mas se expandem para teorias de dominação global, manipulação e crises apocalípticas produzidas por uma suposta elite. Este padrão evidencia como as teorias antivacinas funcionam como uma porta de entrada para uma vasta rede de desinformação, onde cada narrativa reforça as demais, criando um ciclo contínuo de radicalização.

**Figura 05.** Fluxo de links de convites entre comunidades de medicamentos *off label*

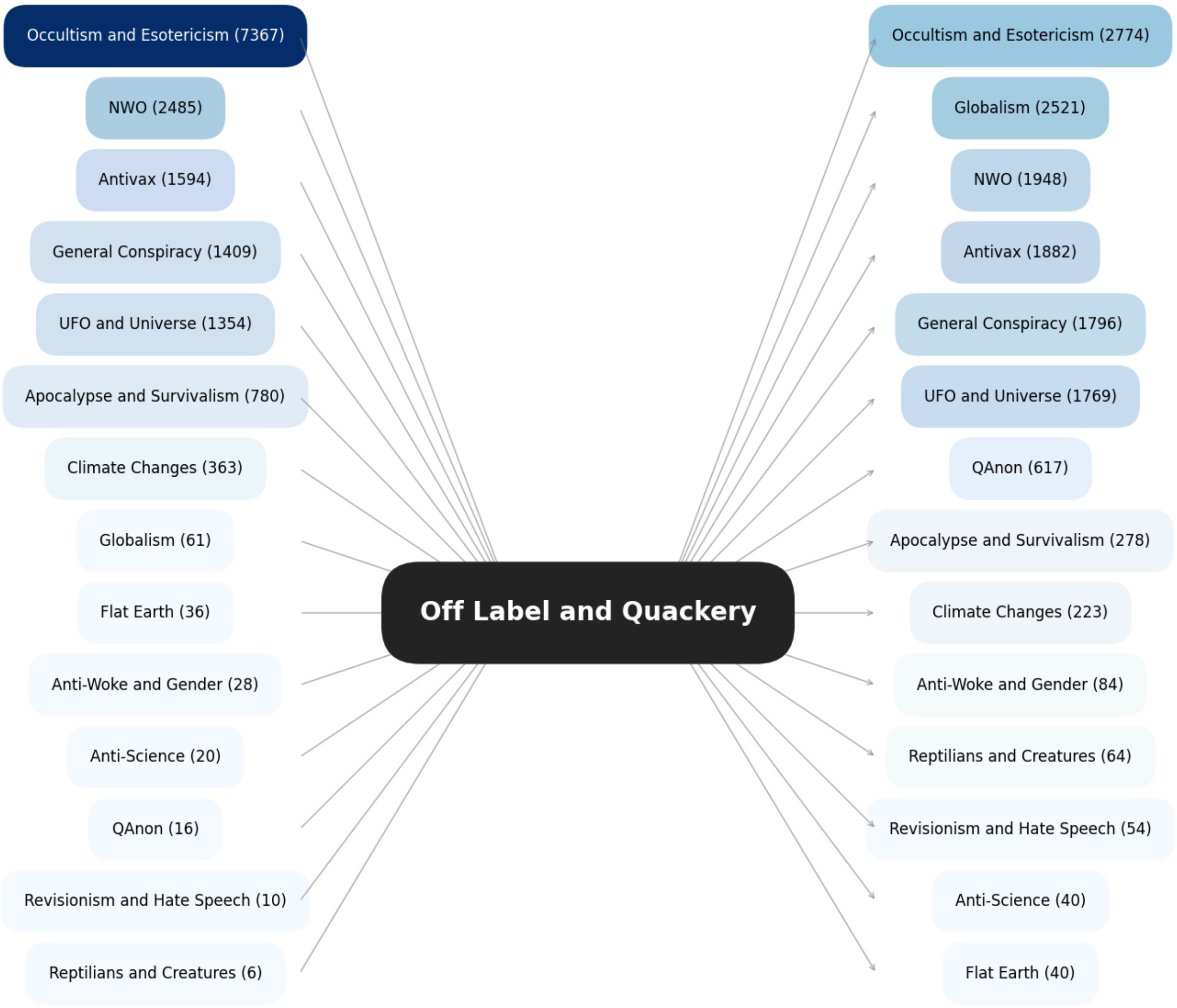

Fonte: Elaboração própria (2024).

Assim como na figura anterior, os medicamentos *off label* são um tema central que conecta várias outras narrativas conspiratórias. À esquerda, Ocultismo e Esoterismo surge como a comunidade mais fortemente conectada com 7.367 links, sugerindo uma associação entre a promoção de tratamentos alternativos e crenças esotéricas. Isso pode indicar uma tendência entre os membros dessas comunidades em buscar "curas" alternativas que muitas vezes são fundamentadas em práticas não científicas ou esotéricas, até mesmo nocivas à saúde da sociedade. Outras conexões significativas incluem NOM com 2.485 links e Antivacinas com 1.594, mostrando que aqueles que são atraídos para a discussão sobre medicamentos *off* *label* também são propensos a serem envolvidos em debates antivacinas e teorias de conspiração global. À direita, as comunidades que mais recebem convites de grupos discutindo medicamentos *off label* incluem Ocultismo e Esoterismo com 2.774 links, Globalismo com 2.521 links e NOM com 1.948 links. Este fluxo de convites reforça a interconexão entre a desinformação sobre saúde e uma gama de outras teorias conspiratórias, especialmente aquelas relacionadas a uma ordem mundial oculta ou manipulação global. O padrão de links sugere que a promoção de medicamentos *off label* é parte de uma estratégia mais ampla de desinformação, onde diferentes narrativas se reforçam mutuamente, aumentando a credibilidade dessas crenças entre os membros dessas comunidades.

### 3.2. Séries temporais

As próximas figuras apresentam uma análise abrangente da evolução das discussões relacionadas a temáticas antivacinas e medicamentos *off label* ao longo dos últimos anos, destacando, em especial, o período marcado pela Pandemia da COVID-19. Por meio de diferentes representações gráficas — incluindo gráficos de linhas, áreas absolutas e áreas relativas —, é possível observar como esses assuntos ganharam proeminência no discurso público, refletindo tensões sociais, desinformação e a influência de eventos globais na percepção coletiva sobre saúde e ciência. Essas visualizações evidenciam não apenas o aumento substancial no volume de debates e compartilhamentos sobre essas temáticas durante os momentos mais críticos da Pandemia, mas também ilustram como essas narrativas se mantiveram persistentes e adaptáveis mesmo após o auge da crise sanitária.

**Figura 06.** Gráfico de linhas do período

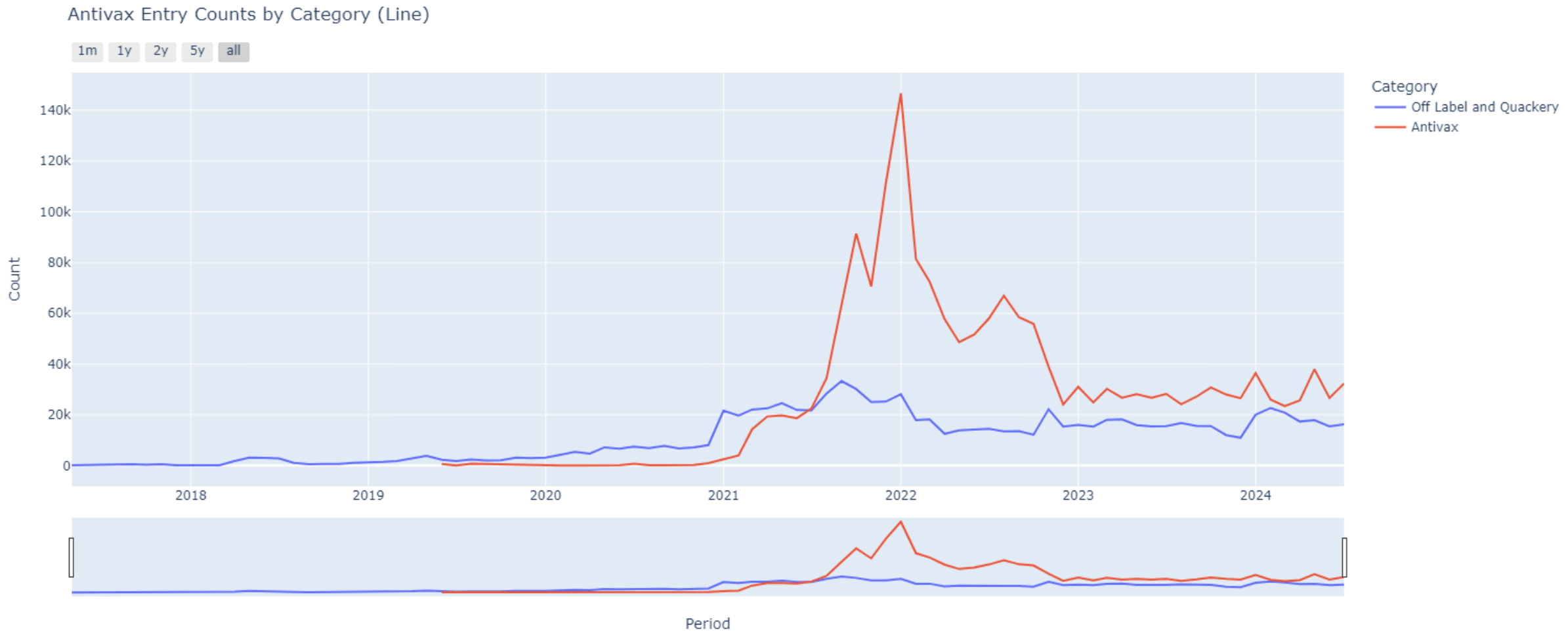

Fonte: Elaboração própria (2024).

O gráfico de linhas do período revela as flutuações nas discussões sobre antivacinas e medicamentos *off label* ao longo dos anos, destacando-se especialmente o impacto profundo que a Pandemia da COVID-19 teve na propagação dessas narrativas. Antes de 2020, ambas as temáticas mantinham um fluxo relativamente estável, com poucas variações significativas. No entanto, a partir do início de 2020, com o avanço global do coronavírus e as subsequentes medidas de saúde pública, como a vacinação em massa, observa-se um aumento abrupto nas menções à temática antivacinas, culminando em um pico notável em 2021 e 2022. Esse aumento pode ser relacionado à proliferação de desinformação em torno das vacinas contra a COVID-19. As teorias da conspiração relacionadas à eficácia e segurança das vacinas ganharam força nesse período, alimentadas por movimentos antivacinas já existentes fora das redes sociais. Paralelamente, o uso de medicamentos *off label*, como a hidroxicloroquina e a ivermectina, também cresceu, refletindo o desespero de uma parcela da população e a promoção irresponsável de tratamentos não comprovados por figuras públicas. Após o pico, o gráfico indica uma tendência de estabilização, mas ainda com níveis elevados em comparação ao período pré-Pandemia, o que sugere que essas narrativas enraizaram-se em certos segmentos da sociedade, mantendo um discurso persistente contra a vacinação e pró-*off label*.

**Figura 07.** Gráfico de área absoluta do período

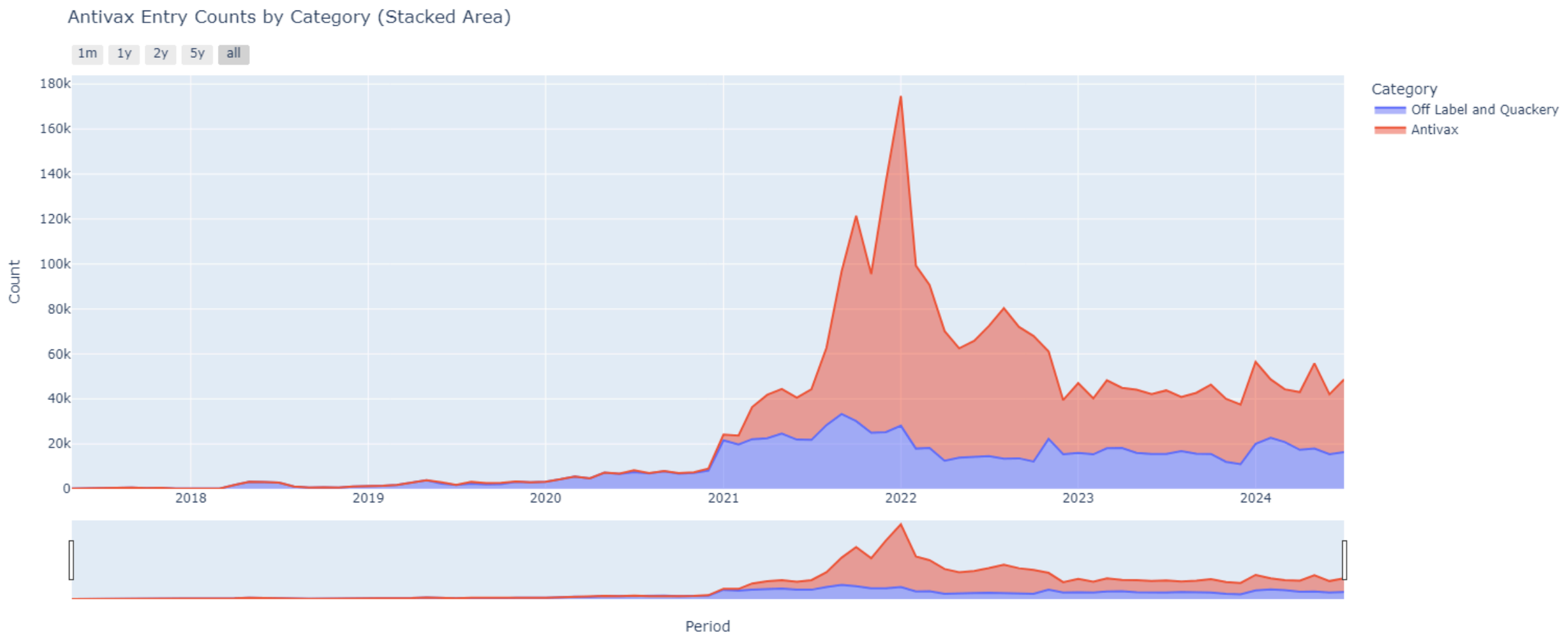

Fonte: Elaboração própria (2024).

O gráfico de área absoluta do período expande a análise anterior, visualizando de forma clara o volume total de discussões ao longo do tempo, dividido entre antivacinas e medicamentos *off label*. A área preenchida de cada categoria ilustra não apenas o aumento drástico nas menções durante a Pandemia da COVID-19, mas também o impacto duradouro dessas narrativas na esfera pública. A partir de 2020, a área referente ao antivacinas aumenta significativamente, evidenciando como a resistência às vacinas se tornou uma das principais preocupações durante a crise sanitária. O gráfico também mostra um crescimento paralelo, embora em menor escala, nas discussões sobre medicamentos *off label*, que se tornaram populares à medida que as pessoas procuravam alternativas aos tratamentos convencionais recomendados por autoridades de saúde. O pico em 2022 sugere um período de intensa controvérsia e polarização, quando essas narrativas atingiram seu ápice de popularidade, antes de começar a declinar. No entanto, a área residual considerável após o pico indica que, embora a intensidade das discussões tenha diminuído, essas temáticas continuam a ocupar um espaço significativo no discurso público, demonstrando o poder de permanência dessas crenças e sua capacidade de resistir à correção factual.

**Figura 08.** Gráfico de área relativa do período

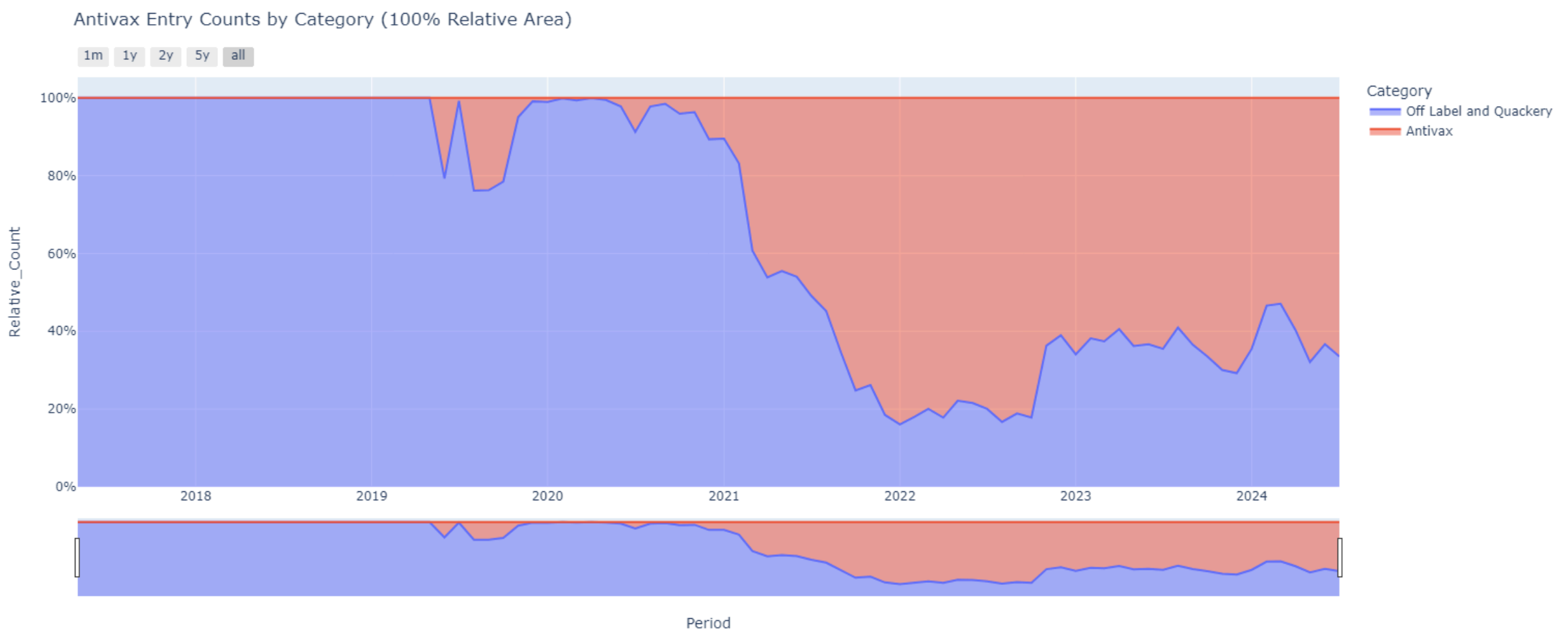

Fonte: Elaboração própria (2024).

O gráfico de área relativa do período oferece uma perspectiva sobre a dominância relativa de cada temática ao longo do tempo, mostrando como a narrativa antivacinas ganhou proeminência durante a Pandemia da COVID-19, superando significativamente as discussões sobre medicamentos *off label*. Antes de 2020, ambas as temáticas ocupavam proporções relativamente equilibradas, com uma leve predominância de discussões sobre tratamentos alternativos. Contudo, com o advento da Pandemia e a subsequente corrida global pela vacinação, as menções a antivacinas rapidamente ultrapassaram as de medicamentos *off label*, atingindo um domínio quase absoluto no período de 2021 até 2022. Este domínio reflete a centralidade da discussão sobre vacinas nas controvérsias públicas e a mobilização de movimentos antivacinas, que conseguiram canalizar temores e desconfianças em relação às vacinas para um público mais amplo. Esse gráfico ilustra nitidamente as mudanças nas prioridades discursivas ao longo do tempo, mostrando como crises globais podem reconfigurar a paisagem das narrativas conspiratórias e de desinformação.

### 3.3. Análise de conteúdo

As nuvens de palavras a seguir proporcionam uma análise visual das principais narrativas emergentes em comunidades online que discutem antivacinas e medicamentos *off label* ao longo dos anos. Essas representações revelam como certas palavras-chave se consolidaram e evoluíram dentro desses discursos, refletindo a interseção entre saúde, política e crenças individuais. A análise dessas palavras não apenas expõe os temas centrais que dominaram as conversas durante a Pandemia da COVID-19, mas também ilustra como essas narrativas se enraizaram e se adaptaram ao contexto social e político em constante mudança. A persistência de termos relacionados à "verdade", "liberdade" e "vacina" sugere uma batalha contínua por controle narrativo, onde as discussões sobre saúde pública se entrelaçam com questões de identidade, poder e resistência a medidas institucionais.

**Figura 09.** Nuvem de palavras consolidadas de antivacinas e medicamentos *off label*

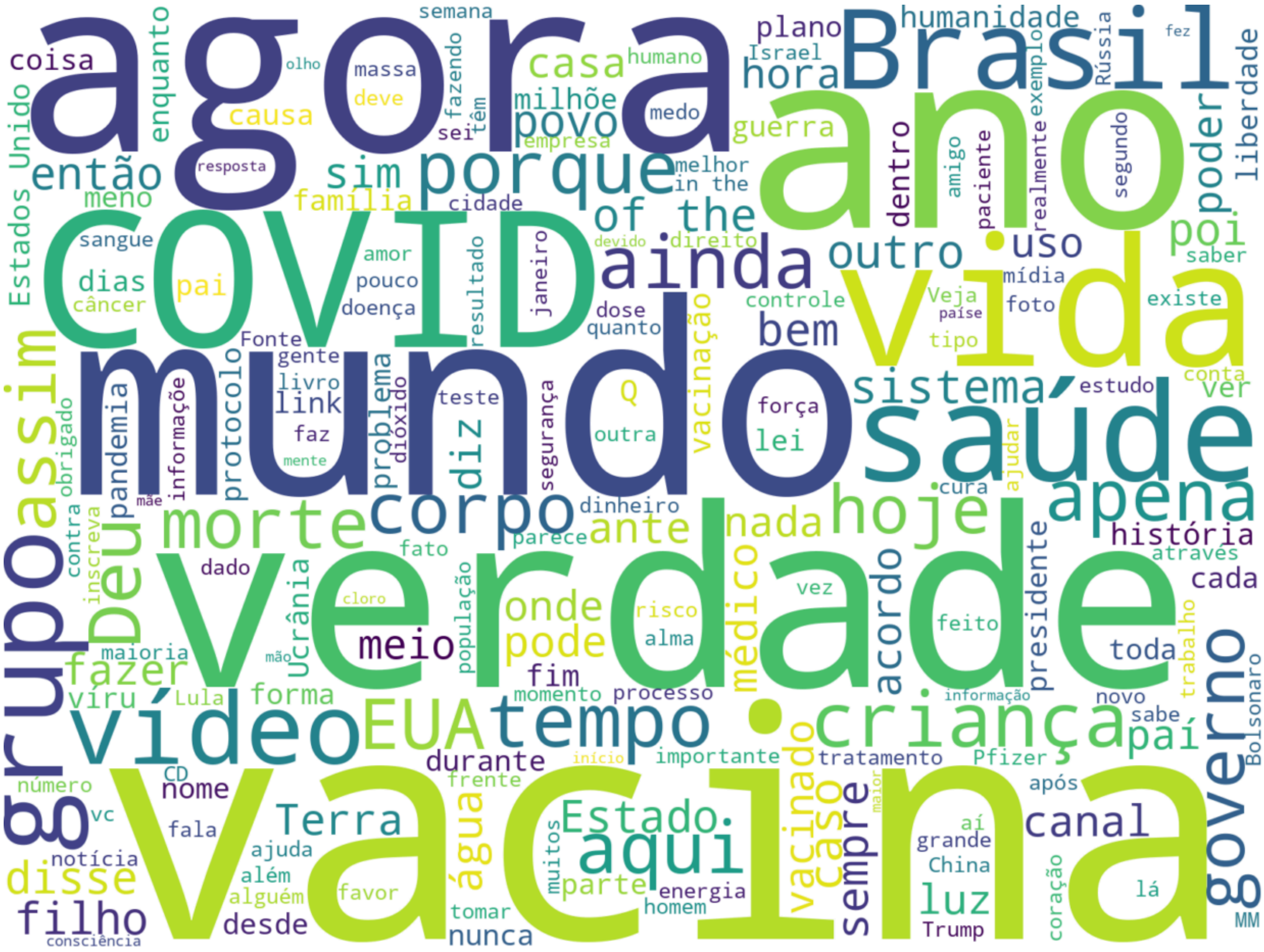

Fonte: Elaboração própria (2024).

A nuvem de palavras consolidadas revela a predominância e a inter-relação das narrativas em torno de antivacinas e medicamentos *off label*. Termos como "agora", "verdade", "vacina", "mundo", "vida" e "covid" destacam-se, refletindo como essas palavras servem como pilares de uma narrativa que mistura urgência, desconfiança e desafios à autoridade científica. Ao observarmos as publicações para maior contexto, vemos que a palavra "agora" sugere uma sensação de imediatismo e pressão temporal, frequentemente utilizada em publicações com apelos emocionais ou apelos à ação, enquanto "verdade" reflete a busca ou a afirmação de uma realidade alternativa, típica em discussões conspiratórias. A recorrência de "vacina" e "covid" evidencia o impacto profundo da Pandemia na polarização do debate sobre saúde, onde "vida" e "morte" aparecem como polos simbólicos. A menção a "sistema", "governo", e "Brasil" indica a inserção dessas discussões em um contexto político mais amplo, onde as medidas de saúde pública são frequentemente vistas como parte de uma agenda maior, alimentando a desconfiança em relação às instituições. Essa nuvem de palavras sugere que a narrativa antivacinas não é isolada, mas interligada a um discurso mais abrangente sobre poder, argumentos sobre liberdades individuais e/ou coletivas, e identidade nacional, onde a saúde pública torna-se um campo de batalha ideológico.

**Quadro 01.** Nuvem de palavras em série temporal de antivacinas

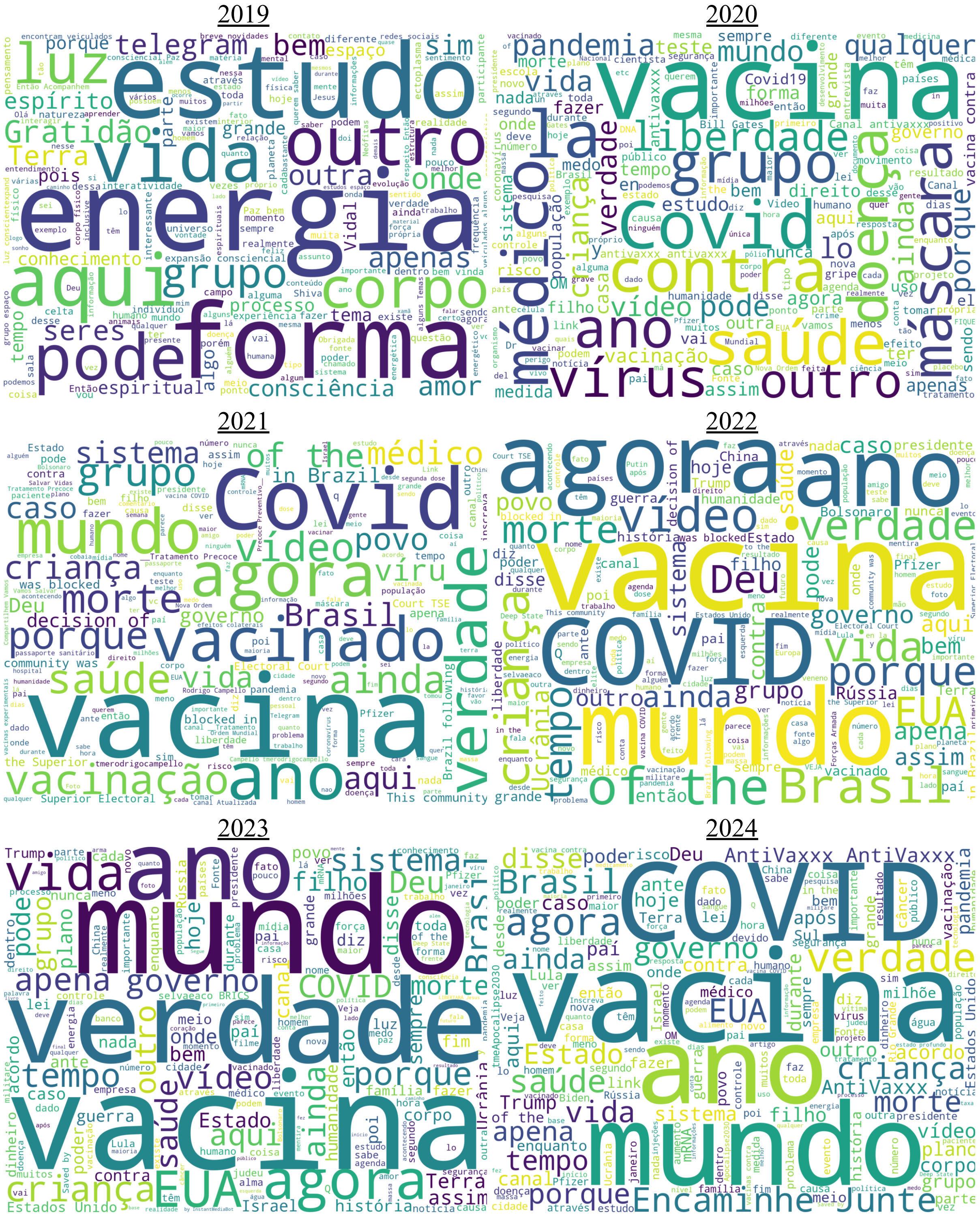

Fonte: Elaboração própria (2024).

A série temporal de nuvens de palavras para antivacinas revela a evolução das discussões ao longo dos anos. Em 2019, palavras como "energia" e "forma" sugerem um foco em temas mais esotéricos ou alternativos, que então se expandem dramaticamente para "vacina", "covid" e "liberdade" em 2020, com o advento da Pandemia. A palavra "controle", em 2020, reflete o medo amplamente disseminado de que as vacinas fossem utilizadas como uma forma de manipulação ou dominação, uma teoria que ganhou força com a apresentação

de propostas como de quarentena e *lockdowns*. Em 2021, “agora” e “verdade” tornam-se centrais, indicando uma intensificação na disputa narrativa, onde a antivacinação é apresentada não apenas como uma escolha de saúde, mas como uma luta pela verdade contra um sistema percebido como supostamente opressor ao propor a saúde coletiva via vacinação. Em 2022 e 2023, palavras como “morte”, “mundo” e “Brasil” indicam uma nacionalização do debate, onde a resistência às vacinas é vinculada a identidades políticas e culturais. A constância de “vacina” ao longo dos anos, especialmente em 2024, mostra que, mesmo após o pico da Pandemia, a temática continua a ser uma preocupação central, revelando a persistência da desinformação e a dificuldade de erradicar essas crenças profundamente enraizadas.

**Quadro 02.** Nuvem de palavras em série temporal de medicamentos *off label*

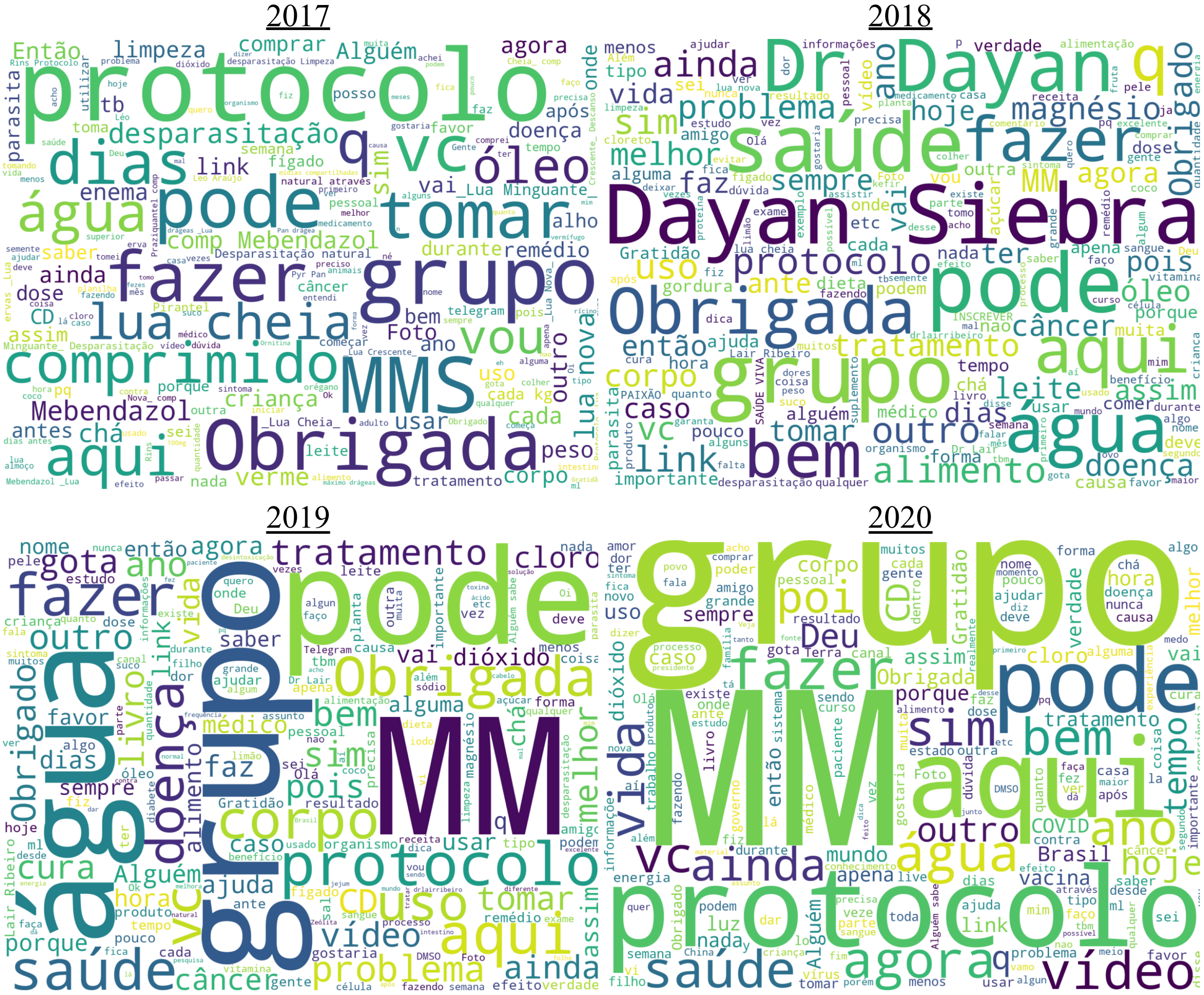

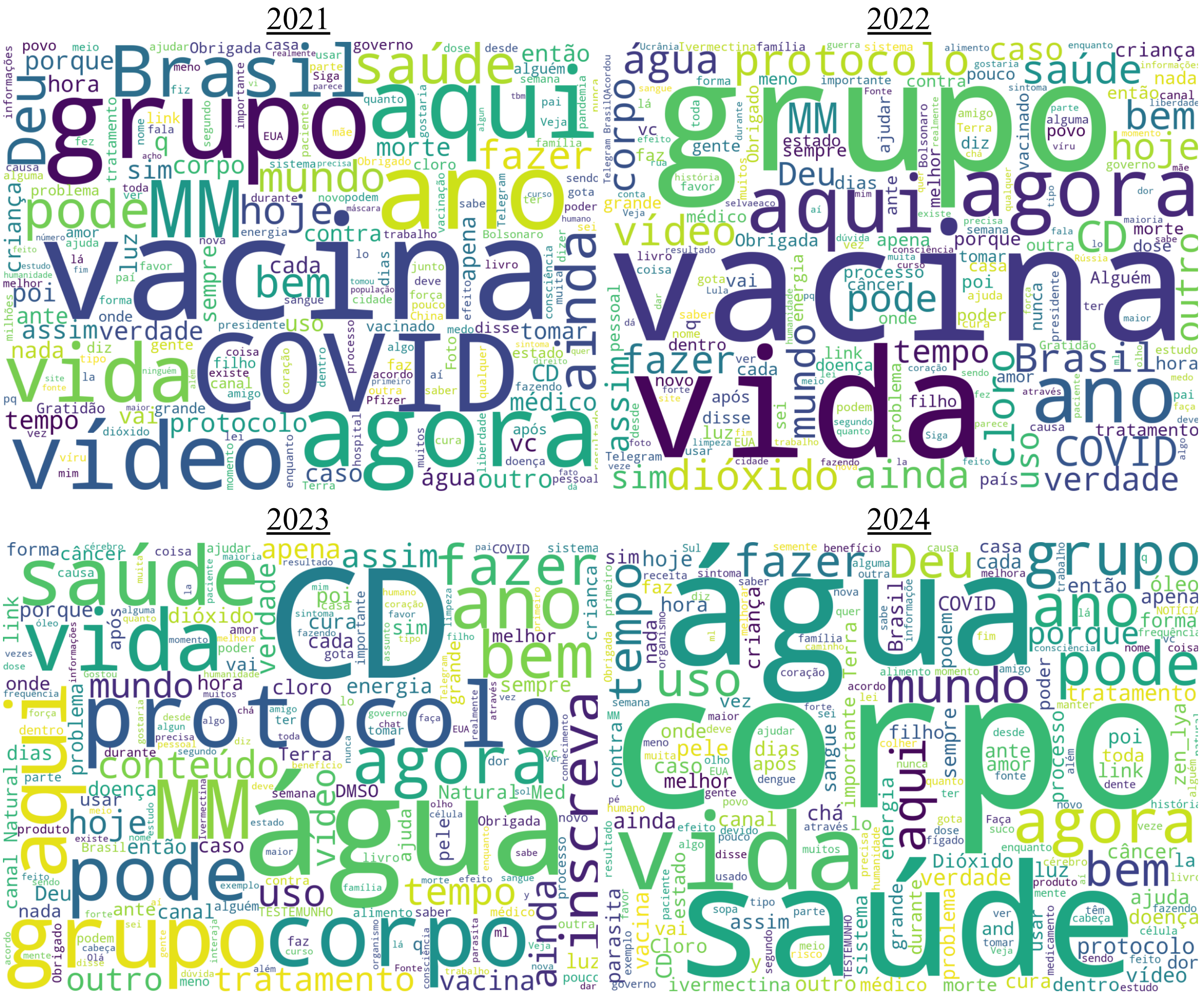

Fonte: Elaboração própria (2024).

A série temporal das nuvens de palavras sobre medicamentos *off label* destaca a evolução das discussões em torno de tratamentos alternativos, especialmente aqueles relacionados ao dióxido de cloro (CDS) e MMS (Solução Mineral Milagrosa), que foram amplamente promovidos como “detox vacinal” durante a Pandemia da COVID-19. É importante destacar que muitos produtos são comercializados nas comunidades como supostos medicamentos milagrosos. Além de lucrarem com a venda de frascos potencialmente nocivos à saúde das pessoas, também ocorre a monetização de infoprodutos, como e-books e cursos onlines sobre práticas de produção de “detox vacinal”, por exemplo.

Nos primeiros anos, como 2017 e 2018, palavras como “Dayan Siebra” e “MMS” já indicam a influência de figuras públicas e a popularização desses produtos, que ganharam tração entre seguidores que buscavam alternativas aos tratamentos convencionais. A partir de 2019, termos como “grupo” e “protocolo” sugerem a formação de comunidades dedicadas ao compartilhamento de informações sobre esses tratamentos, estabelecendo redes de apoio que promovem o uso de CDS como uma solução supostamente eficaz contra uma variedade de doenças. Durante a Pandemia da COVID-19, em 2020 e 2021, as menções a “vacina” e “covid” aumentam significativamente, integrando essas substâncias ao debate antivacinal, onde o dióxido de cloro foi frequentemente promovido como uma alternativa para “desintoxicar” o corpo dos supostos efeitos nocivos das vacinas. Em 2022 e 2023, “MMS”, “grupo” e “protocolo” permanecem centrais, refletindo a persistência dessas redes e o foco contínuo em tratamentos que prometem “purificar” o organismo. A constância desses termos em 2024 sugere que, mesmo com a diminuição da Pandemia, a crença equivocada na eficácia do MMS e CDS continua viva em diversas comunidades, mostrando a resiliência dessas narrativas de desinformação e a dificuldade de combatê-las dentro dessas bolhas de crença.

### 3.4. Sobreposição de agenda temática

As figuras a seguir mostram não apenas a centralidade das discussões sobre saúde dentro dessas comunidades, mas também como esses tópicos se interconectam com uma ampla gama de outras teorias conspiratórias e disputas culturais. A análise visual das figuras permite identificar como cada temática atua como um ponto de convergência para diferentes narrativas de desinformação, amplificando o alcance dessas comunidades e reforçando suas crenças centrais. Ao integrar temas variados como geopolítica, esoterismo, e guerras culturais, essas comunidades conseguem criar um discurso coeso que atrai e mantém seus membros, dificultando a intervenção e a correção factual.

**Figura 10.** Temáticas de vacinação e Pandemia

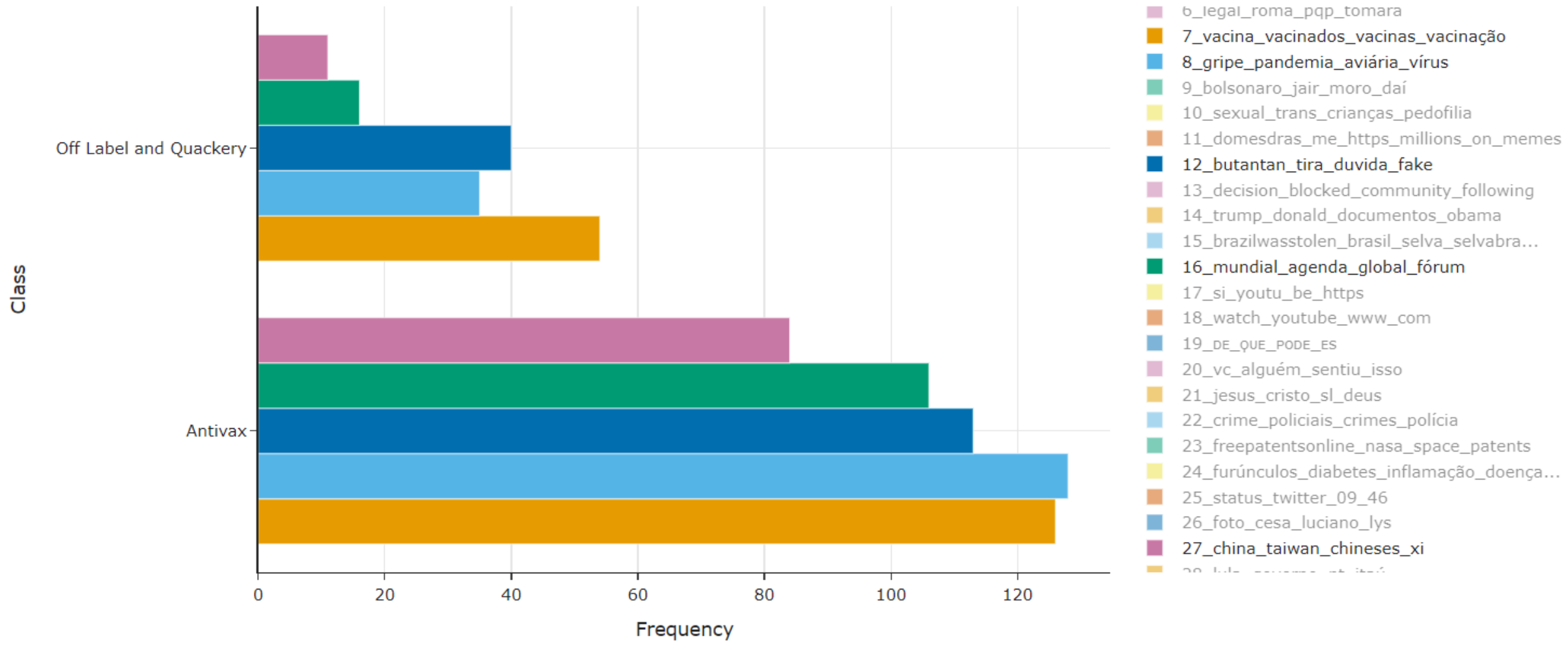

Fonte: Elaboração própria (2024).

A figura ilustra como as temáticas relacionadas à vacinação e à Pandemia estão interligadas com outras narrativas, ampliando o escopo do discurso antivacinas. Notavelmente, tópicos como “Butantan”, “fake” e “vacina”, “vacinados”, “vacinas”, “vacinação” destacam-se, sugerindo que essas comunidades não apenas questionam a segurança e eficácia das vacinas, mas também promovem uma desconfiança generalizada em relação às instituições de saúde pública, como o Instituto Butantan. A referência à narrativa “mundial”, “agenda”, “global”, “fórum” indica uma sobreposição com teorias de conspiração global, onde a vacinação é vista como parte de uma agenda oculta para o controle populacional. Essa integração de narrativas externas reflete a estratégia dessas comunidades de vincular suas crenças centrais a temas amplamente discutidos, utilizando-se da Pandemia como um catalisador para reforçar e disseminar teorias de desinformação.

**Figura 11.** Temáticas de disputas geopolíticas

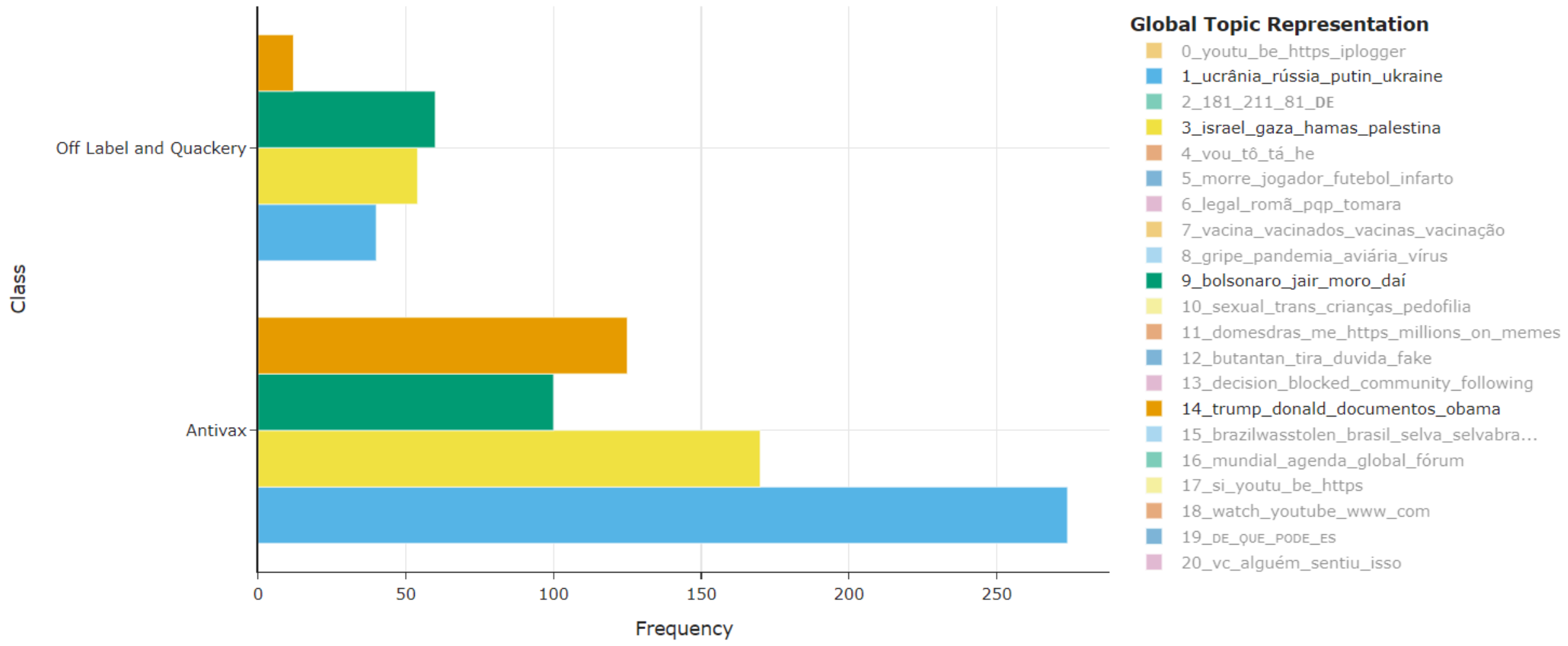

Fonte: Elaboração própria (2024).

Esta figura evidencia como as discussões antivacinas e de medicamentos *off label* são frequentemente associadas a narrativas geopolíticas, refletindo a politização do discurso. O destaque para o tópico "Ucrânia", "Rússia", "Putin", "Ukraine" sugere que essas comunidades utilizam conflitos geopolíticos como elementos e evidências para validar suas teorias de conspiração, relacionando temáticas. A inclusão de "Bolsonaro", "Jair" e "Moro" mostra a politização interna dessas discussões, onde políticos locais são posicionados como figuras centrais que influenciam ou são influenciados por essa suposta conspiração global. Essas sobreposições indicam que as comunidades antivacinas não se limitam a temas de saúde, mas incorporam disputas políticas e internacionais para fortalecer suas narrativas e ampliar seu alcance, utilizando eventos globais como ferramentas para reforçar a desconfiança em relação a instituições e políticas de saúde.

**Figura 12.** Temáticas de medicamentos *off label*

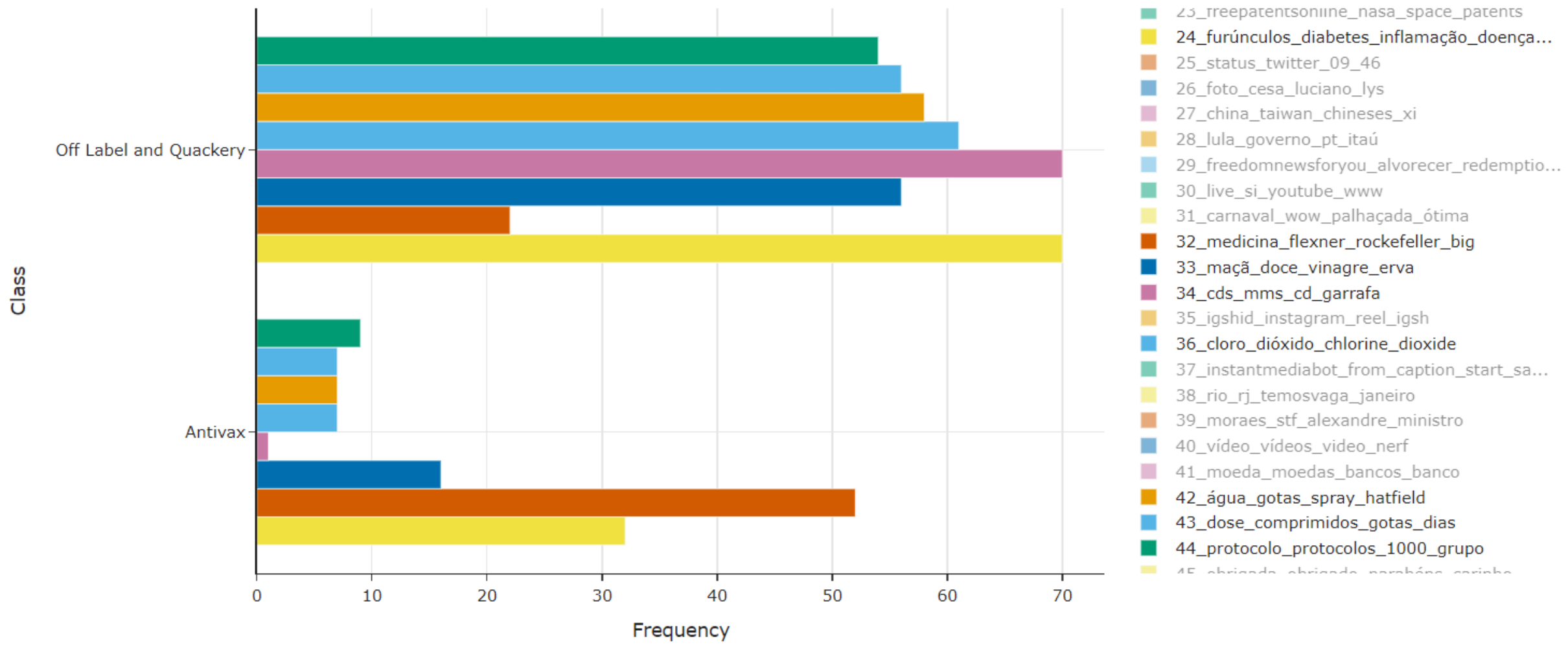

Fonte: Elaboração própria (2024).

Nesta figura, os tópicos de medicamentos *off label* mostram como a promoção de tratamentos alternativos, como MMS e CDS, está conectada a outras narrativas que questionam a medicina convencional. O tópico “medicina”, “Flexner”, “Rockefeller”, “Big” reflete uma crítica às fundações, frequentemente retratada nessas comunidades como ferramenta de controle pelas elites. “Cloro”, “dióxido”, “chlorine”, “dioxide” é diretamente ligado à promoção do CDS como uma solução alternativa às vacinas, reforçando a desconfiança na ciência tradicional. Essas sobreposições de narrativas mostram como as comunidades antivacinas e de medicamentos *off label* constroem uma visão de mundo onde tratamentos convencionais são vistos como perigosos ou manipuladores, e alternativas “naturais” são promovidas como as verdadeiras soluções, mesmo sem bases científicas.

**Figura 13.** Temáticas de guerras culturais e intersecção com vacinação

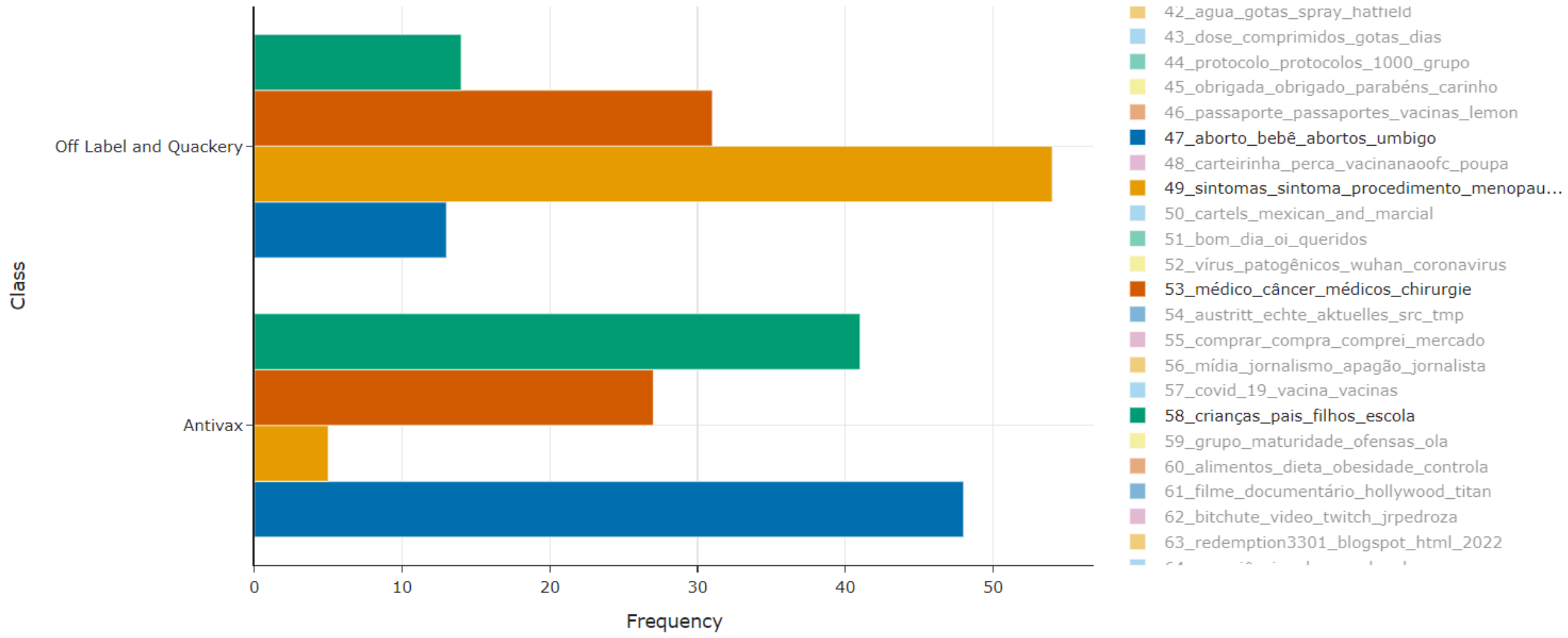

Fonte: Elaboração própria (2024).

Esta figura destaca como temas de guerras culturais estão entrelaçados com as narrativas antivacinas e de medicamentos *off label*. Os tópicos “aborto”, “bebê”, “abortos”, “umbigo” e “crianças”, “pais”, “filhos”, “escola” indicam como essas comunidades frequentemente integram narrativas de guerra cultural, como o debate sobre direitos reprodutivos e educação infantil, ao discurso antivacinas. Em alguns casos, chegam a afirmar que estariam sendo usadas células de fetos abortados para fazer vacinas, ou ainda que vacinas poderiam causar aborto espontâneo em mulheres. Isso não só amplia o escopo da desinformação, mas também reforça a ideia de que as vacinas são parte de uma agenda mais ampla que visa corromper ou controlar valores tradicionais e familiares. A interseção dessas narrativas permite que a resistência às vacinas seja vista como parte de uma defesa mais ampla de valores morais e culturais, criando um campo comum entre diferentes grupos que se opõem às mudanças sociais e políticas percebidas como ameaças. Assim, as discussões sobre saúde se tornam inseparáveis de debates culturais, solidificando o compromisso dos membros dessas comunidades com a causa antivacinas, enquanto também se engajam em uma cruzada.

**Figura 14.** Temáticas de outras teorias da conspiração gerais

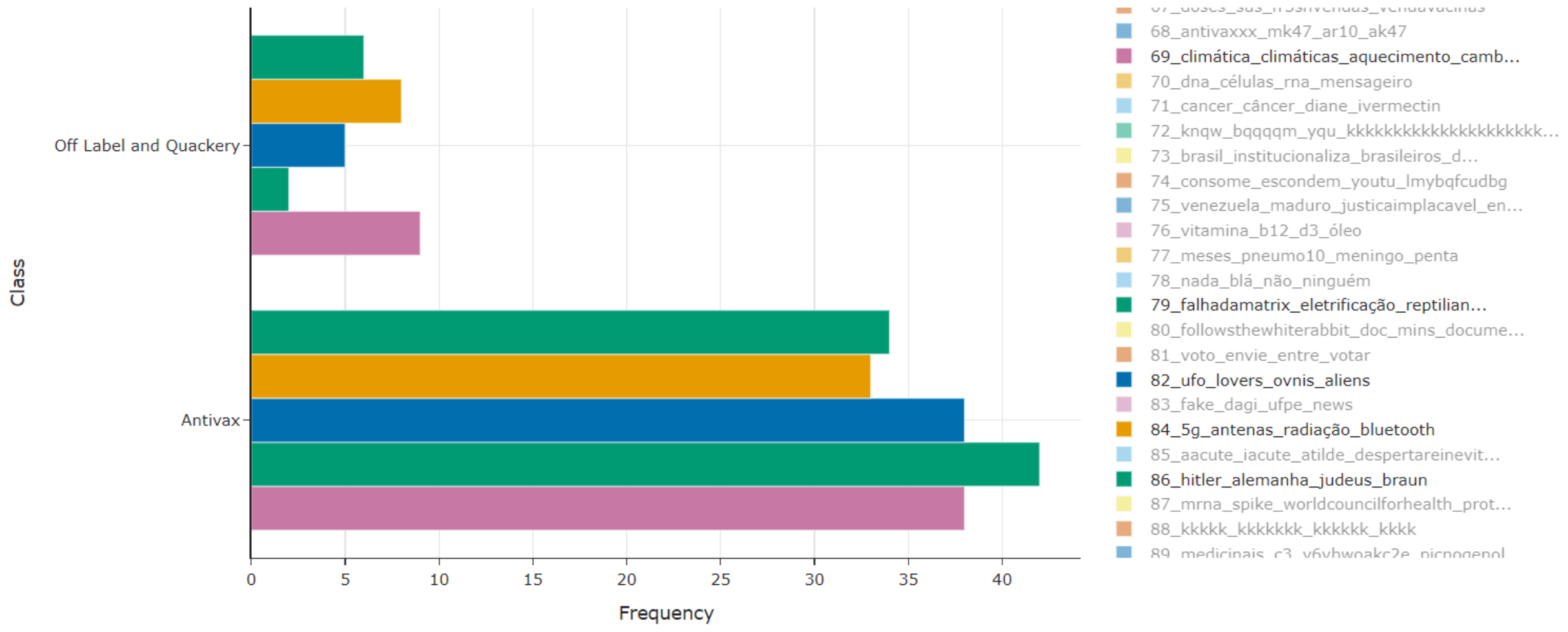

Fonte: Elaboração própria (2024).

A figura sobre teorias da conspiração gerais revela como as comunidades antivacinas e de medicamentos *off label* se entrelaçam com uma ampla gama de outras teorias. O destaque para o tópico “falhamatrix”, “eletrificação”, “reptilian” indica a inclusão de teorias mais extremas, como a crença em reptilianos ou a existência de uma matrix, no discurso antivacinas. Isso demonstra como essas comunidades ampliam suas narrativas centrais ao integrar teorias que, embora aparentemente desconectadas, reforçam uma visão de mundo onde as vacinas são parte de um plano mais vasto de dominação global. Ao associar as vacinas a ideias de controle por elites ocultas, essas comunidades conseguem atrair indivíduos que já estão predispostos a acreditar em outras formas de conspiração, criando uma rede interligada de desinformação que se sustenta e se reforça mutuamente.

**Figura 15.** Temáticas de uma falsa relação entre vacinas e autismo

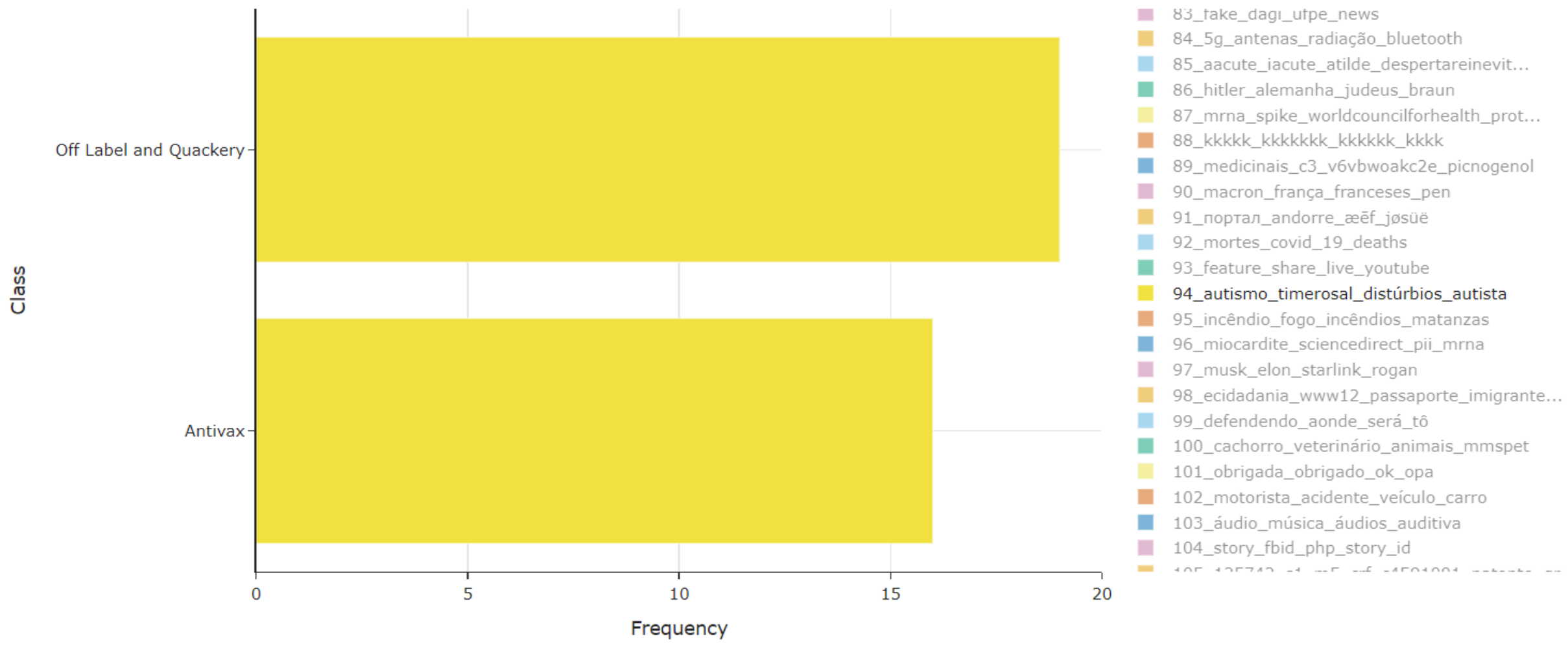

Fonte: Elaboração própria (2024).

A figura que trata das temáticas de uma falsa relação entre vacinas e autismo evidencia como essa narrativa específica continua a ser uma peça retórica dentro das comunidades antivacinas e de medicamentos *off label*. O tópico “autismo”, “timerosal”, “distúrbios”, “autista” aparece com frequência significativa tanto nas discussões antivacinas quanto nas de tratamentos alternativos. Essa persistência reflete a resiliência dessa teoria conspiratória, que tem sido amplamente desmentida pela comunidade científica, mas continua a ser promovida por essas comunidades como uma verdade oculta. A ideia de que as vacinas causam autismo serve como um ponto de convergência para diferentes narrativas de desinformação, criando uma base comum que conecta temas de saúde a questões mais amplas de desconfiança em relação à medicina convencional e às autoridades. Essa sobreposição temática mostra como as comunidades antivacinas e de medicamentos *off label* utilizam teorias antigas e desmentidas para sustentar suas crenças centrais, ao mesmo tempo em que recrutam novos membros para suas fileiras, ampliando o ciclo de desinformação.

## 4. Reflexões e trabalhos futuros

Para responder a pergunta de pesquisa “**como são caracterizadas e articuladas as comunidades de teorias da conspiração brasileiras sobre temáticas de antivacinas (*antivax*) e tratamentos e medicamentos alternativos (*off label*) no Telegram?**”, este estudo adotou técnicas espelhadas em uma série de sete publicações que buscam caracterizar e descrever o fenômeno das teorias da conspiração no Telegram, adotando o Brasil como estudo de caso. Após meses de investigação, foi possível extrair um total de 195 comunidades de teorias da conspiração brasileiras no Telegram sobre temáticas de antivacinas (*antivax*) e tratamentos e medicamentos alternativos (*off label*), estas somando 5.406.762 de conteúdos publicados entre maio de 2017 (primeiras publicações) até agosto de 2024 (realização deste estudo), com 440.651 usuários somados dentre as comunidades.

Foram adotadas quatro abordagens principais: **(i)** Rede, que envolveu a criação de um algoritmo para mapear as conexões entre as comunidades por meio de convites circulados entre grupos e canais; **(ii)** Séries temporais, que utilizou bibliotecas como "Pandas" (McKinney, 2010) e "Plotly" (Plotly Technologies Inc., 2015) para analisar a evolução das publicações e engajamentos ao longo do tempo; **(iii)** Análise de conteúdo, sendo aplicadas técnicas de análise textual para identificar padrões e frequências de palavras nas comunidades ao longo dos semestres; e **(iv)** Sobreposição de agenda temática, que utilizou o modelo BERTopic (Grootendorst, 2020) para agrupar e interpretar grandes volumes de textos, gerando tópicos coerentes a partir das publicações analisadas. A seguir, as principais reflexões são detalhadas, sendo seguidas por sugestões para trabalhos futuros.

### 4.1. Principais reflexões

**Temáticas como Nova Ordem Mundial e Apocalipse e Survivalism são as principais portas de entrada para narrativas antivacinas:** As principais comunidades que atuam como portas de entrada para as narrativas antivacinas incluem Nova Ordem Mundial e Apocalipse e Sobrevivência. Essas comunidades direcionaram um número significativo de links para as comunidades antivacinas, com a Nova Ordem Mundial gerando 6.091 links e Apocalipse e Survivalism 1.920 links. Isso indica que essas temáticas conspiratórias alimentam a desconfiança em vacinas, ao conectá-las com ideias de controle global;

**Globalism, General Conspiracy e Nova Ordem Mundial são as principais comunidades que recebem convites das comunidades antivacinas:** Entre as comunidades que mais recebem convites a partir das comunidades antivacinas estão Globalism, com 10.066 links, General Conspiracy com 9.403 links, e Nova Ordem Mundial com 6.260 links. Esses números mostram que uma vez dentro das comunidades antivacinas, os membros são rapidamente expostos a uma rede mais ampla de desinformação, onde teorias relacionadas ao globalismo e a conspirações generalizadas são fortemente promovidas, fortalecendo o discurso antivacinas com outras teorias conspiratórias;

**Ocultismo e Esoterismo são as maiores fontes de convites para comunidades de medicamentos *off label*:** No contexto das comunidades de medicamentos *off label*, Ocultismo e Esoterismo se destacam como a principal fonte de convites, gerando 7.367 links para essas comunidades. Isso sugere uma forte interconexão entre crenças esotéricas e a promoção de medicamentos *off label*, onde práticas alternativas e não científicas encontram terreno fértil para propagação a partir da manipulação da crença esotérica de usuários;

**Ocultismo e Esoterismo, Globalism e Nova Ordem Mundial recebem o maior número de convites a partir das comunidades de medicamentos *off label*:** As principais comunidades que recebem convites das comunidades de medicamentos *off label* incluem Ocultismo e Esoterismo com 2.774 links, Globalism com 2.521 links, e Nova Ordem Mundial com 1.948 links. Esses dados indicam que a promoção de medicamentos *off label* está intimamente ligada a narrativas esotéricas e teorias de conspiração global, criando uma rede de desinformação coesa e interconectada;

**As narrativas antivacinas experimentaram um aumento de 290% durante a Pandemia, evidenciando uma interconectividade crescente com outras teorias**

**conspiratórias:** Durante o auge da Pandemia da COVID-19, as discussões sobre antivacinas experimentaram um crescimento de 290% em relação ao período pré-Pandemia. Essa expansão não se limitou às temáticas de saúde, mas também reforçou a interconectividade com outras teorias, como Globalismo e Nova Ordem Mundial, demonstrando como a crise sanitária foi utilizada para promover uma agenda mais ampla de desinformação;

**A forte sobreposição de pautas entre antivacinas e outras teorias de conspiração alimenta uma rede interdependente de desinformação:** Comunidades antivacinas e de medicamentos *off label* não operam de forma isolada; há uma significativa sobreposição de pautas com outras teorias de conspiração, como Globalismo e Nova Ordem Mundial. Por exemplo, Globalismo foi responsável por 10.066 links de convites para grupos antivacinas, evidenciando uma rede interdependente onde diferentes narrativas se reforçam mutuamente, criando um ciclo contínuo de desinformação;

**A mentirosa e desonesta narrativa que associa vacinas ao autismo permanece uma das mais persistentes e frequentemente reintroduzidas nas discussões:** Mesmo após repetidos desmentidos científicos, a falsa narrativa que liga vacinas ao autismo continua sendo uma das mais discutidas nas comunidades antivacinas e de medicamentos *off label*, onde são apresentadas supostas curas ao autismo. A análise revelou que essa narrativa foi constantemente reintroduzida nas discussões para reforçar a desconfiança generalizada em relação à medicina convencional, destacando sua persistência nas comunidades analisadas;

**As comunidades antivacinas atuam como *hubs* centrais na disseminação de desinformação, conectando e amplificando múltiplas narrativas:** O estudo identificou que algumas comunidades antivacinas funcionam como *hubs* centrais dentro da rede de desinformação, com uma capacidade desproporcional de influenciar o discurso. Essas comunidades conectam e amplificam múltiplas narrativas conspiratórias, funcionando como pontos de convergência onde diferentes teorias se encontram e se reforçam mutuamente;

**A forte interconectividade entre discussões antivacinas e de medicamentos *off label* revela uma bolha ideológica coesa e resistente:** A interconectividade entre as discussões antivacinas e de medicamentos *off label*, como MMS e CDS, é notável. As mesmas comunidades que promovem a desinformação sobre vacinas também estão frequentemente na vanguarda da promoção de tratamentos alternativos perigosos. Isso cria uma sobreposição de crenças dentro de uma bolha ideológica coesa, facilitando a disseminação de desinformação;

**A interseção entre desinformação sobre saúde e narrativas esotéricas cria uma rede altamente influente e atraente para novos membro e monetiza com a venda de fármacos irregulares e produtos químicos com o dióxido de cloro:** O estudo revela que a combinação de desinformação sobre saúde, como as narrativas antivacinas e *off label*, com temas esotéricos como Ocultismo e Esoterismo, não apenas fortalece a coesão dentro dessas comunidades, mas também as torna mais atrativas para novos membros. Ao misturar teorias de conspiração globais com crenças alternativas, essas comunidades conseguem criar uma rede de desinformação que é ao mesmo tempo complexa e atraente, aumentando seu alcance e a dificuldade de desmantelamento dessas narrativas.

### 4.2. Trabalhos futuros

Com base nos principais achados deste estudo, várias direções podem ser sugeridas para futuros estudos. Primeiramente, a exploração de pontes conspiratórias entre temas globais e narrativas antivacinas merece atenção. Sabendo que temáticas como Nova Ordem Mundial e Apocalipse e Sobrevivência são portas de entrada significativas para as narrativas antivacinas, investigações futuras poderiam examinar como essas teorias evoluem e se entrelaçam com narrativas de saúde em diferentes contextos culturais. Além disso, seria interessante explorar como novas teorias conspiratórias emergentes, como "Great Reset" ou "5G" como ondas de suposta dominação de mentes globalmente, pois poderiam desempenhar papéis semelhantes na perpetuação de desinformação sobre vacinas;

Outro ponto relevante é o mapeamento das estruturas de rede e a disseminação de desinformação. Dado que comunidades como Globalismo, Conspirações Gerais e Nova Ordem Mundial recebem um grande número de convites das comunidades antivacinas, futuros estudos poderiam se concentrar em mapear mais detalhadamente as dinâmicas internas dessas comunidades. Compreender como essas estruturas facilitam a disseminação de desinformação e identificar os indivíduos ou grupos que agem como "*super spreaders*" dessas narrativas pode fornecer *insights* valiosos para a interrupção dessas redes.

A interseção entre esoterismo e medicamentos *off label* também é uma área promissora para futuras pesquisas. Considerando que Ocultismo e Esoterismo são as principais fontes de convites para comunidades de medicamentos *off label*, investigações poderiam se focar na psicologia por trás da atração por práticas esotéricas e alternativas. Estudos que examinem a eficácia das campanhas de correção factual nesses contextos específicos seriam essenciais para desenvolver estratégias mais eficientes de combate à desinformação entre populações que aderem a práticas não científicas.

Futuros estudos também poderiam explorar a resistência e a persistência de narrativas falsas ao longo do tempo. A narrativa falsa que associa vacinas ao autismo, que continua a ser persistentemente reintroduzida nas discussões, indica a necessidade de estratégias para desmantelar essas crenças profundamente enraizadas. Pesquisas poderiam se concentrar em como essas narrativas, mesmo desmentidas, conseguem ressurgir e ganhar tração novamente, proporcionando subsídios para intervenções mais eficazes nas plataformas digitais.

Além disso, a análise da interconectividade e a criação de bolsões ideológicos coesos são aspectos críticos para entender como as comunidades antivacinas e de medicamentos *off label* interagem. Futuros estudos poderiam investigar como essas bolhas se formam e se mantêm, além de explorar como diferentes comunidades colaboram ou competem por influência dentro da rede maior de desinformação. Compreender as barreiras que dificultam a entrada de informação científica nesses bolsões seria fundamental para o desenvolvimento de intervenções mais eficazes no enfrentamento à desinformação nociva.

Estudar as estruturas de retroalimentação entre narrativas conspiratórias também é essencial. A sobreposição de pautas entre antivacinas e outras teorias de conspiração sugere que há um ciclo de retroalimentação de desinformação. Pesquisas futuras poderiam se concentrar em como essas narrativas se reforçam mutuamente e como esses ciclos podem ser interrompidos, oferecendo estratégias para desacelerar a propagação dessas crenças errôneas.

Por fim, é importante explorar novos métodos para identificar e neutralizar *hubs* de desinformação. Futuros estudos poderiam se concentrar em desenvolver técnicas para detectar e limitar a influência desses *hubs* antes que eles ganhem tração. Isso poderia incluir a criação de algoritmos ou ferramentas que permitam monitorar e intervir em tempo real, especialmente durante crises globais, onde a desinformação pode ter consequências graves.

## 5. Referências

## 6. Biografia do autor

**Ergon Cugler de Moraes Silva** possui mestrado em Administração Pública e Governo (FGV), MBA pós-graduação em Ciência de Dados e Análise (USP) e bacharelado em Gestão de Políticas Públicas (USP). Ele está associado ao Núcleo de Estudos da Burocracia (NEB FGV), colabora com o Observatório Interdisciplinar de Políticas Públicas (OIPP USP), com o Grupo de Estudos em Tecnologia e Inovações na Gestão Pública (GETIP USP), com o Monitor de Debate Político no Meio Digital (Monitor USP) e com o Grupo de Trabalho sobre Estratégia, Dados e Soberania do Grupo de Estudo e Pesquisa sobre Segurança Internacional do Instituto de Relações Internacionais da Universidade de Brasília (GEPSI UnB). É também pesquisador no Instituto Brasileiro de Informação em Ciência e Tecnologia (IBICT), onde trabalha para o Governo Federal em estratégias contra a desinformação. Brasília, Distrito Federal, Brasil. Site: https://ergoncugler.com/.